\documentclass[
prd,preprint,aps,endfloats,tightenlines,
showpacs,superscriptaddress,nofootinbib,
amssymb
]{revtex4}
\usepackage{graphicx}
\usepackage{ulem}
\usepackage{color}
\usepackage{bm}
\begin{document}
\setlength{\baselineskip}{16pt}
%
%---------------------------------------------------------------------------
%
\title{
Non-perturbative evaluation for anomalous dimension in 2-dimensional $O(3)$ sigma model
}
%
%----------------------------------
\author{
Sergio Calle Jimenez
}
\affiliation{
Department of Physics, Tokyo Institute of Technology,
Tokyo 152-8551, Japan
}
%
%----------------------------------
\author{
Makoto Oka
}
\affiliation{
Department of Physics, Tokyo Institute of Technology,
Tokyo 152-8551, Japan
}
\affiliation{
Advanced Science Research Center, Japan Atomic Energy Agency, 
Tokai, Ibaraki 319-1195, Japan
}
%
%----------------------------------
\author{
Kiyoshi Sasaki
}
\affiliation{
Department of Physics, Tokyo Institute of Technology,
Tokyo 152-8551, Japan
}
%
%----------------------------------
%
% @@ =======================================================================
%
\begin{abstract}
We calculate 
the wave-function renormalization in 2-dimensional $O(3)$ sigma model, 
non-perturbatively. 
It is evaluated in a box with a finite spatial extent. 
We determine the anomalous dimension in the finite volume scheme 
through an analysis of the step scaling function. 
Results are compared with a perturbative evaluation, 
and the reasonable behavior is confirmed. 
\end  {abstract}
\pacs{ 11.10.Hi, 11.10Kk, 11.10.Lm, 11.15.Ha }
\maketitle
%
% @@ =======================================================================
%
\section  {Introduction}
\label{sec:Introduction}
In the late '80s and the early '90s, 
some people have developed methods to extract information at infinite volume 
from the information on a system in a finite box. 
One     is to calculate the scattering phase shift    \cite{Luscher.ps}, and 
another is to determine the running coupling constant \cite{Luscher.rc}. 
The latter prescribes 
a renormalization group (RG) evolution of the renormalized coupling 
as a response to change of the box size. 
The workability of this method might give us 
a foresight of the existence of RG equation 
which describes the box-size dependence of $N$-point Green function. 
It will be useful to describe a spatially extended object in a finite box. 
An example  is the deuteron, and 
another one is the Efimov system \cite{Efimov:1970zz}. 
For the latter, 
the infinity of the object size is essential, 
so that we cannot avoid a discussion of the box-size dependence 
in a study with a lattice simulation.

The end goal of our study is to establish such a RG equation. 
For this purpose, we use the 2-dimensional $O(n)$ sigma model. 
Until now, however, 
the wave-function renormalization has not been discussed 
in the context where the system has been put in a finite box. 
For the establishment of the RG equation, 
one might need 
information on the scale dependence of wave-function renormalization, 
which is called anomalous dimension, 
The present work aims to give a non-perturbative evaluation 
for the anomalous dimension as a first step for constructing the RG equation.

Before entering the main issue, 
we summarize the basic properties on the $2$-dimensional $O(n)$ sigma model 
\cite{GellMann:1960np} 
from the standpoint of perturbative studies. 
In the middle '70s, 
the renormalizability, and 
the asymptotic freedom 
have been established \cite{Polyakov:1975rr,Migdal:1975zf}. 
The asymptotic freedom is important 
to guarantee the applicability of the perturbative expansion 
in the high-energy region. 
Then, the renormalization above $2$ dimension 
has been also discussed \cite{Brezin_Zinn-Justin,Brezin:1976ap}. 
The infrared (IR) divergence is regularized with a magnetic field 
in these studies.  
A vanishment of the magnetic field activates the IR divergence, again. 
However, 
the renormalization in minimal subtraction (MS) scheme 
\cite{tHooft:1973mfk} requires 
only the cancellation of ultraviolet (UV) divergence. 
If we restrict ourselves to evaluate the renormalization factor, 
the IR divergence is not a difficulty. 
On the other hands, if we focus on a physical quantity like a mass gap, 
the IR divergence is a serious problem.

The IR divergence originates from the low dimensionality of the system. 
Mermin-Wagner's theorem forbids 
spontaneous symmetry breaking in $2$ dimension \cite{Mermin:1966fe}. 
The perturbative expansion with a fixed direction of the magnetization 
cannot be justified. 
As the results, 
the IR divergence remains in the the Green function of the pion mode. 
It has been conjectured by Elitzur \cite{Elitzur:1978ww}
and proved by David \cite{David:1980rr}
that 
the $O(n)$-invariant Green function including the sigma mode is IR finite. 
Then, 
the $O(n)$-invariant Green function in a finite box has been discussed, 
and the value of the mass gap has been evaluated 
\cite{Luscher:1982uv,Hasenfratz:1984jk,Floratos:1984bz,Brezin:1985xx,Luscher.rc}. 
In these studies, 
the particular attention is payed for the treatment of the zero mode. 
There are several methods. 
One is to add compact extra dimensions 
\cite{Luscher:1982uv}. 
Another is to separate the collective motion of the magnetization 
\cite{Hasenfratz:1984jk}. 
Another is to use a background field 
\cite{Floratos:1984bz,Brezin:1985xx}. 
The other is to introduce the IR-cutoff mass 
by using the lattice regularization \cite{Luscher.rc}. 
We add that 
some works are relatively recently performed besides these works 
\cite{Shin:1996gi,Niedermayer:2010mx,Niedermayer:2016yll,Niedermayer:2016ilf}.

At present, 
we have no difficulty to evaluate 
the mass gap of the $2$-dimensional $O(n)$ sigma model in a finite box 
regardless of perturbatively or non-perturbatively. 
The mass gap can be used to define the renormalized coupling, 
and discuss the scale dependence of it \cite{Luscher.rc}. 
On the other hands, 
having the $O(n)$-invariant $2$-point Green function, 
it is also easy to evaluate the amplitude. 
Then, 
we can define the wave-function renormalization from the amplitude, 
and discuss the scale dependence. 
This is just the theme addressed in the present study.

This article is organized as follows. 
In Sec. \ref{sec:Model__renormalization_shceme_and_scale_dependence}, 
we introduce the $2$-dimensional $O(n)$ sigma model. 
Then, we briefly describe 
a renormalization in the finite volume scheme and 
the procedure to determine the evolution of parameters of the theory. 
In Sec. \ref{sec:Details_of_Monte_Carlo_simulation}, 
the details of Monte Carlo simulation are explained. 
In Sec. \ref{sec:Numerical_results}, 
we show the results for the wave-function renormalization
in addition to results for the renormalized coupling. 
We also discuss their scale dependence. 
In Sec. \ref{sec:Conclusion}, our conclusions are given. 
%
% @@ ==================================================================
%
\section  {Model, renormalization scheme and scale dependence}
\label{sec:Model__renormalization_shceme_and_scale_dependence}
%
% @@@ =========================================================================
%
\subsection{2-dimensional $O(n)$ sigma model}
\label{ssec:2-dimensional_On_sigma_model}
The 2-dimensional $O(n)$ sigma model 
is formally prescribed by the euclidean action 
\begin{equation}
  S[{\bm \phi}]
  =
  \frac{1}{2g^2}\,\int d^2 x\ 
  \partial_\mu {\bm \phi}(x)\cdot\partial_\mu {\bm \phi}(x)\quad
  (\mu=0,1)\ .
\label{eqn:2d_action}
\end  {equation}
Here ${\bm \phi}(x)=(\,\phi_i(x),\,i=1,\cdots,n\,)$ 
is the $n$-component field with the constraint 
\begin{equation}
  {\bm \phi}(x)\cdot {\bm \phi}(x) = 1\ . 
\end  {equation}
The bullet point symbol denotes 
the scalar product of $n$-component vectors. 
The system is put on a finite box 
with the temporal extent $T$ and the spatial extent $L$. 
We assume that $T$ is sufficiently large compared to $L$. 
In the present work, 
we impose 
the Neumann  boundary condition (NBC) for the temporal direction, and 
the periodic boundary condition (PBC) for the spatial  direction, 
\begin{equation}
  \frac{\partial}{\partial x_0}\,{\bm \phi}(x_0,x_1) = 0\quad
  (x_0 \in \partial\Lambda_\tau)\ ,\quad
  {\bm \phi}(x_0,x_1+Ln) = {\bm \phi}(x_0,x_1)\quad(n\in\mathbb{Z})\ ,
\end  {equation}
where $\partial\Lambda_\tau$ represents the temporal boundary. 
The reason why we use the NBC in the temporal direction 
is to realize the $O(n)$-invariant state at $\partial\Lambda_\tau\,$. 
As the results, 
states other than the spin-$1$ state 
do not contribute the $2$-point Green function. 
The NBC is called also the free boundary condition.

As we have said in Sec. \ref{sec:Introduction}, 
the $O(n)$-invariant Green function should be used to avoid the IR divergence. 
The $O(n)$-invariant $2$-point Green function is defined by 
\begin{equation}
  G_{\rm inv}(x;y) = \langle\,{\bm \phi}(x)\cdot{\bm \phi}(y)\,\rangle\ ,
\label{eqn:def_inv_Green}
\end  {equation}
where the angle blacket refers to the expectation values 
over the configurations of ${\bm \phi}$ field. 
We consider the zero-momentum projected Green function, 
\begin{equation}
  G_{\rm inv}(x_0;y_0)
  =
  \frac{1}{L^2}\int dx_1 dy_1\ 
  {\rm e}^{-ip_1(y_1-x_1)}\,G_{\rm inv}(x;y)\,\Big|_{p_1=0}\ .
\label{eqn:def_0_mom_projected_Green_func}
\end  {equation}
With the NBC for the temporal direction, it can be written as 
\begin{equation}
  G_{\rm inv}(x_0;y_0)
  =
  A\,{\rm e}^{-M|y_0-x_0|}
  +
  {\cal O}({\rm e}^{-(4\pi/L)|y_0-x_0|})\ .
\label{eqn:0_mom_projected_Green_func}
\end  {equation}
For our purpose, 
the mass gap $M$ and the amplitude $A$ are needed. 
Energies of the excited states are known to be at least $4\pi/L$. 
With a finite $L$, 
they are large enough to ignore the contributions to $G_{\rm inv}(x_0;y_0)$. 
%
% @@@ =========================================================================
%
\subsection{Renormalization scheme}
\label{ssec:Renormalization_scheme}
The renormalization of the $2$-dimensional $O(n)$ sigma model 
is done by the replacement 
\begin{eqnarray}
  g^2
  &=&
  Z_{\rm R}^g\,g_{\rm R}^2\ ,
\\
  {\bm \phi}(z)
  &=&
 (Z_{\rm R}^\phi)^{1/2}\,
  {\bm \phi}_{\rm R}(z)\ . 
\end  {eqnarray}
Here     $g_{\rm R}^2$  is the renormalized coupling, and 
${\bm \phi}_{\rm R}(z)$ is the renormalized field. 
We hava a finite arbitrariness for the choice of them. 
This arbitrariness can be removed by setting values of 
$g_{\rm R}^2$ and the wave-function renormalization $Z_{\rm R}^\phi$ 
at an energy scale $\mu$. 
Such conditions are called renormalization conditions. 
The definition of $\mu$ in the MS scheme 
is given in Sec. \ref{app:Perturbative_evaluation}.

In the present study, 
we consider the renormalization conditions at $\mu=1/L$ as 
\begin{eqnarray}
  \frac{n-1}{2L}\, 
  g_{\rm FV}^2   (\mu)\,|_{\mu=1/L}
  &=&
  M\ ,
\label{eqn:def_FS1}
\\
  Z_{\rm FV}^\phi(\mu)\,|_{\mu=1/L}
  &=&
  A\ .
\label{eqn:def_FS2}
\end  {eqnarray}
$M$ and $A$ are the mass gap and amplitude, respectively, 
which are determined from the measured value of 
the $O(n)$-invariant $2$-point Green function. 
The renormalized coupling in Eq. (\ref{eqn:def_FS1}) 
has been first proposed in Ref. \cite{Luscher.rc}, 
and called the finite volume (FV) coupling. 
In the following, 
we refer Eqs. (\ref{eqn:def_FS1}) and (\ref{eqn:def_FS2}) 
by renormalization in the FV scheme. 
As long as there is no confusion, 
we write the argument of $g_{\rm FV}^2$ and $Z_{\rm FV}^\phi$
by $L$, but not $1/L$.

The $\beta$ function and the anomalous dimension describe 
$\mu$ dependence of the renormalized parameters 
with the fixed bare parameters $g^2$ and ${\bm \phi}(x)$. 
They are defined as 
\begin{eqnarray}
  \beta_{\rm R} (g_{\rm R}^2)
  &\equiv &
  \mu\,\frac{d}{d\mu}\,  g_{\rm R}^2   (\mu)\ ,
\label{eqn:RG_bt}
\\
  \gamma_{\rm R}(g_{\rm R}^2)
  &\equiv &
  \mu\,\frac{d}{d\mu}\ln Z_{\rm R}^\phi(\mu)\ ,
\label{eqn:RG_gm}
\end  {eqnarray}
respectively. 
In the FV scheme, they are written as  
\begin{eqnarray}
  \beta_{\rm FV} (g_{\rm FV}^2)
  &=&
  -L\,\frac{d}{dL}\,  g_{\rm FV}^2   (L)\ ,
\label{eqn:RG2_bt}
\\
  \gamma_{\rm FV}(g_{\rm FV}^2)
  &=&
  -L\,\frac{d}{dL}\ln Z_{\rm FV}^\phi(L)\ ,
\label{eqn:RG2_gm}
\end  {eqnarray}
by using $\mu=1/L$ to Eqs. (\ref{eqn:RG_bt}) and (\ref{eqn:RG_gm}). 
%
% @@@ =========================================================================
%
\subsection{Scale dependence}
\label{ssec:Scale_dependence}
The step scaling function (SSF) describes 
how parameters of a theory evolves when the scale is changed. 
We consider two types of SSFs, $\sigma^g$ and $\sigma^\phi$. 
They are defined through 
\begin{eqnarray}
  g_{\rm FV}^2(sL) 
  &=& 
  \sigma^g   (s,g_{\rm FV}^2(L))\ ,
\label{eqn:def_SSF_g2}
\\
  Z_{\rm FV}^\phi(sL)
  &=&
  \sigma^\phi(s,g_{\rm FV}^2(L))\ Z_{\rm FV}^\phi(L)\ ,
\label{eqn:def_SSF_ph}
\end  {eqnarray}
with a scaling factor $s$. 
$\sigma^g   (s,g_{\rm FV}^2)$ is proposed in 
Ref. \cite{Luscher.rc}, and 
$\sigma^\phi(s,g_{\rm FV}^2)$ is motivated from 
Ref. \cite{Capitani:1998mq}. 
They are related to 
$\beta_{\rm FV}(g_{\rm FV}^2)$ and $\gamma_{\rm FV}(g_{\rm FV}^2)$ by 
\begin{eqnarray}
   \beta_{\rm FV}(\sigma^g(s,g_{\rm FV}^2))
  &=&
  - s\,\frac{\partial\,   \sigma^g   (s,g_{\rm FV}^2)}{\partial s}\ ,
\label{eqn:rel1}
\\
  \gamma_{\rm FV}(\sigma^g(s,g_{\rm FV}^2))
  &=&
  - s\,\frac{\partial\,\ln\sigma^\phi(s,g_{\rm FV}^2)}{\partial s}\ ,
\label{eqn:rel2}
\end  {eqnarray}
respectively. 
However, the SSFs are directly related with values measured 
in a Monte Carlo simulation in contrast to 
$\beta_{\rm FV}(g_{\rm FV}^2)$ and $\gamma_{\rm FV}(g_{\rm FV}^2)$\,.

If $\beta_{\rm FV}(g_{\rm FV}^2)$ and $\gamma_{\rm FV}(g_{\rm FV}^2)$ 
are perturbatively known, 
$\sigma^g(s,g_{\rm FV}^2)$ and $\sigma^\phi(s,g_{\rm FV}^2)$ 
can be evaluated. 
Using an abbreviation $u\equiv g_{\rm FV}^2$ 
to simplify the expression, 
we consider the perturbative expansions of 
$\beta_{\rm FV}(u)$ and $\gamma_{\rm FV}(u)$ 
\begin{equation}
   \beta_{\rm FV}(u)
  =
  -u^2\,\sum_{i=0}^\infty\, \beta_{{\rm FV},i}\, u^i\ ,\quad
  \gamma_{\rm FV}(u)
  =
  -u  \,\sum_{i=0}^\infty\,\gamma_{{\rm FV},i}\, u^i\ ,
\label{eqn:exp1}
\end  {equation}
and ones of the SSFs 
\begin{equation}
  \sigma^g   (s,u) 
  = u + u\,\sum_{i=0}^\infty\,\sigma^g_i   (s)\, u^{i+1}\ ,\quad
  \sigma^\phi(s,u) 
  = 1 +    \sum_{i=0}^\infty\,\sigma^\phi_i(s)\, u^{i+1}\ .
\label{eqn:exp2}
\end  {equation}
By substituting 
Eqs. (\ref{eqn:exp1}) and (\ref{eqn:exp2}) to 
Eqs. (\ref{eqn:rel1}) and (\ref{eqn:rel2}), 
and by comparing the coefficients at the same order of $u$, 
we obtain 
\begin{eqnarray}
  & &
  \sigma^g_0(s)
  =
    \beta_{{\rm FV},0}\, \ln s\ ,\quad
  \sigma^g_1(s)
  =
    \beta_{{\rm FV},1}\,    \ln s
  + \beta_{{\rm FV},0}^2\, (\ln s)^2\ ,
\nonumber
\\
  & &
  \sigma^g_2(s)
  =
               \beta_{{\rm FV},2}\,                     \ln s
  + \frac{5}{2}\beta_{{\rm FV},0} \beta_{{\rm FV},1}\, (\ln s)^2
  +            \beta_{{\rm FV},0}^3\,                  (\ln s)^3\ ,
\label{eqn:coeff1}
\end  {eqnarray}
and 
\begin{eqnarray}
  & &
  \sigma^\phi_0(s)
  =
    \gamma_{{\rm FV},0}\, \ln s\ ,\quad 
  \sigma^\phi_1(s)
  =
    \gamma_{{\rm FV},1}\, \ln s
  + \frac{1}{2}(\beta_{{\rm FV},0}+\gamma_{{\rm FV},0})
    \gamma_{{\rm FV},0}\,(\ln s)^2\ ,
\nonumber
\\
  & &
  \sigma^\phi_2(s)
  =
     \gamma_{{\rm FV},2}\,
    \ln s
  + ( \beta_{{\rm FV},0}\gamma_{{\rm FV},1}
    +\gamma_{{\rm FV},0}\gamma_{{\rm FV},1}
    + \beta_{{\rm FV},1}\gamma_{{\rm FV},0}/2)\, 
    (\ln s)^2
\nonumber
\\
  & &
  \qquad\qquad
  + \frac{1}{3}(\beta_{{\rm FV},0}+\gamma_{{\rm FV},0})
               (\beta_{{\rm FV},0}+\gamma_{{\rm FV},0}/2)\gamma_{{\rm FV},0}\, 
    (\ln s)^3\ ,
\label{eqn:coeff2}
\end  {eqnarray}
up to the 3-loop order. 
Note that all the coefficients in $\sigma^g_i(s)$ and $\sigma^\phi_i(s)$ 
are not independent due to the constraint 
$\sigma^{g,\phi}(s_2s_1,u)=\sigma^{g,\phi}(s_2,\sigma^g(s_1,u))$\,.

Later, we need 
the perturbative evaluation of $\sigma^{g,\phi}(s,u)$. 
They are evaluated by Eq. (\ref{eqn:exp2})
with Eqs. (\ref{eqn:coeff1}) and (\ref{eqn:coeff2}) obtained from 
\begin{eqnarray}
  & &  \beta_{{\rm FV},0} =   \frac{        n-2 }{2\pi  }\ ,\quad 
       \beta_{{\rm FV},1} =   \frac{        n-2 }{4\pi^2}\ ,\quad 
       \beta_{{\rm FV},2} =   \frac{(n-1)  (n-2)}{8\pi^3}\ ,\quad 
\nonumber  \\
  & & \gamma_{{\rm FV},0} = - \frac{ n-1        }{2\pi  }\ ,\quad
      \gamma_{{\rm FV},1} =                             0\ ,\quad
      \gamma_{{\rm FV},2} =                             0\ , 
\label{eqn:coeff_FV}
\end  {eqnarray}
and $\sigma^{g,\phi}_i(s)=0$ for $i\ge 3$\,.
We refer these SSFs by $\sigma_{\rm P}^{g,\phi}(s)$\,. 
The derivation of Eq. (\ref{eqn:coeff_FV}) is given 
in Sec. \ref{app:Perturbative_evaluation}. 
%
% @@ ==================================================================
%
\section  {Details of Monte Carlo simulation}
\label{sec:Details_of_Monte_Carlo_simulation}
%
% @@@ =========================================================================
%
\subsection{Setup}
\label{ssec:Setup}
We set $n=3$. 
The calculation is performed on the $(T/a)\times (L/a)$ lattice with $T=5L$.
Here $a$ is the lattice spacing, 
and is determined from the bare coupling $g^2$. 
Due to the discretization, 
the action is changed to 
\begin{equation}
  S_{\rm lat}[{\bm \phi}]
  =
  -\frac{1}{g^2}\,\sum_{x,\,\mu}\,{\bm \phi}(x)\cdot{\bm \phi}(x+\hat{\mu})\ ,
\end  {equation}
where $x$ moves all the lattice space-time points, 
and $\hat{\mu}$ is a unit vector in the $\mu$ direction. 
The NBC is imposed for the temporal direction, and 
the PBC            for the spatial  direction. 
In Table \ref{tbl:parameters}, 
we list $(1/g^2,L/a)$ which are used in the present calculation. 
As we will mention in Sec. \ref{ssec:Data_of_SSF}, 
we classify them into five sets 
(``A'', ``B'', ``C'', ``D'' and ``E'') 
depending on the value of renormalized coupling. 
For updating of ${\bm \phi}$ configurations, 
the heat bath algorithm is used, 
and even sites and odd sites are alternately updated. 
After the thermalization by $5000$ sweeps, 
we calculate 
\begin{equation}
  G_{\rm inv}^{(i)}(t)
  =
  \frac{1}{L^2}\,
  \sum_{x_1,\,y_1}\,
  {\bm \phi}(t_{\rm src},x_1)
  \cdot 
  {\bm \phi}(t          ,y_1)
\label{eqn:lat_Ginv}
\end  {equation}
on the $i$-th configuration at every $100$ sweeps. 
We set $t_{\rm src}/a=L/a$. 
The total number of samples is $999950$ for each parameter set.
The expectation value of Eq. (\ref{eqn:lat_Ginv}) 
is nothing less than the $O(n)$-invariant $2$-point Green function. 
%
%---------------------------------
%
\begin{table}[p]
\begin{center}
\caption{
A list of $(1/g^2,L/a)$ which are used in the present calculation. 
}
\medskip
\begin{tabular}{ccccc}
\hline
\hline
set & $\qquad$ & $1/g^2$ & $\qquad$ & $L/a$                              \\
\hline
A & & $2.0786$ & & $ 6$, $ 7$, $ 8$, $ 9$, $10$, $11$, $12$              \\
  & & $2.1043$ & & $ 7$, $ 8$, $ 9$, $10$, $11$, $12$, $13$, $14$        \\
  & & $2.1275$ & & $ 8$, $ 9$, $10$, $11$, $12$, $13$, $14$, $15$, $16$  \\
  & & $2.1625$ & & $10$, $12$, $14$, $16$, $18$, $20$                    \\
  & & $2.1954$ & & $12$, $14$, $16$, $18$, $20$, $22$, $24$              \\
  & & $2.2403$ & & $16$, $18$, $20$, $22$, $24$, $26$, $28$, $30$, $32$  \\
\hline
B & & $1.9637$ & & $ 6$, $ 7$, $ 8$, $ 9$, $10$, $11$, $12$              \\
  & & $1.9875$ & & $ 7$, $ 8$, $ 9$, $10$, $11$, $12$, $13$, $14$        \\
  & & $2.0100$ & & $ 8$, $ 9$, $10$, $11$, $12$, $13$, $14$, $15$, $16$  \\
  & & $2.0489$ & & $10$, $12$, $14$, $16$, $18$, $20$                    \\
  & & $2.0794$ & & $12$, $14$, $16$, $18$, $20$, $22$, $24$              \\
  & & $2.1260$ & & $16$, $18$, $20$, $22$, $24$, $26$, $28$, $30$, $32$  \\
\hline
C & & $1.8439$ & & $ 6$, $ 7$, $ 8$, $ 9$, $10$, $11$, $12$              \\
  & & $1.8711$ & & $ 7$, $ 8$, $ 9$, $10$, $11$, $12$, $13$, $14$        \\
  & & $1.8947$ & & $ 8$, $ 9$, $10$, $11$, $12$, $13$, $14$, $15$, $16$  \\
  & & $1.9319$ & & $10$, $12$, $14$, $16$, $18$, $20$                    \\
  & & $1.9637$ & & $12$, $14$, $16$, $18$, $20$, $22$, $24$              \\
  & & $2.0100$ & & $16$, $18$, $20$, $22$, $24$, $26$, $28$, $30$, $32$  \\
\hline
D & & $1.7276$ & & $ 6$, $ 7$, $ 8$, $ 9$, $10$, $11$, $12$              \\
  & & $1.7553$ & & $ 7$, $ 8$, $ 9$, $10$, $11$, $12$, $13$, $14$        \\
  & & $1.7791$ & & $ 8$, $ 9$, $10$, $11$, $12$, $13$, $14$, $15$, $16$  \\
  & & $1.8171$ & & $10$, $12$, $14$, $16$, $18$, $20$                    \\
  & & $1.8497$ & & $12$, $14$, $16$, $18$, $20$, $22$, $24$              \\
  & & $1.8965$ & & $16$, $18$, $20$, $22$, $24$, $26$, $28$, $30$, $32$  \\
\hline
E & & $1.6050$ & & $ 6$, $ 7$, $ 8$, $ 9$, $10$, $11$, $12$              \\
  & & $1.6346$ & & $ 7$, $ 8$, $ 9$, $10$, $11$, $12$, $13$, $14$        \\
  & & $1.6589$ & & $ 8$, $ 9$, $10$, $11$, $12$, $13$, $14$, $15$, $16$  \\
  & & $1.6982$ & & $10$, $12$, $14$, $16$, $18$, $20$                    \\
  & & $1.7306$ & & $12$, $14$, $16$, $18$, $20$, $22$, $24$              \\
  & & $1.7800$ & & $16$, $18$, $20$, $22$, $24$, $26$, $28$, $30$, $32$  \\
\hline
\hline
\end  {tabular}
\label{tbl:parameters}
\end  {center}
\end  {table}
%
%---------------------------------
%
% @@@ =========================================================================
%
\subsection{Autocorrelation}
\label{ssec:Autocorrelation}
We consider the autocorrelation function of $G_{\rm inv}^{(i)}(t)$, 
\begin{equation}
  A(j) 
  \equiv 
  \frac{1}{N}\,\sum_{i=1}^N
  \left[\, 
    \left(\,
      G_{\rm inv}^{(i  )}(t) - \langle\,G_{\rm inv}(t)\,\rangle_0\,
    \right)
    \left(\,
      G_{\rm inv}^{(i+j)}(t) - \langle\,G_{\rm inv}(t)\,\rangle_j\,
    \right)\,
  \right]\ ,
\end  {equation}
where the angle bracket denotes the expectation value as 
\begin{equation}
  \langle\,G_{\rm inv}(t)\,\rangle_j
  \equiv
  \frac{1}{N}\,\sum_{i=1}^N\,G_{\rm inv}^{(i+j)}(t)\ .
\end  {equation}
The function $A(j)$ represents the correlation 
between $G_{\rm inv}^{(i)}(t)$'s separated by the $j$-time measurements. 
For a precise analysis, we introduce the integrated autocorrelation time, 
\begin{equation}
  \tau_{\rm int}(j)
  =
  \frac{1}{2} + \sum_{i=1}^j\,\frac{A(i)}{A(0)}\ .
\end  {equation}
$2\tau_{\rm int}(\infty)$ will indicate the separation 
where the measurements can be regarded to be independent.

\begin{figure}[p]
\begin{center}
\includegraphics[width=160mm]{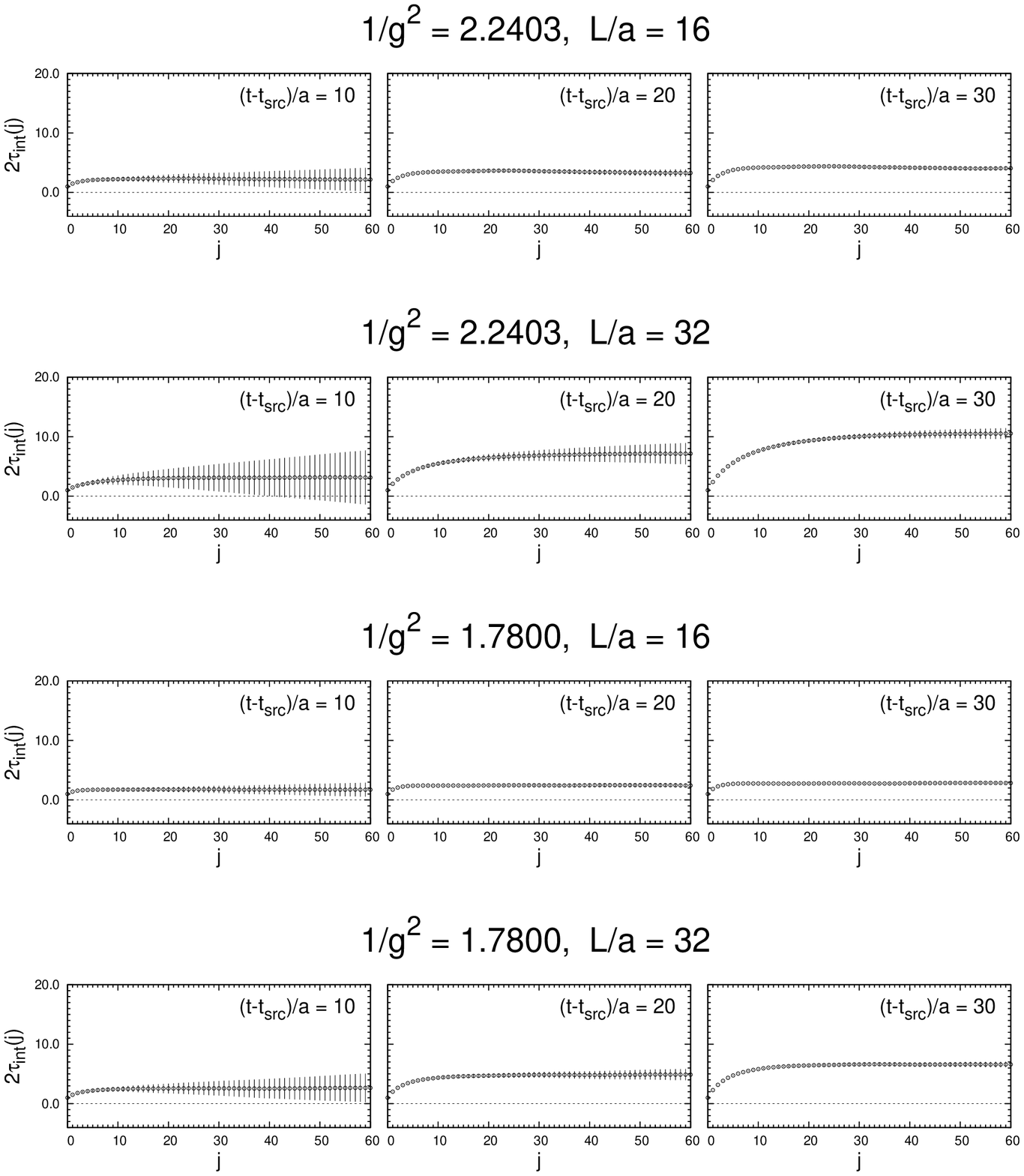}
\caption{
$2\tau_{\rm int}(j)$ for some $(1/g^2,L/a)$'s. 
We show the data for $1/g^2$ 
which give the smallest lattice spacing from the sets A and E. 
For each $(1/g^2,L/a)$\,, 
the situations with $(t-t_{\rm src})/a=10$, $20$ and $30$ are shown. 
}
\label{fig:tau_inte}
\end  {center}
\end  {figure}

In Fig. \ref{fig:tau_inte}, 
we give $2\tau_{\rm int}(j)$ for some $(1/g^2,L/a)$'s. 
To guarantee a reliable analysis, 
$N\gg j$ is required, and we adopt $N=100000$. 
The statistical error is evaluated by the single-eliminated jackknife method. 
The autocorrelation time becomes large near the continuum limit, 
so that we show the data for $1/g^2$ 
which gives the smallest lattice spacing from the sets A and E. 
For each $(1/g^2,L/a)$\,, 
the situations with $(t-t_{\rm src})/a=10$, $20$ and $30$ are shown. 
The data with large $(t-t_{\rm src})/a$ 
do not give a significant contribution 
for the evaluation of the mass gap and amplitude. 
We consider that 
the verification at $(t-t_{\rm src})/a=10-30$ is sufficient. 
We confirm, from Fig. \ref{fig:tau_inte}, 
that $2\tau_{\rm int}(\infty)$ is at most about $10$ 
with our simulation parameters. 
For the sake of safety, 
we evaluate the statistical errors on the mass gap and amplitude 
by the jackknife method with the bin size of $50$ samples 
in the following analysis. 
%
%---------------------------------
%
%
%---------------------------------
%

% @@@ =========================================================================
%
\subsection{Fit range}
\label{ssec:Fit_range}
\begin{figure}[p]
\begin{center}
\includegraphics[width=160mm]{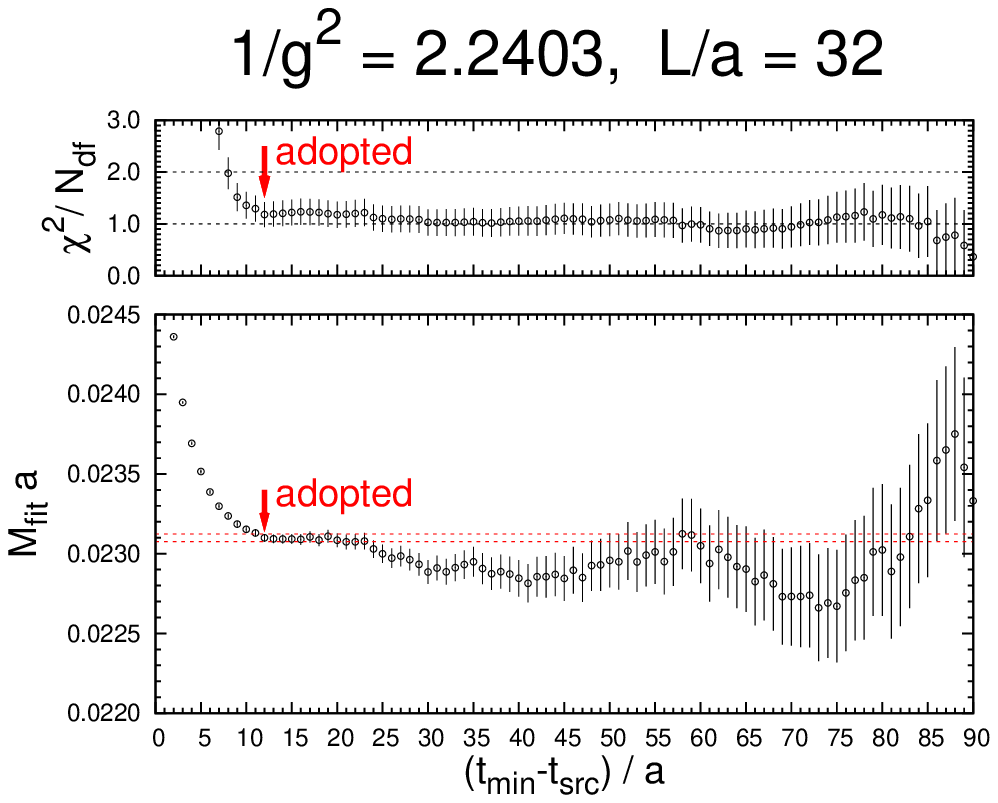}
\caption{
An example of the fit-range dependence. 
$\chi^2/N_{\rm df}$ in the fit is shown     in the top    panel, 
and the fitted mass gap, $M_{\rm fit}a$, is in the bottom panel. 
For this parameter set, $(t_{\rm min}-t_{\rm src})/a=12$ is adopted. 
}
\label{fig:tmin-dep}
\end  {center}
\end  {figure}

For the $O(N)$-invariant $2$-point Green function, 
we carry out the fit 
considering the correlation between the different time slices 
with the variance-covariance matrix. 
We refer the fit range by $[\,t_{\rm min}:t_{\rm max}\,]$\,.
For all the parameter sets, 
$(t_{\rm max}-t_{\rm src})/a=3(L/a)-1$ is chosen 
to avoid the contamination from the temporal boundary. 
On the other hand, 
$t_{\rm min}$ should be determined from the behavior of $\chi^2/N_{\rm df}$\,. 
We increase $t_{\rm min}$ from $1$, 
and adopt the value 
when $\chi^2/N_{\rm df}$ falls to the vicinity of $1$. 
An example for $(1/g^2,L/a)=(2.2403,32)$ 
is shown in Fig. \ref{fig:tmin-dep}. 
For this parameter set, $(t_{\rm min}-t_{\rm src})/a=12$ is adopted, 
and the fitted value is shown by the horizontal dotted lines. 
%
%---------------------------------
%
%
%---------------------------------
%

%
% @@ ==================================================================
%
\section  {Numerical results}
\label{sec:Numerical_results}
%
% @@@ =========================================================================
%
\subsection{Data of SSF}
\label{ssec:Data_of_SSF}
We obtain the mass gap in the lattice unit $Ma$ and the amplitude $A$ 
by fitting Eq. (\ref{eqn:0_mom_projected_Green_func}) to 
the data of $O(n)$-invariant $2$-point Green function. 
The renormalized coupling         $g_{\rm FV}^2$ can be extracted 
from $Ma$ by Eq. (\ref{eqn:def_FS1}), and 
the wave-function renormalization $Z_{\rm FV}^\phi$ 
from $A$ by Eq. (\ref{eqn:def_FS2}). 
On a lattice, 
$g_{\rm FV}^2$ and $Z_{\rm FV}^\phi$ depend on $L/a$ 
in addition to the physical extent $L$. 
To take the continuum limit later, 
we classify the numerical data into five sets 
(``A'', ``B'', ``C'', ``D'' and ``E''). 
The classification is based on 
the value of $g_{\rm FV}^2$ in the smallest $L$ for each $1/g^2$. 
We refer the smallest $L$ by $L_{0}$. 
In Tables \ref{tbl:raw_dat1_A}, 
          \ref{tbl:raw_dat1_B}, 
          \ref{tbl:raw_dat1_C}, 
          \ref{tbl:raw_dat1_D} and 
          \ref{tbl:raw_dat1_E}, 
we show $g_{\rm FV}^2(L,L/a)$ and $Z_{\rm FV}^\phi(L,L/a)$ 
measured with various $(1/g^2,L/a)$. 
We also list $\chi^2/N_{\rm df}$ in the fit. 
%
%---------------------------------
%
%  Set A
%
\begin{table}[p]
\begin{center}
\caption{$g_{\rm FV}^2(L,L/a)$ and $Z_{\rm FV}^\phi(L,L/a)$ for various $(1/g^2,L/a)$ of set A.}
\medskip
\begin{tabular}{ccccccccc}
\hline
\hline
\multicolumn{4}{c}{$1/g^2=2.0786$} & \qquad\qquad & 
\multicolumn{4}{c}{$1/g^2=2.1043$}                               \\
\hline
$L/a$ & $g_{\rm FV}^2$& $Z_{\rm FV}^\phi$ & $\chi^2/N_{\rm df}$ & & 
$L/a$ & $g_{\rm FV}^2$& $Z_{\rm FV}^\phi$ & $\chi^2/N_{\rm df}$  \\
\hline
$ 6$ & $0.67562(44)$ & $0.736987(89)$ & $0.90(51)$ & & $ 7$ & $0.67546(52)$ & $0.71670 (12)$ & $0.83(45)$ \\
$ 7$ & $0.68841(53)$ & $0.71269 (12)$ & $0.92(48)$ & & $ 8$ & $0.68627(72)$ & $0.69617 (22)$ & $1.92(67)$ \\
$ 8$ & $0.70089(50)$ & $0.692435(98)$ & $1.11(48)$ & & $ 9$ & $0.69730(46)$ & $0.678804(79)$ & $1.06(44)$ \\
$ 9$ & $0.71109(47)$ & $0.674314(80)$ & $1.03(43)$ & & $10$ & $0.70611(53)$ & $0.66287 (10)$ & $1.00(41)$ \\
$10$ & $0.72053(54)$ & $0.65817 (10)$ & $1.04(42)$ & & $11$ & $0.71506(51)$ & $0.648882(87)$ & $1.38(45)$ \\
$11$ & $0.73019(52)$ & $0.644050(89)$ & $0.97(38)$ & & $12$ & $0.72442(49)$ & $0.636262(75)$ & $0.81(33)$ \\
$12$ & $0.73917(59)$ & $0.63109 (11)$ & $0.58(28)$ & & $13$ & $0.73101(56)$ & $0.624345(94)$ & $0.85(33)$ \\
     &               &                &            & & $14$ & $0.73940(54)$ & $0.613753(83)$ & $0.98(33)$ \\
\hline
\hline
\multicolumn{4}{c}{$1/g^2=2.1275$} & \qquad\qquad & 
\multicolumn{4}{c}{$1/g^2=2.1625$}                               \\
\hline
$L/a$ & $g_{\rm FV}^2$& $Z_{\rm FV}^\phi$ & $\chi^2/N_{\rm df}$ & & 
$L/a$ & $g_{\rm FV}^2$& $Z_{\rm FV}^\phi$ & $\chi^2/N_{\rm df}$  \\
\hline
$ 8$ & $0.67564(48)$ & $0.700346(95)$ & $1.30(52)$ & & $10$ & $0.67532(50)$ & $0.672916(98)$ & $0.94(40)$ \\
$ 9$ & $0.68502(45)$ & $0.682692(78)$ & $1.18(46)$ & & $12$ & $0.69241(47)$ & $0.647087(74)$ & $0.54(27)$ \\
$10$ & $0.69352(52)$ & $0.66698 (10)$ & $1.08(42)$ & & $14$ & $0.70607(51)$ & $0.625249(80)$ & $1.41(40)$ \\
$11$ & $0.70200(50)$ & $0.653087(87)$ & $1.63(49)$ & & $16$ & $0.71807(48)$ & $0.606667(65)$ & $1.49(38)$ \\
$12$ & $0.71110(48)$ & $0.640655(75)$ & $0.59(28)$ & & $18$ & $0.72879(61)$ & $0.590271(93)$ & $0.68(25)$ \\
$13$ & $0.71862(55)$ & $0.629024(93)$ & $1.09(37)$ & & $20$ & $0.74184(51)$ & $0.576403(61)$ & $1.44(33)$ \\
$14$ & $0.72579(53)$ & $0.618446(82)$ & $1.01(34)$ & &      &               &                &            \\
$15$ & $0.73228(51)$ & $0.608579(73)$ & $0.96(32)$ & &      &               &                &            \\
$16$ & $0.73815(50)$ & $0.599474(67)$ & $1.29(36)$ & &      &               &                &            \\
\hline
\hline
\multicolumn{4}{c}{$1/g^2=2.1954$} & \qquad\qquad & 
\multicolumn{4}{c}{$1/g^2=2.2403$}                               \\
\hline
$L/a$ & $g_{\rm FV}^2$& $Z_{\rm FV}^\phi$ & $\chi^2/N_{\rm df}$ & & 
$L/a$ & $g_{\rm FV}^2$& $Z_{\rm FV}^\phi$ & $\chi^2/N_{\rm df}$  \\
\hline
$12$ & $0.67551(46)$ & $0.652903(72)$ & $0.55(27)$ & & $16$ & $0.67526(68)$ & $0.62115 (14)$ & $1.11(34)$ \\
$14$ & $0.68986(43)$ & $0.631629(58)$ & $1.02(34)$ & & $18$ & $0.68645(58)$ & $0.605781(89)$ & $0.65(24)$ \\
$16$ & $0.70072(47)$ & $0.613205(64)$ & $2.01(44)$ & & $20$ & $0.69597(62)$ & $0.591990(94)$ & $0.99(28)$ \\
$18$ & $0.71007(60)$ & $0.597037(92)$ & $0.75(26)$ & & $22$ & $0.70598(53)$ & $0.579779(65)$ & $1.25(30)$ \\
$20$ & $0.72075(64)$ & $0.582953(96)$ & $0.85(26)$ & & $24$ & $0.71253(64)$ & $0.568376(86)$ & $1.08(27)$ \\
$22$ & $0.73000(62)$ & $0.570317(82)$ & $0.97(26)$ & & $26$ & $0.71991(62)$ & $0.558049(75)$ & $0.90(23)$ \\
$24$ & $0.73715(74)$ & $0.55872 (11)$ & $0.91(25)$ & & $28$ & $0.72725(74)$ & $0.548606(95)$ & $0.82(22)$ \\
     &               &                &            & & $30$ & $0.73428(72)$ & $0.539910(85)$ & $1.17(25)$ \\
     &               &                &            & & $32$ & $0.73919(76)$ & $0.531715(89)$ & $1.18(24)$ \\
\hline
\hline
\end  {tabular}
\label{tbl:raw_dat1_A}
\end  {center}
\end  {table}
%
%---------------------------------

%---------------------------------
%
%  Set B
%
\begin{table}[p]
\begin{center}
\caption{$g_{\rm FV}^2(L,L/a)$ and $Z_{\rm FV}^\phi(L,L/a)$ for various $(1/g^2,L/a)$ of set B.}
\medskip
\begin{tabular}{ccccccccc}
\hline
\hline
\multicolumn{4}{c}{$1/g^2=1.9637$} & \qquad\qquad & 
\multicolumn{4}{c}{$1/g^2=1.9875$}                               \\
\hline
$L/a$ & $g_{\rm FV}^2$& $Z_{\rm FV}^\phi$ & $\chi^2/N_{\rm df}$ & & 
$L/a$ & $g_{\rm FV}^2$& $Z_{\rm FV}^\phi$ & $\chi^2/N_{\rm df}$  \\
\hline
$ 6$ & $0.73747(63)$ & $0.71909 (17)$ & $0.54(41)$ & & $ 7$ & $0.73929(57)$ & $0.69777 (13)$ & $1.12(53)$ \\
$ 7$ & $0.75339(58)$ & $0.69344 (13)$ & $0.50(35)$ & & $ 8$ & $0.75316(65)$ & $0.67613 (16)$ & $0.96(46)$ \\
$ 8$ & $0.76891(55)$ & $0.67183 (10)$ & $0.82(42)$ & & $ 9$ & $0.76653(51)$ & $0.657341(85)$ & $1.56(53)$ \\
$ 9$ & $0.78230(52)$ & $0.652625(87)$ & $0.77(37)$ & & $10$ & $0.77661(58)$ & $0.64026 (11)$ & $0.80(36)$ \\
$10$ & $0.79236(71)$ & $0.63504 (16)$ & $0.80(37)$ & & $11$ & $0.78893(56)$ & $0.625564(94)$ & $0.77(34)$ \\
$11$ & $0.80587(57)$ & $0.620335(95)$ & $1.45(46)$ & & $12$ & $0.79912(64)$ & $0.61190 (11)$ & $0.79(33)$ \\
$12$ & $0.81642(66)$ & $0.60642 (12)$ & $0.75(32)$ & & $13$ & $0.80946(62)$ & $0.59952 (10)$ & $1.05(36)$ \\
     &               &                &            & & $14$ & $0.81904(60)$ & $0.588191(89)$ & $0.97(33)$ \\
\hline
\hline
\multicolumn{4}{c}{$1/g^2=2.0100$} & \qquad\qquad & 
\multicolumn{4}{c}{$1/g^2=2.0489$}                               \\
\hline
$L/a$ & $g_{\rm FV}^2$& $Z_{\rm FV}^\phi$ & $\chi^2/N_{\rm df}$ & & 
$L/a$ & $g_{\rm FV}^2$& $Z_{\rm FV}^\phi$ & $\chi^2/N_{\rm df}$  \\
\hline
$ 8$ & $0.74019(53)$ & $0.68055 (10)$ & $1.29(52)$ & & $10$ & $0.73761(66)$ & $0.65247 (15)$ & $0.72(35)$ \\
$ 9$ & $0.75190(49)$ & $0.661754(84)$ & $0.79(38)$ & & $12$ & $0.75740(60)$ & $0.62503 (11)$ & $0.76(33)$ \\
$10$ & $0.76117(68)$ & $0.64466 (16)$ & $0.58(32)$ & & $14$ & $0.77468(57)$ & $0.602081(86)$ & $1.23(38)$ \\
$11$ & $0.77292(55)$ & $0.630250(92)$ & $1.07(40)$ & & $16$ & $0.78772(80)$ & $0.58195 (15)$ & $0.74(28)$ \\
$12$ & $0.78414(53)$ & $0.616994(80)$ & $0.80(33)$ & & $18$ & $0.80339(68)$ & $0.565226(98)$ & $0.87(28)$ \\
$13$ & $0.79270(60)$ & $0.604552(99)$ & $0.82(32)$ & & $20$ & $0.81694(73)$ & $0.55012 (10)$ & $1.08(30)$ \\
$14$ & $0.80228(59)$ & $0.593399(88)$ & $1.30(39)$ & &      &               &                &            \\
$15$ & $0.80917(76)$ & $0.58254 (14)$ & $0.98(33)$ & &      &               &                &            \\
$16$ & $0.81636(83)$ & $0.57284 (16)$ & $1.30(37)$ & &      &               &                &            \\
\hline
\hline
\multicolumn{4}{c}{$1/g^2=2.0794$} & \qquad\qquad & 
\multicolumn{4}{c}{$1/g^2=2.1260$}                               \\
\hline
$L/a$ & $g_{\rm FV}^2$& $Z_{\rm FV}^\phi$ & $\chi^2/N_{\rm df}$ & & 
$L/a$ & $g_{\rm FV}^2$& $Z_{\rm FV}^\phi$ & $\chi^2/N_{\rm df}$  \\
\hline
$12$ & $0.73826(59)$ & $0.63119 (11)$ & $0.55(28)$ & & $16$ & $0.73845(58)$ & $0.599035(88)$ & $1.00(32)$ \\
$14$ & $0.75515(55)$ & $0.608666(84)$ & $0.95(33)$ & & $18$ & $0.75124(63)$ & $0.582660(95)$ & $1.05(31)$ \\
$16$ & $0.76813(78)$ & $0.58894 (15)$ & $0.94(32)$ & & $20$ & $0.76170(67)$ & $0.567950(98)$ & $0.93(27)$ \\
$18$ & $0.78243(66)$ & $0.572381(96)$ & $0.80(27)$ & & $22$ & $0.77400(66)$ & $0.555121(85)$ & $1.25(30)$ \\
$20$ & $0.79442(71)$ & $0.55745 (10)$ & $1.05(29)$ & & $24$ & $0.78255(78)$ & $0.54314 (11)$ & $0.94(25)$ \\
$22$ & $0.80650(69)$ & $0.544151(87)$ & $1.02(27)$ & & $26$ & $0.79209(76)$ & $0.532309(95)$ & $1.01(25)$ \\
$24$ & $0.81634(81)$ & $0.53195 (11)$ & $0.75(23)$ & & $28$ & $0.80000(90)$ & $0.52232 (12)$ & $1.10(25)$ \\
     &               &                &            & & $30$ & $0.80918(87)$ & $0.51320 (11)$ & $0.86(21)$ \\
     &               &                &            & & $32$ & $0.81889(84)$ & $0.505102(95)$ & $0.83(20)$ \\
\hline
\hline
\end  {tabular}
\label{tbl:raw_dat1_B}
\end  {center}
\end  {table}
%
%---------------------------------

%---------------------------------
%
%  Set C
%
\begin{table}[p]
\begin{center}
\caption{$g_{\rm FV}^2(L,L/a)$ and $Z_{\rm FV}^\phi(L,L/a)$ for various $(1/g^2,L/a)$ of set C.}
\medskip
\begin{tabular}{ccccccccc}
\hline
\hline
\multicolumn{4}{c}{$1/g^2=1.8439$} & \qquad\qquad & 
\multicolumn{4}{c}{$1/g^2=1.8711$}                               \\
\hline
$L/a$ & $g_{\rm FV}^2$& $Z_{\rm FV}^\phi$ & $\chi^2/N_{\rm df}$ & & 
$L/a$ & $g_{\rm FV}^2$& $Z_{\rm FV}^\phi$ & $\chi^2/N_{\rm df}$  \\
\hline
$ 6$ & $0.81779(72)$ & $0.69774 (19)$ & $0.59(42)$ & & $ 7$ & $0.81674(80)$ & $0.67550 (23)$ & $0.67(42)$ \\
$ 7$ & $0.8368 (10)$ & $0.66928 (36)$ & $1.30(61)$ & & $ 8$ & $0.83601(60)$ & $0.65282 (11)$ & $1.14(49)$ \\
$ 8$ & $0.85636(91)$ & $0.64610 (27)$ & $1.09(51)$ & & $ 9$ & $0.85232(57)$ & $0.632517(92)$ & $0.70(36)$ \\
$ 9$ & $0.87577(58)$ & $0.626062(94)$ & $1.01(43)$ & & $10$ & $0.86582(78)$ & $0.61411 (17)$ & $1.05(43)$ \\
$10$ & $0.89178(68)$ & $0.60762 (12)$ & $0.95(40)$ & & $11$ & $0.88066(75)$ & $0.59823 (14)$ & $0.72(33)$ \\
$11$ & $0.90702(65)$ & $0.59134 (10)$ & $1.12(41)$ & & $12$ & $0.89595(72)$ & $0.58385 (12)$ & $0.56(28)$ \\
$12$ & $0.92297(74)$ & $0.57669 (13)$ & $0.93(36)$ & & $13$ & $0.90939(70)$ & $0.57061 (11)$ & $0.73(30)$ \\
     &               &                &            & & $14$ & $0.92210(68)$ & $0.558491(97)$ & $1.06(35)$ \\
\hline
\hline
\multicolumn{4}{c}{$1/g^2=1.8947$} & \qquad\qquad & 
\multicolumn{4}{c}{$1/g^2=1.9319$}                               \\
\hline
$L/a$ & $g_{\rm FV}^2$& $Z_{\rm FV}^\phi$ & $\chi^2/N_{\rm df}$ & & 
$L/a$ & $g_{\rm FV}^2$& $Z_{\rm FV}^\phi$ & $\chi^2/N_{\rm df}$  \\
\hline
$ 8$ & $0.81744(59)$ & $0.65790 (11)$ & $0.93(44)$ & & $10$ & $0.81599(74)$ & $0.62809 (17)$ & $1.10(44)$ \\
$ 9$ & $0.83262(55)$ & $0.637830(91)$ & $1.01(43)$ & & $12$ & $0.84231(67)$ & $0.59917 (12)$ & $0.72(32)$ \\
$10$ & $0.84622(76)$ & $0.61976 (17)$ & $0.84(38)$ & & $14$ & $0.86368(73)$ & $0.57445 (12)$ & $1.23(38)$ \\
$11$ & $0.86124(62)$ & $0.60431 (10)$ & $1.22(43)$ & & $16$ & $0.88138(89)$ & $0.55311 (16)$ & $0.95(32)$ \\
$12$ & $0.87416(70)$ & $0.58991 (12)$ & $0.81(33)$ & & $18$ & $0.90137(76)$ & $0.53532 (11)$ & $1.06(31)$ \\
$13$ & $0.88743(68)$ & $0.57696 (11)$ & $1.15(38)$ & & $20$ & $0.92031(82)$ & $0.51936 (11)$ & $0.91(27)$ \\
$14$ & $0.89816(66)$ & $0.564864(94)$ & $1.02(34)$ & &      &               &                &            \\
$15$ & $0.90795(85)$ & $0.55333 (15)$ & $1.00(33)$ & &      &               &                &            \\
$16$ & $0.91855(93)$ & $0.54310 (17)$ & $0.86(30)$ & &      &               &                &            \\
\hline
\hline
\multicolumn{4}{c}{$1/g^2=1.9637$} & \qquad\qquad & 
\multicolumn{4}{c}{$1/g^2=2.0100$}                               \\
\hline
$L/a$ & $g_{\rm FV}^2$& $Z_{\rm FV}^\phi$ & $\chi^2/N_{\rm df}$ & & 
$L/a$ & $g_{\rm FV}^2$& $Z_{\rm FV}^\phi$ & $\chi^2/N_{\rm df}$  \\
\hline
$12$ & $0.81642(66)$ & $0.60642 (12)$ & $0.75(32)$ & & $16$ & $0.81636(83)$ & $0.57284 (16)$ & $1.30(37)$ \\
$14$ & $0.83773(61)$ & $0.582484(90)$ & $0.80(30)$ & & $18$ & $0.83360(70)$ & $0.55586 (10)$ & $0.76(26)$ \\
$16$ & $0.85333(86)$ & $0.56138 (16)$ & $1.10(34)$ & & $20$ & $0.84805(75)$ & $0.54043 (11)$ & $0.82(26)$ \\
$18$ & $0.87205(73)$ & $0.54388 (10)$ & $0.78(26)$ & & $22$ & $0.86180(74)$ & $0.526669(91)$ & $1.21(29)$ \\
$20$ & $0.88871(80)$ & $0.52819 (11)$ & $0.93(27)$ & & $24$ & $0.87336(88)$ & $0.51403 (12)$ & $1.00(26)$ \\
$22$ & $0.90454(77)$ & $0.514099(93)$ & $1.16(29)$ & & $26$ & $0.88646(85)$ & $0.50271 (10)$ & $0.95(24)$ \\
$24$ & $0.91761(93)$ & $0.50112 (12)$ & $0.92(25)$ & & $28$ & $0.8965 (10)$ & $0.49211 (13)$ & $1.17(26)$ \\
     &               &                &            & & $30$ & $0.90739(98)$ & $0.48235 (11)$ & $0.87(21)$ \\
     &               &                &            & & $32$ & $0.92037(95)$ & $0.47385 (10)$ & $0.92(21)$ \\
\hline
\hline
\end  {tabular}
\label{tbl:raw_dat1_C}
\end  {center}
\end  {table}
%
%---------------------------------

%---------------------------------
%
%  Set D
%
\begin{table}[p]
\begin{center}
\caption{$g_{\rm FV}^2(L,L/a)$ and $Z_{\rm FV}^\phi(L,L/a)$ for various $(1/g^2,L/a)$ of set D.}
\medskip
\begin{tabular}{ccccccccc}
\hline
\hline
\multicolumn{4}{c}{$1/g^2=1.7276$} & \qquad\qquad & 
\multicolumn{4}{c}{$1/g^2=1.7553$}                               \\
\hline
$L/a$ & $g_{\rm FV}^2$& $Z_{\rm FV}^\phi$ & $\chi^2/N_{\rm df}$ & & 
$L/a$ & $g_{\rm FV}^2$& $Z_{\rm FV}^\phi$ & $\chi^2/N_{\rm df}$  \\
\hline
$ 6$ & $0.91813(82)$ & $0.67343 (21)$ & $0.96(54)$ & & $ 7$ & $0.91732(73)$ & $0.65013 (16)$ & $1.76(66)$ \\
$ 7$ & $0.94527(76)$ & $0.64314 (16)$ & $1.27(56)$ & & $ 8$ & $0.94251(68)$ & $0.62534 (12)$ & $0.88(43)$ \\
$ 8$ & $0.97343(71)$ & $0.61815 (13)$ & $1.36(53)$ & & $ 9$ & $0.96594(65)$ & $0.60345 (10)$ & $1.01(43)$ \\
$ 9$ & $0.99753(66)$ & $0.59562 (10)$ & $0.53(31)$ & & $10$ & $0.98462(90)$ & $0.58341 (19)$ & $0.99(42)$ \\
$10$ & $1.01974(78)$ & $0.57552 (13)$ & $0.70(34)$ & & $11$ & $1.00482(86)$ & $0.56624 (16)$ & $1.17(42)$ \\
$11$ & $1.04285(76)$ & $0.55802 (11)$ & $0.93(37)$ & & $12$ & $1.02561(83)$ & $0.55070 (14)$ & $0.92(36)$ \\
$12$ & $1.06385(86)$ & $0.54171 (14)$ & $0.94(36)$ & & $13$ & $0.10457(81)$ & $0.53651 (12)$ & $0.84(32)$ \\
     &               &                &            & & $14$ & $1.06333(79)$ & $0.52324 (11)$ & $1.90(47)$ \\
\hline
\hline
\multicolumn{4}{c}{$1/g^2=1.7791$} & \qquad\qquad & 
\multicolumn{4}{c}{$1/g^2=1.8171$}                               \\
\hline
$L/a$ & $g_{\rm FV}^2$& $Z_{\rm FV}^\phi$ & $\chi^2/N_{\rm df}$ & & 
$L/a$ & $g_{\rm FV}^2$& $Z_{\rm FV}^\phi$ & $\chi^2/N_{\rm df}$  \\
\hline
$ 8$ & $0.91775(66)$ & $0.63133 (12)$ & $1.10(48)$ & & $10$ & $0.91626(83)$ & $0.60044 (18)$ & $1.08(43)$ \\
$ 9$ & $0.93942(63)$ & $0.609855(99)$ & $0.89(40)$ & & $12$ & $0.95094(77)$ & $0.56913 (13)$ & $0.82(34)$ \\
$10$ & $0.95722(87)$ & $0.59030 (19)$ & $1.00(42)$ & & $14$ & $0.98082(97)$ & $0.54265 (18)$ & $0.67(29)$ \\
$11$ & $0.97698(70)$ & $0.57354 (11)$ & $0.97(38)$ & & $16$ & $1.00748(80)$ & $0.52021 (11)$ & $1.04(32)$ \\
$12$ & $0.99454(81)$ & $0.55794 (13)$ & $0.84(34)$ & & $18$ & $1.03295(88)$ & $0.50068 (11)$ & $0.72(25)$ \\
$13$ & $1.01269(78)$ & $0.54394 (12)$ & $1.15(38)$ & & $20$ & $1.05874(95)$ & $0.48347 (12)$ & $0.92(27)$ \\
$14$ & $1.02894(77)$ & $0.53101 (10)$ & $1.38(40)$ & &      &               &                &            \\
$15$ & $1.04460(98)$ & $0.51874 (16)$ & $1.07(34)$ & &      &               &                &            \\
$16$ & $1.05967(84)$ & $0.50782 (11)$ & $1.00(32)$ & &      &               &                &            \\
\hline
\hline
\multicolumn{4}{c}{$1/g^2=1.8497$} & \qquad\qquad & 
\multicolumn{4}{c}{$1/g^2=1.8965$}                               \\
\hline
$L/a$ & $g_{\rm FV}^2$& $Z_{\rm FV}^\phi$ & $\chi^2/N_{\rm df}$ & & 
$L/a$ & $g_{\rm FV}^2$& $Z_{\rm FV}^\phi$ & $\chi^2/N_{\rm df}$  \\
\hline
$12$ & $0.91781(74)$ & $0.57834 (13)$ & $0.79(33)$ & & $16$ & $0.91747(94)$ & $0.54374 (17)$ & $1.39(38)$ \\
$14$ & $0.94411(69)$ & $0.552425(97)$ & $0.82(31)$ & & $18$ & $0.93771(79)$ & $0.52531 (11)$ & $0.85(28)$ \\
$16$ & $0.96708(99)$ & $0.52991 (18)$ & $0.92(31)$ & & $20$ & $0.95696(86)$ & $0.50888 (11)$ & $1.23(31)$ \\
$18$ & $0.99091(83)$ & $0.51113 (11)$ & $0.74(26)$ & & $22$ & $0.97658(84)$ & $0.494351(98)$ & $1.14(29)$ \\
$20$ & $1.01343(91)$ & $0.49428 (12)$ & $1.01(29)$ & & $24$ & $0.9918 (10)$ & $0.48081 (13)$ & $0.88(24)$ \\
$22$ & $1.03716(89)$ & $0.47935 (10)$ & $1.04(27)$ & & $26$ & $1.00912(98)$ & $0.46880 (10)$ & $1.03(25)$ \\
$24$ & $1.0539 (11)$ & $0.46540 (13)$ & $1.15(28)$ & & $28$ & $1.0244 (11)$ & $0.45765 (12)$ & $0.90(23)$ \\
     &               &                &            & & $30$ & $1.0401 (11)$ & $0.44730 (12)$ & $1.05(24)$ \\
     &               &                &            & & $32$ & $1.0560 (11)$ & $0.43810 (11)$ & $1.47(27)$ \\
\hline
\hline
\end  {tabular}
\label{tbl:raw_dat1_D}
\end  {center}
\end  {table}
%
%---------------------------------

%---------------------------------
%
%  Set E
%
\begin{table}[p]
\begin{center}
\caption{$g_{\rm FV}^2(L,L/a)$ and $Z_{\rm FV}^\phi(L,L/a)$ for various $(1/g^2,L/a)$ of set E.}
\medskip
\begin{tabular}{ccccccccc}
\hline
\hline
\multicolumn{4}{c}{$1/g^2=1.6050$} & \qquad\qquad & 
\multicolumn{4}{c}{$1/g^2=1.6346$}                               \\
\hline
$L/a$ & $g_{\rm FV}^2$& $Z_{\rm FV}^\phi$ & $\chi^2/N_{\rm df}$ & & 
$L/a$ & $g_{\rm FV}^2$& $Z_{\rm FV}^\phi$ & $\chi^2/N_{\rm df}$  \\
\hline
$ 6$ & $1.05866(98)$ & $0.64252 (24)$ & $1.58(70)$ & & $ 7$ & $1.0574 (11)$ & $0.61760 (29)$ & $0.36(31)$ \\
$ 7$ & $1.10194(90)$ & $0.60934 (18)$ & $1.16(54)$ & & $ 8$ & $1.09602(81)$ & $0.59098 (14)$ & $1.14(49)$ \\
$ 8$ & $1.14332(85)$ & $0.58142 (14)$ & $1.78(61)$ & & $ 9$ & $1.12986(76)$ & $0.56664 (11)$ & $1.28(48)$ \\
$ 9$ & $1.18096(80)$ & $0.55636 (12)$ & $0.64(34)$ & & $10$ & $1.16218(90)$ & $0.54500 (15)$ & $0.77(36)$ \\
$10$ & $1.21769(95)$ & $0.53412 (15)$ & $1.12(43)$ & & $11$ & $1.19398(87)$ & $0.52598 (13)$ & $0.72(33)$ \\
$11$ & $1.2535 (11)$ & $0.51437 (19)$ & $0.72(33)$ & & $12$ & $1.2261 (10)$ & $0.50844 (16)$ & $1.32(43)$ \\
$12$ & $1.2896 (11)$ & $0.49645 (16)$ & $1.37(44)$ & & $13$ & $1.25752(99)$ & $0.49268 (14)$ & $0.92(34)$ \\
     &               &                &            & & $14$ & $1.28756(97)$ & $0.47814 (12)$ & $1.26(38)$ \\
\hline
\hline
\multicolumn{4}{c}{$1/g^2=1.6589$} & \qquad\qquad & 
\multicolumn{4}{c}{$1/g^2=1.6982$}                               \\
\hline
$L/a$ & $g_{\rm FV}^2$& $Z_{\rm FV}^\phi$ & $\chi^2/N_{\rm df}$ & & 
$L/a$ & $g_{\rm FV}^2$& $Z_{\rm FV}^\phi$ & $\chi^2/N_{\rm df}$  \\
\hline
$ 8$ & $1.06073(78)$ & $0.59846 (13)$ & $0.80(41)$ & & $10$ & $1.05950(82)$ & $0.56631 (14)$ & $0.78(36)$ \\
$ 9$ & $1.09148(73)$ & $0.57462 (11)$ & $0.66(35)$ & & $12$ & $1.10810(90)$ & $0.53170 (14)$ & $1.42(44)$ \\
$10$ & $1.12050(87)$ & $0.55344 (14)$ & $0.66(33)$ & & $14$ & $1.15522(86)$ & $0.50309 (11)$ & $0.65(27)$ \\
$11$ & $1.14998(84)$ & $0.53491 (12)$ & $1.21(42)$ & & $16$ & $1.19819(97)$ & $0.47839 (12)$ & $1.19(35)$ \\
$12$ & $1.17791(97)$ & $0.51778 (15)$ & $0.69(31)$ & & $18$ & $1.2397 (11)$ & $0.45698 (13)$ & $0.91(29)$ \\
$13$ & $1.20647(94)$ & $0.50241 (13)$ & $1.12(38)$ & & $20$ & $1.2813 (12)$ & $1.43811 (14)$ & $0.90(27)$ \\
$14$ & $1.23363(92)$ & $0.48817 (12)$ & $1.47(41)$ & &      &               &                &            \\
$15$ & $1.2582 (12)$ & $0.47455 (19)$ & $1.25(37)$ & &      &               &                &            \\
$16$ & $1.2810 (14)$ & $0.46181 (22)$ & $1.26(37)$ & &      &               &                &            \\
\hline
\hline
\multicolumn{4}{c}{$1/g^2=1.7306$} & \qquad\qquad & 
\multicolumn{4}{c}{$1/g^2=1.7800$}                               \\
\hline
$L/a$ & $g_{\rm FV}^2$& $Z_{\rm FV}^\phi$ & $\chi^2/N_{\rm df}$ & & 
$L/a$ & $g_{\rm FV}^2$& $Z_{\rm FV}^\phi$ & $\chi^2/N_{\rm df}$  \\
\hline
$12$ & $1.05944(86)$ & $0.54277 (14)$ & $0.86(35)$ & & $16$ & $1.05797(84)$ & $0.50805 (11)$ & $1.18(34)$ \\
$14$ & $1.10256(82)$ & $0.51503 (11)$ & $0.84(31)$ & & $18$ & $1.08760(93)$ & $0.48799 (12)$ & $0.77(26)$ \\
$16$ & $1.1338 (12)$ & $0.48998 (20)$ & $0.70(27)$ & & $20$ & $1.1172 (10)$ & $0.47038 (13)$ & $0.91(27)$ \\
$18$ & $1.1731 (10)$ & $0.46981 (12)$ & $1.09(31)$ & & $22$ & $1.14777(87)$ & $0.454904(84)$ & $1.01(27)$ \\
$20$ & $1.2083 (11)$ & $0.45137 (13)$ & $0.99(28)$ & & $24$ & $1.1707 (12)$ & $0.44014 (14)$ & $0.93(25)$ \\
$22$ & $1.2470 (11)$ & $0.43537 (11)$ & $0.75(23)$ & & $26$ & $1.1976 (12)$ & $0.42709 (12)$ & $1.03(25)$ \\
$24$ & $1.2772 (13)$ & $0.42008 (15)$ & $0.92(25)$ & & $28$ & $1.2213 (13)$ & $0.41507 (13)$ & $1.36(28)$ \\
     &               &                &            & & $30$ & $1.2472 (12)$ & $0.40391 (11)$ & $1.11(24)$ \\
     &               &                &            & & $32$ & $1.2712 (13)$ & $0.39383 (12)$ & $0.81(20)$ \\
\hline
\hline
\end  {tabular}
\label{tbl:raw_dat1_E}
\end  {center}
\end  {table}
%
%---------------------------------

We can determine SSFs by using 
Eqs. (\ref{eqn:def_SSF_g2}) and (\ref{eqn:def_SSF_ph}). 
However, they are the SSFs on a lattice. 
We need to extrapolate the lattice SSFs to the continuum ones. 
As the preparation, in each set, 
we line up $g_{\rm FV}^2(L_0,L_0/a)$ to the specific value. 
As an example, we consider the set A in Table \ref{tbl:raw_dat1_A}. 
We adopt $u_0'\equiv 0.6755$ as the specific value. 
By using SSFs, we can evolve 
$g_{\rm FV}^2(L_0,L_0/a)=0.6752-0.6756$ 
to $g_{\rm FV}^2(s_0L_0,s_0L_0/a)=u_0'$ with some factor $s_0$. 
In this situation, $s_0$ is nearly equal to $1$, so that we can safely use 
the perturbative expression of the continuum SSFs. 
By solving 
$u_0' = \sigma^g_{\rm P}(s_0,g_{\rm FV}^2(L_0,L_0/a))$
numerically with Newton's method, we determine $s_0$. 
Then, we evaluate the lattice SSFs by 
\begin{eqnarray}
  \Sigma^g        (s,u_0',a/L_0)
  &=&
  \sigma^g_{\rm P}(s_0,g_{\rm FV}^2(sL_0,sL_0/a))\ ,
\\
  \Sigma^\phi     (s,u_0',a/L_0)
  &=&
  \frac{\sigma^\phi_{\rm P}(s_0,g_{\rm FV}^2   (sL_0,sL_0/a))\,
                                Z_{\rm FV}^\phi(sL_0,sL_0/a)}
       {\sigma^\phi_{\rm P}(s_0,g_{\rm FV}^2   ( L_0, L_0/a))\,
                                Z_{\rm FV}^\phi( L_0, L_0/a)}\ . 
\end  {eqnarray}
In the present study, 
$s\equiv L/L_0$ is a factor greater than $1$ but not more than $2$.  
The statistical errors are roughly estimated from 
\begin{eqnarray}
  \Delta (\Sigma^g(s,u_0',a/L_0))
  &=&
  \sqrt{\,
    \left[\,
      \frac{\partial \sigma^g_{\rm P}(s  ,u_0'  )}{\partial u}\,
      \frac{\partial \sigma^g_{\rm P}(s_0,u_0   )}{\partial u}\,\Delta(u_0)\,
    \right]^2
    + 
    \left[\,
      \frac{\partial \sigma^g_{\rm P}(s_0,u_1   )}{\partial u}\,\Delta(u_1)\,
    \right]^2
  }\ ,\quad
\\
  \Delta (\Sigma^\phi(s,u_0',a/L_0))
  &=&
  \sqrt{\,
    \left[\,
      \Delta
      \left(
        \frac{\sigma^\phi_{\rm P}(s_0,u_1)}{\sigma^\phi_{\rm P}(s_0,u_0)}
      \right)
      \frac{v_1}{v_0}\,
    \right]^2
    + 
    \left[\,
      \frac{\sigma^\phi_{\rm P}(s_0,u_1)}{\sigma^\phi_{\rm P}(s_0,u_0)}\ 
      \Delta
      \left(
        \frac{v_1}{v_0}
      \right)\,
    \right]^2
  }\ .
\end  {eqnarray}
The symbol $\Delta$ denotes the statistical error. 
We refer 
$g_{\rm FV}^2   ( L_0, L_0/a)$ by $u_0$, 
$g_{\rm FV}^2   (sL_0,sL_0/a)$ by $u_1$, 
$Z_{\rm FV}^\phi( L_0, L_0/a)$ by $v_0$, and 
$Z_{\rm FV}^\phi(sL_0,sL_0/a)$ by $v_1$ to simplify the expressions. 
$\Delta(\sigma^\phi_{\rm P}(s_0,u_1)/\sigma^\phi_{\rm P}(s_0,u_0))$ 
is estimated from 
\begin{equation}
  \Delta
  \left(
    \frac{\sigma^\phi_{\rm P}(s_0,u_1)}{\sigma^\phi_{\rm P}(s_0,u_0)}
  \right)
  =
  \sqrt{\,
    \left[\,
      E\,\Delta(u_0)\,
    \right]^2
    + 
    \left[\,
      \frac{1}{\sigma^\phi_{\rm P}(s_0,u_0)}\,
      \frac{\partial \sigma^\phi_{\rm P}(s_0,u_1)}{\partial u}\,
      \Delta(u_1)\,
    \right]^2
  }\ ,
\end  {equation}
where the coefficient $E$ is defined as 
\begin{eqnarray}
  E
  &\equiv & 
  -
  \frac{         \sigma^\phi_{\rm P}(s_0,u_1)}{\sigma^\phi_{\rm P}(s_0,u_0)^2}\,
  \frac{\partial \sigma^\phi_{\rm P}(s_0,u_0)}{\partial u}
\nonumber
\\
  & &
  -
  \left(\,
  \frac{                                    1}{\sigma^\phi_{\rm P}(s_0,u_0)  }\,
  \frac{\partial \sigma^\phi_{\rm P}(s_0,u_1)}{\partial s}
  -
  \frac{         \sigma^\phi_{\rm P}(s_0,u_1)}{\sigma^\phi_{\rm P}(s_0,u_0)^2}\,
  \frac{\partial \sigma^\phi_{\rm P}(s_0,u_0)}{\partial s}\,
  \right)
  \frac{\partial \sigma^g_{\rm P}   (s_0,u_0)/ \partial u}
       {\partial \sigma^g_{\rm P}   (s_0,u_0)/ \partial s}\,\ .
\end  {eqnarray}
In Tables \ref{tbl:raw_dat2_A}, 
          \ref{tbl:raw_dat2_B}, 
          \ref{tbl:raw_dat2_C}, 
          \ref{tbl:raw_dat2_D} and 
          \ref{tbl:raw_dat2_E}, 
we give 
$\Sigma^g   (s,u_0',a/L_0)$ and 
$\Sigma^\phi(s,u_0',a/L_0)$ for various $(L_0/a,s)$ of each set. 
Each set corresponds to 
$u_0'=0.6755$ for set A, 
$u_0'=0.7383$ for set B, 
$u_0'=0.8166$ for set C, 
$u_0'=0.9176$ for set D, and 
$u_0'=1.0595$ for set E. 
%
%---------------------------------
%
%  Set A
%
\begin{table}[p]
\begin{center}
\caption{$\Sigma^g(s,u_0',a/L_0)$ and $\Sigma^\phi(s,u_0',a/L_0)$ for set A $(u_0'=0.6755)$.}
\medskip
\begin{tabular}{ccccccc}
\hline
\hline
\multicolumn{3}{c}{$L_0/a= 6$} & \qquad\qquad & 
\multicolumn{3}{c}{$L_0/a= 7$}                                                       \\
\hline
$s$ & $\Sigma^g$& $\Sigma^\phi$ & & 
$s$ & $\Sigma^g$& $\Sigma^\phi$                                                      \\
\hline
$ 7/ 6$ & $0.68829(75)$ & $0.96704(20)$ & & $ 8/ 7$ & $0.68631(95)$ & $0.97135(35)$  \\
$ 8/ 6$ & $0.70076(73)$ & $0.93956(18)$ & & $ 9/ 7$ & $0.69735(78)$ & $0.94712(20)$  \\
$ 9/ 6$ & $0.71095(71)$ & $0.91498(17)$ & & $10/ 7$ & $0.70616(82)$ & $0.92488(22)$  \\
$10/ 6$ & $0.72039(76)$ & $0.89308(19)$ & & $11/ 7$ & $0.71512(80)$ & $0.90537(21)$  \\
$11/ 6$ & $0.73004(75)$ & $0.87392(18)$ & & $12/ 7$ & $0.72447(79)$ & $0.88776(20)$  \\
$12/ 6$ & $0.73902(80)$ & $0.85634(20)$ & & $13/ 7$ & $0.73106(84)$ & $0.87113(22)$  \\
        &               &               & & $14/ 7$ & $0.73945(83)$ & $0.85635(21)$  \\
\hline
\hline
\multicolumn{3}{c}{$L_0/a= 7$} & \qquad\qquad & 
\multicolumn{3}{c}{$L_0/a=10$}                                                       \\
\hline
$s$ & $\Sigma^g$& $\Sigma^\phi$ & & 
$s$ & $\Sigma^g$& $\Sigma^\phi$                                                      \\
\hline
$ 9/ 8$ & $0.68488(73)$ & $0.97480(17)$ & & $12/10$ & $0.69260(77)$ & $0.96161(18)$  \\
$10/ 8$ & $0.69337(78)$ & $0.95236(20)$ & & $14/10$ & $0.70628(80)$ & $0.92914(19)$  \\
$11/ 8$ & $0.70185(76)$ & $0.93253(18)$ & & $16/10$ & $0.71828(78)$ & $0.90152(18)$  \\
$12/ 8$ & $0.71094(75)$ & $0.91479(17)$ & & $18/10$ & $0.72900(86)$ & $0.87715(21)$  \\
$13/ 8$ & $0.71847(80)$ & $0.89818(19)$ & & $20/10$ & $0.74207(80)$ & $0.85653(19)$  \\
$14/ 8$ & $0.72562(78)$ & $0.88308(19)$ & &         &               &                \\
$15/ 8$ & $0.73212(77)$ & $0.86900(18)$ & &         &               &                \\
$16/ 8$ & $0.73799(76)$ & $0.85600(18)$ & &         &               &                \\
\hline
\hline
\multicolumn{3}{c}{$L_0/a=12$} & \qquad\qquad & 
\multicolumn{3}{c}{$L_0/a=16$}                                                       \\
\hline
$s$ & $\Sigma^g$& $\Sigma^\phi$ & & 
$s$ & $\Sigma^g$& $\Sigma^\phi$                                                      \\
\hline
$14/12$ & $0.68985(70)$ & $0.96742(14)$ & & $18/16$ & $0.6867 (10)$ & $0.97525(26)$  \\
$16/12$ & $0.70071(73)$ & $0.93920(15)$ & & $20/16$ & $0.6962 (10)$ & $0.95304(27)$  \\
$18/12$ & $0.71007(81)$ & $0.91444(18)$ & & $22/16$ & $0.70624(98)$ & $0.93337(25)$  \\
$20/12$ & $0.72074(84)$ & $0.89286(19)$ & & $24/16$ & $0.7128 (10)$ & $0.91501(26)$  \\
$22/12$ & $0.73000(83)$ & $0.87351(18)$ & & $26/16$ & $0.7202 (10)$ & $0.89838(26)$  \\
$24/12$ & $0.73714(92)$ & $0.85574(21)$ & & $28/16$ & $0.7275 (11)$ & $0.88317(28)$  \\
        &               &               & & $30/16$ & $0.7346 (11)$ & $0.86917(27)$  \\
        &               &               & & $32/16$ & $0.7395 (11)$ & $0.85597(28)$  \\
\hline
\hline
\end  {tabular}
\label{tbl:raw_dat2_A}
\end  {center}
\end  {table}
%
%---------------------------------

%---------------------------------
%
%  Set B
%
\begin{table}[p]
\begin{center}
\caption{$\Sigma^g(s,u_0',a/L_0)$ and $\Sigma^\phi(s,u_0',a/L_0)$ for set B $(u_0'=0.7383)$.}
\medskip
\begin{tabular}{ccccccc}
\hline
\hline
\multicolumn{3}{c}{$L_0/a= 6$} & \qquad\qquad & 
\multicolumn{3}{c}{$L_0/a= 7$}                                                       \\
\hline
$s$ & $\Sigma^g$& $\Sigma^\phi$ & & 
$s$ & $\Sigma^g$& $\Sigma^\phi$                                                      \\
\hline
$ 7/ 6$ & $0.75426(98)$ & $0.96428(30)$ & & $ 8/ 7$ & $0.75214(96)$ & $0.96903(29)$  \\
$ 8/ 6$ & $0.76981(96)$ & $0.93420(28)$ & & $ 9/ 7$ & $0.76547(86)$ & $0.94215(22)$  \\
$ 9/ 6$ & $0.78323(94)$ & $0.90746(26)$ & & $10/ 7$ & $0.77552(91)$ & $0.91770(24)$  \\
$10/ 6$ & $0.7933 (11)$ & $0.88298(33)$ & & $11/ 7$ & $0.78780(90)$ & $0.89666(23)$  \\
$11/ 6$ & $0.80687(97)$ & $0.86251(28)$ & & $12/ 7$ & $0.79796(95)$ & $0.87710(25)$  \\
$12/ 6$ & $0.8175 (10)$ & $0.84313(30)$ & & $13/ 7$ & $0.80826(93)$ & $0.85938(24)$  \\
        &               &               & & $14/ 7$ & $0.81781(92)$ & $0.84318(24)$  \\
\hline
\hline
\multicolumn{3}{c}{$L_0/a= 8$} & \qquad\qquad & 
\multicolumn{3}{c}{$L_0/a=10$}                                                       \\
\hline
$s$ & $\Sigma^g$& $\Sigma^\phi$ & & 
$s$ & $\Sigma^g$& $\Sigma^\phi$                                                      \\
\hline
$ 9/ 8$ & $0.74995(81)$ & $0.97245(19)$ & & $12/10$ & $0.7581 (10)$ & $0.95790(28)$  \\
$10/ 8$ & $0.75916(94)$ & $0.94738(27)$ & & $14/10$ & $0.77545(99)$ & $0.92270(26)$  \\
$11/ 8$ & $0.77085(85)$ & $0.92627(20)$ & & $16/10$ & $0.7885 (11)$ & $0.89182(33)$  \\
$12/ 8$ & $0.78200(84)$ & $0.90685(19)$ & & $18/10$ & $0.8042 (11)$ & $0.86616(28)$  \\
$13/ 8$ & $0.79052(88)$ & $0.88861(21)$ & & $20/10$ & $0.8178 (11)$ & $0.84299(29)$  \\
$14/ 8$ & $0.80004(87)$ & $0.87227(20)$ & &         &               &                \\
$15/ 8$ & $0.80689(99)$ & $0.85635(26)$ & &         &               &                \\
$16/ 8$ & $0.8140 (11)$ & $0.84212(28)$ & &         &               &                \\
\hline
\hline
\multicolumn{3}{c}{$L_0/a=12$} & \qquad\qquad & 
\multicolumn{3}{c}{$L_0/a=16$}                                                       \\
\hline
$s$ & $\Sigma^g$& $\Sigma^\phi$ & & 
$s$ & $\Sigma^g$& $\Sigma^\phi$                                                      \\
\hline
$14/12$ & $0.75519(91)$ & $0.96431(21)$ & & $18/16$ & $0.75109(95)$ & $0.97267(21)$  \\
$16/12$ & $0.7682 (11)$ & $0.93305(29)$ & & $20/16$ & $0.76155(97)$ & $0.94812(22)$  \\
$18/12$ & $0.78247(98)$ & $0.90682(23)$ & & $22/16$ & $0.77384(97)$ & $0.92671(21)$  \\
$20/12$ & $0.7945 (10)$ & $0.88316(24)$ & & $24/16$ & $0.7824 (11)$ & $0.90672(24)$  \\
$22/12$ & $0.8065 (10)$ & $0.86210(23)$ & & $26/16$ & $0.7919 (10)$ & $0.88863(22)$  \\
$24/12$ & $0.8164 (11)$ & $0.84277(26)$ & & $28/16$ & $0.7998 (11)$ & $0.87196(26)$  \\
        &               &               & & $30/16$ & $0.8090 (11)$ & $0.85674(24)$  \\
        &               &               & & $32/16$ & $0.8187 (11)$ & $0.84323(24)$  \\
\hline
\hline
\end  {tabular}
\label{tbl:raw_dat2_B}
\end  {center}
\end  {table}
%
%---------------------------------

%---------------------------------
%
%  Set C
%
\begin{table}[p]
\begin{center}
\caption{$\Sigma^g(s,u_0',a/L_0)$ and $\Sigma^\phi(s,u_0',a/L_0)$ for set C $(u_0'=0.8166)$.}
\medskip
\begin{tabular}{ccccccc}
\hline
\hline
\multicolumn{3}{c}{$L_0/a= 6$} & \qquad\qquad & 
\multicolumn{3}{c}{$L_0/a= 7$}                                                       \\
\hline
$s$ & $\Sigma^g$& $\Sigma^\phi$ & & 
$s$ & $\Sigma^g$& $\Sigma^\phi$                                                      \\
\hline
$ 7/ 6$ & $0.8355 (13)$ & $0.95926(58)$ & & $ 8/ 7$ & $0.8359 (12)$ & $0.96644(37)$  \\
$ 8/ 6$ & $0.8550 (13)$ & $0.92610(47)$ & & $ 9/ 7$ & $0.8522 (12)$ & $0.93639(35)$  \\
$ 9/ 6$ & $0.8744 (11)$ & $0.89743(30)$ & & $10/ 7$ & $0.8657 (13)$ & $0.90914(41)$  \\
$10/ 6$ & $0.8903 (11)$ & $0.87103(32)$ & & $11/ 7$ & $0.8805 (13)$ & $0.88563(39)$  \\
$11/ 6$ & $0.9055 (11)$ & $0.84774(31)$ & & $12/ 7$ & $0.8958 (12)$ & $0.86435(37)$  \\
$12/ 6$ & $0.9214 (12)$ & $0.82678(33)$ & & $13/ 7$ & $0.9092 (12)$ & $0.84476(37)$  \\
        &               &               & & $14/ 7$ & $0.9219 (12)$ & $0.82682(36)$  \\
\hline
\hline
\multicolumn{3}{c}{$L_0/a= 8$} & \qquad\qquad & 
\multicolumn{3}{c}{$L_0/a=10$}                                                       \\
\hline
$s$ & $\Sigma^g$& $\Sigma^\phi$ & & 
$s$ & $\Sigma^g$& $\Sigma^\phi$                                                      \\
\hline
$ 9/ 8$ & $0.83175(92)$ & $0.96953(21)$ & & $12/10$ & $0.8430 (12)$ & $0.95392(32)$  \\
$10/ 8$ & $0.8453 (11)$ & $0.94209(30)$ & & $14/10$ & $0.8644 (12)$ & $0.91453(32)$  \\
$11/ 8$ & $0.86030(97)$ & $0.91863(22)$ & & $16/10$ & $0.8821 (13)$ & $0.88053(37)$  \\
$12/ 8$ & $0.8732 (10)$ & $0.89678(25)$ & & $18/10$ & $0.9021 (12)$ & $0.85218(31)$  \\
$13/ 8$ & $0.8864 (10)$ & $0.87710(24)$ & & $20/10$ & $0.9211 (12)$ & $0.82675(33)$  \\
$14/ 8$ & $0.89713(99)$ & $0.85874(23)$ & &         &               &                \\
$15/ 8$ & $0.9069 (11)$ & $0.84123(29)$ & &         &               &                \\
$16/ 8$ & $0.9175 (12)$ & $0.82569(32)$ & &         &               &                \\
\hline
\hline
\multicolumn{3}{c}{$L_0/a=12$} & \qquad\qquad & 
\multicolumn{3}{c}{$L_0/a=16$}                                                       \\
\hline
$s$ & $\Sigma^g$& $\Sigma^\phi$ & & 
$s$ & $\Sigma^g$& $\Sigma^\phi$                                                      \\
\hline
$14/12$ & $0.8379 (10)$ & $0.96052(24)$ & & $18/16$ & $0.8338 (13)$ & $0.97034(32)$  \\
$16/12$ & $0.8535 (12)$ & $0.92571(32)$ & & $20/16$ & $0.8483 (13)$ & $0.94341(32)$  \\
$18/12$ & $0.8723 (11)$ & $0.89685(26)$ & & $22/16$ & $0.8621 (13)$ & $0.91937(31)$  \\
$20/12$ & $0.8889 (12)$ & $0.87096(27)$ & & $24/16$ & $0.8736 (14)$ & $0.89731(34)$  \\
$22/12$ & $0.9048 (11)$ & $0.84773(26)$ & & $26/16$ & $0.8867 (14)$ & $0.87753(33)$  \\
$24/12$ & $0.9178 (12)$ & $0.82632(29)$ & & $28/16$ & $0.8967 (15)$ & $0.85903(36)$  \\
        &               &               & & $30/16$ & $0.9077 (14)$ & $0.84198(35)$  \\
        &               &               & & $32/16$ & $0.9207 (14)$ & $0.82714(34)$  \\
\hline
\hline
\end  {tabular}
\label{tbl:raw_dat2_C}
\end  {center}
\end  {table}
%
%---------------------------------

%---------------------------------
%
%  Set D
%
\begin{table}[p]
\begin{center}
\caption{$\Sigma^g(s,u_0',a/L_0)$ and $\Sigma^\phi(s,u_0',a/L_0)$ for set D $(u_0'=0.9176)$.}
\medskip
\begin{tabular}{ccccccc}
\hline
\hline
\multicolumn{3}{c}{$L_0/a= 6$} & \qquad\qquad & 
\multicolumn{3}{c}{$L_0/a= 7$}                                                       \\
\hline
$s$ & $\Sigma^g$& $\Sigma^\phi$ & & 
$s$ & $\Sigma^g$& $\Sigma^\phi$                                                      \\
\hline
$ 7/ 6$ & $0.9447 (13)$ & $0.95505(39)$ & & $ 8/ 7$ & $0.9428 (12)$ & $0.96185(30)$  \\
$ 8/ 6$ & $0.9728 (13)$ & $0.91796(36)$ & & $ 9/ 7$ & $0.9662 (12)$ & $0.92817(28)$  \\
$ 9/ 6$ & $0.9969 (13)$ & $0.88453(34)$ & & $10/ 7$ & $0.9849 (13)$ & $0.89734(38)$  \\
$10/ 6$ & $1.0191 (13)$ & $0.85469(36)$ & & $11/ 7$ & $1.0052 (13)$ & $0.87092(34)$  \\
$11/ 6$ & $1.0422 (13)$ & $0.82873(36)$ & & $12/ 7$ & $1.0260 (13)$ & $0.84700(32)$  \\
$12/ 6$ & $1.0631 (14)$ & $0.80452(38)$ & & $13/ 7$ & $1.0461 (13)$ & $0.82517(31)$  \\
        &               &               & & $14/ 7$ & $1.0637 (12)$ & $0.80476(30)$  \\
\hline
\hline
\multicolumn{3}{c}{$L_0/a= 8$} & \qquad\qquad & 
\multicolumn{3}{c}{$L_0/a=10$}                                                       \\
\hline
$s$ & $\Sigma^g$& $\Sigma^\phi$ & & 
$s$ & $\Sigma^g$& $\Sigma^\phi$                                                      \\
\hline
$ 9/ 8$ & $0.9393 (11)$ & $0.96599(24)$ & & $12/10$ & $0.9524 (13)$ & $0.94778(36)$  \\
$10/ 8$ & $0.9571 (12)$ & $0.93501(35)$ & & $14/10$ & $0.9824 (15)$ & $0.90361(42)$  \\
$11/ 8$ & $0.9768 (11)$ & $0.90847(25)$ & & $16/10$ & $1.0091 (14)$ & $0.86618(34)$  \\
$12/ 8$ & $0.9944 (12)$ & $0.88376(29)$ & & $18/10$ & $1.0347 (14)$ & $0.83360(36)$  \\
$13/ 8$ & $1.0125 (12)$ & $0.86160(27)$ & & $20/10$ & $1.0606 (15)$ & $0.80490(37)$  \\
$14/ 8$ & $1.0288 (12)$ & $0.84112(26)$ & &         &               &                \\
$15/ 8$ & $1.0444 (13)$ & $0.82170(33)$ & &         &               &                \\
$16/ 8$ & $1.0595 (12)$ & $0.80439(28)$ & &         &               &                \\
\hline
\hline
\multicolumn{3}{c}{$L_0/a=12$} & \qquad\qquad & 
\multicolumn{3}{c}{$L_0/a=16$}                                                       \\
\hline
$s$ & $\Sigma^g$& $\Sigma^\phi$ & & 
$s$ & $\Sigma^g$& $\Sigma^\phi$                                                      \\
\hline
$14/12$ & $0.9439 (12)$ & $0.95521(27)$ & & $18/16$ & $0.9378 (15)$ & $0.96610(36)$  \\
$16/12$ & $0.9668 (14)$ & $0.91629(37)$ & & $20/16$ & $0.9571 (15)$ & $0.93587(37)$  \\
$18/12$ & $0.9907 (13)$ & $0.88382(29)$ & & $22/16$ & $0.9767 (15)$ & $0.90915(35)$  \\
$20/12$ & $1.0132 (13)$ & $0.85469(30)$ & & $24/16$ & $0.9919 (16)$ & $0.88424(38)$  \\
$22/12$ & $1.0369 (13)$ & $0.82889(29)$ & & $26/16$ & $1.0093 (16)$ & $0.86215(37)$  \\
$24/12$ & $1.0536 (14)$ & $0.80476(33)$ & & $28/16$ & $1.0246 (16)$ & $0.84165(38)$  \\
        &               &               & & $30/16$ & $1.0403 (17)$ & $0.82261(39)$  \\
        &               &               & & $32/16$ & $1.0562 (16)$ & $0.80567(38)$  \\
\hline
\hline
\end  {tabular}
\label{tbl:raw_dat2_D}
\end  {center}
\end  {table}
%
%---------------------------------

%---------------------------------
%
%  Set E
%
\begin{table}[p]
\begin{center}
\caption{$\Sigma^g(s,u_0',a/L_0)$ and $\Sigma^\phi(s,u_0',a/L_0)$ for set E $(u_0'=1.0595)$.}
\medskip
\begin{tabular}{ccccccc}
\hline
\hline
\multicolumn{3}{c}{$L_0/a= 6$} & \qquad\qquad & 
\multicolumn{3}{c}{$L_0/a= 7$}                                                       \\
\hline
$s$ & $\Sigma^g$& $\Sigma^\phi$ & & 
$s$ & $\Sigma^g$& $\Sigma^\phi$                                                      \\
\hline
$ 7/ 6$ & $1.1029 (16)$ & $0.94831(46)$ & & $ 8/ 7$ & $1.0983 (17)$ & $0.95677(51)$  \\
$ 8/ 6$ & $1.1443 (16)$ & $0.90480(42)$ & & $ 9/ 7$ & $1.1323 (17)$ & $0.91727(48)$  \\
$ 9/ 6$ & $1.1820 (16)$ & $0.86577(41)$ & & $10/ 7$ & $1.1648 (18)$ & $0.88216(50)$  \\
$10/ 6$ & $1.2188 (17)$ & $0.83113(44)$ & & $11/ 7$ & $1.1968 (17)$ & $0.85129(49)$  \\
$11/ 6$ & $1.2548 (18)$ & $0.80037(48)$ & & $12/ 7$ & $1.2291 (18)$ & $0.82282(51)$  \\
$12/ 6$ & $1.2909 (17)$ & $0.77243(46)$ & & $13/ 7$ & $1.2607 (18)$ & $0.79723(50)$  \\
        &               &               & & $14/ 7$ & $1.2909 (18)$ & $0.77363(50)$  \\
\hline
\hline
\multicolumn{3}{c}{$L_0/a= 8$} & \qquad\qquad & 
\multicolumn{3}{c}{$L_0/a=10$}                                                       \\
\hline
$s$ & $\Sigma^g$& $\Sigma^\phi$ & & 
$s$ & $\Sigma^g$& $\Sigma^\phi$                                                      \\
\hline
$ 9/ 8$ & $1.0902 (13)$ & $0.96023(29)$ & & $12/10$ & $1.1081 (15)$ & $0.93888(35)$  \\
$10/ 8$ & $1.1191 (14)$ & $0.92487(32)$ & & $14/10$ & $1.1552 (14)$ & $0.88836(31)$  \\
$11/ 8$ & $1.1485 (14)$ & $0.89396(30)$ & & $16/10$ & $1.1982 (15)$ & $0.84474(33)$  \\
$12/ 8$ & $1.1764 (14)$ & $0.86538(34)$ & & $18/10$ & $1.2397 (16)$ & $0.80695(35)$  \\
$13/ 8$ & $1.2048 (14)$ & $0.83973(32)$ & & $20/10$ & $1.2813 (16)$ & $0.77362(37)$  \\
$14/ 8$ & $1.2319 (14)$ & $0.81597(31)$ & &         &               &                \\
$15/ 8$ & $1.2564 (16)$ & $0.79323(40)$ & &         &               &                \\
$16/ 8$ & $1.2791 (17)$ & $0.77197(45)$ & &         &               &                \\
\hline
\hline
\multicolumn{3}{c}{$L_0/a=12$} & \qquad\qquad & 
\multicolumn{3}{c}{$L_0/a=16$}                                                       \\
\hline
$s$ & $\Sigma^g$& $\Sigma^\phi$ & & 
$s$ & $\Sigma^g$& $\Sigma^\phi$                                                      \\
\hline
$14/12$ & $1.1026 (15)$ & $0.94887(32)$ & & $18/16$ & $1.0892 (15)$ & $0.96045(32)$  \\
$16/12$ & $1.1339 (17)$ & $0.90272(44)$ & & $20/16$ & $1.1190 (15)$ & $0.92574(33)$  \\
$18/12$ & $1.1731 (16)$ & $0.86557(34)$ & & $22/16$ & $1.1496 (15)$ & $0.89521(28)$  \\
$20/12$ & $1.2083 (16)$ & $0.83158(36)$ & & $24/16$ & $1.1727 (17)$ & $0.86611(35)$  \\
$22/12$ & $1.2471 (16)$ & $0.80210(35)$ & & $26/16$ & $1.1996 (17)$ & $0.84037(33)$  \\
$24/12$ & $1.2773 (18)$ & $0.77393(40)$ & & $28/16$ & $1.2234 (17)$ & $0.81670(35)$  \\
        &               &               & & $30/16$ & $1.2494 (17)$ & $0.79469(34)$  \\
        &               &               & & $32/16$ & $1.2735 (18)$ & $0.77481(35)$  \\
\hline
\hline
\end  {tabular}
\label{tbl:raw_dat2_E}
\end  {center}
\end  {table}
%
%---------------------------------

\begin{figure}[p]
\begin{center}
\includegraphics[width=160mm]{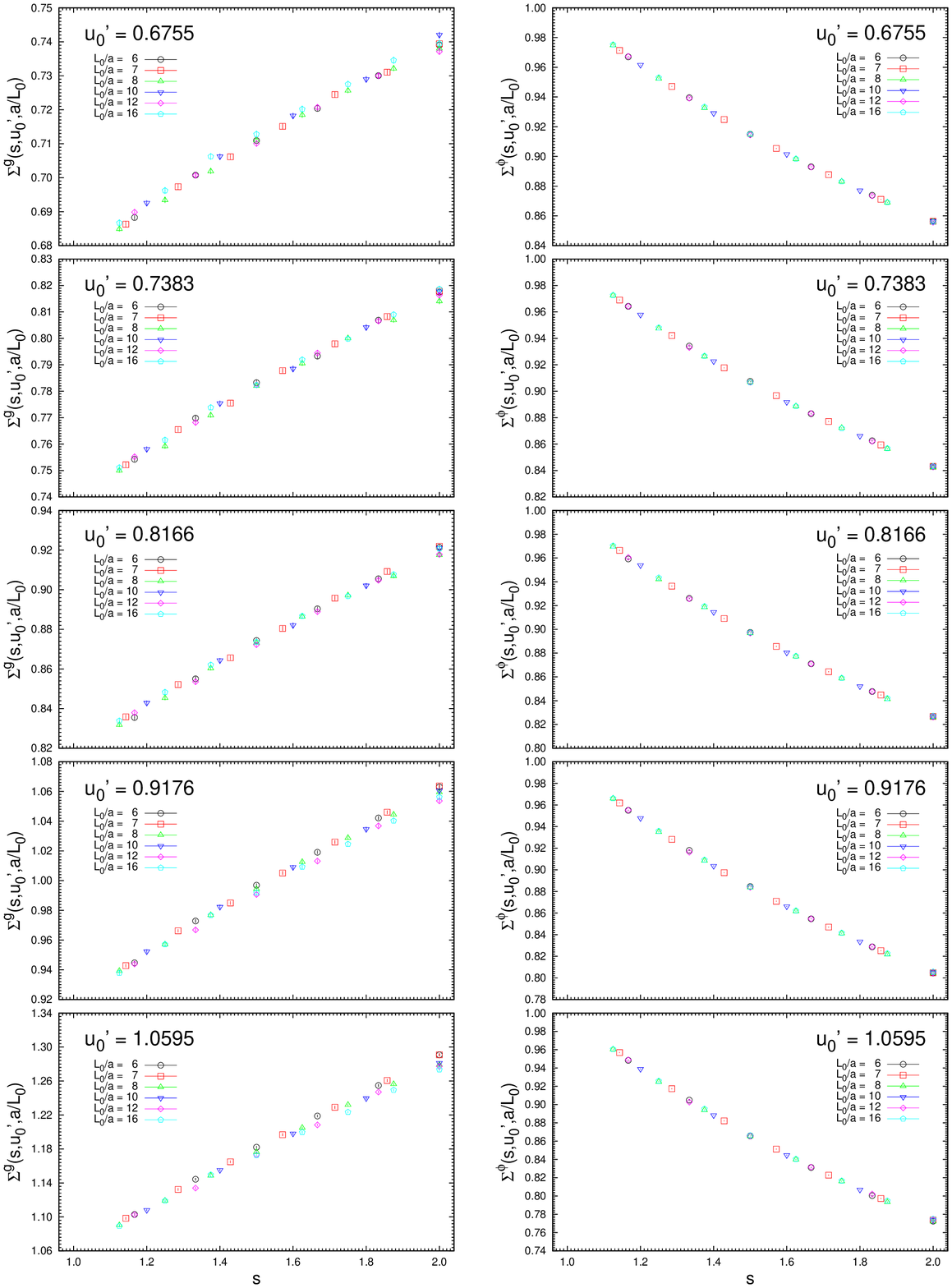}
\caption{$s$-dependence of 
  $\Sigma^g   (s,u_0',a/L_0)$ and 
  $\Sigma^\phi(s,u_0',a/L_0)$ for various $(u_0',L_0/a)$. }
\label{fig:overview}
\end  {center}
\end  {figure}

In Fig. \ref{fig:overview}, 
we show the $s$-dependence of 
$\Sigma^g   (s,u_0',a/L_0)$ and 
$\Sigma^\phi(s,u_0',a/L_0)$ for various $(u_0',L_0/a)$. 
While the data points of 
$\Sigma^\phi(s,u_0',a/L_0)$ are almost located on a single curve 
without depending on $L_0/a$\,, 
the data points of 
$\Sigma^g   (s,u_0',a/L_0)$ show a larger fluctuation. 
One of the reasons is considered as follows. 
The renormalized coupling $g_{\rm FV}^2$ is defined 
by multiplying $L/a$ to the dimensionless mass gap $Ma$. 
As the results, 
the uncertainty of $Ma$ due to the fit-range dependence 
is amplified in the value of $g_{\rm FV}^2$\,. 
However, 
$\Sigma^g   (s,u_0',a/L_0)$ might have the $(L_0/a)$-dependence 
beyond this uncertainty. 
In Sec. \ref{ssec:Continuum_limit}, 
we discuss the $(L_0/a)$-dependence of 
$\Sigma^g   (s,u_0',a/L_0)$  and 
$\Sigma^\phi(s,u_0',a/L_0)$, and 
evaluate the values at $(s,a/L_0)=(2,0)$. 
%
%---------------------------------
%
%
%---------------------------------
%
% @@@ =========================================================================
%
\subsection{Continuum limit}
\label{ssec:Continuum_limit}
The continuum limit is nothing less than the limit of $a/L_0\to 0$\,. 
We extract the values of 
$\Sigma^{g,\phi}(s,u_0',a/L_0)$ at $(s,a/L_0)=(2,0)$ by fit for each $u_0'$\,. 
We use the fitting forms as 
\begin{eqnarray}
  \Sigma^g   (s,u_0',a/L_0)
  &=&
  u_0'
  +
  \sum_k\,W_k(s,a/L_0)\,\Sigma_k^g   (u_0')\ ,
\label{eqn:fit_form_Sigma_g2}
\\
  \Sigma^\phi(s,u_0',a/L_0)
  &=&
  1
  +
  \sum_k\,W_k(s,a/L_0)\,\Sigma_k^\phi(u_0')\ .
\label{eqn:fit_form_Sigma_ph}
\end  {eqnarray}
The function $W_k(s,a/L_0)$ is defined by 
\begin{equation}
  W_{(i-1)\,j_{\rm max}+j}\,(s,a/L_0)
  \equiv 
  (\ln s)^i\,(a/L_0)^{2j}\ ,
\label{eqn:def_W_func}
\end  {equation}
for $i=1,\cdots,i_{\rm max}$ and $j=0,\cdots,j_{\rm max}-1$\,.
These are motivated from 
Eqs. (\ref{eqn:exp2}), 
     (\ref{eqn:coeff1}) and 
     (\ref{eqn:coeff2}). 
$\Sigma_k^g   (u_0')$ and 
$\Sigma_k^\phi(u_0')$ are free parameters. 
They are determined for each $u_0'$ by minimizing $\chi^2$, 
in other words, solving the linear simultaneous equation 
\begin{equation}
  \sum_l\,A^{g,\phi}_{kl}(u_0')\,\Sigma^{g,\phi}_l(u_0') 
  = 
  B^{g,\phi}_k(u_0')\ ,
\end  {equation}
with 
\begin{equation}
  A^{g,\phi}_{kl}(u_0')
  \equiv 
  \sum_{s,\,a/L_0}
  \frac{W_k(s,a/L_0)\,W_l(s,a/L_0)}
       {\Delta\left(\Sigma^{g,\phi}(s,u_0',a/L_0)\right)^2}\ ,
\end  {equation}
\begin{equation}
  B^{g,\phi}_k(u_0')
  \equiv 
  \sum_{s,\,a/L_0}
  \frac{\left(\bar{\Sigma}^{g,\phi}(s,u_0',a/L_0)-\{u_0',1\}\right)
        W_k(s,a/L_0)}
       {\Delta\left(\Sigma^{g,\phi}(s,u_0',a/L_0)\right)^2}\ .
\end  {equation}
Here 
$\bar{\Sigma}^{g,\phi}(s,u_0',a/L_0)$ denotes 
the expectation value of $\Sigma^{g,\phi}(s,u_0',a/L_0)$\,. 
$\{u_0',1\}$ means 
$u_0'$ for $B^g_k   (u_0')$, and 
$1$    for $B^\phi_k(u_0')$\,.
The summation is taken over $(s,a/L_0)$ used in the measurement.

We need an attention for the evaluation of 
the statistical errors of $\Sigma^{g,\phi}(s,u_0',0)$\,. 
When $(\ln s)$ takes non-zero value, 
correlation between the fitting parameters must be considered. 
The variance-covariance matrix 
for the parameter $\Sigma^{g,\phi}_k(u_0')$ can be described by 
$\left(A^{g,\phi}(u_0')\right)^{-1}$\,.
Thus, the statistical errors are evaluated by 
\begin{equation}
  \Delta
  \left(\Sigma^{g,\phi}(s,u_0',0)\right)^2
  =
  \sum_{k,\,l}\,
  \left(A^{g,\phi}(u_0')\right)^{-1}_{kl}\,
  W_k(s,0)\,W_l(s,0)\ .
\label{eqn:stat_err1}
\end  {equation}

Eqs. (\ref{eqn:rel1}) and 
     (\ref{eqn:rel2}) 
suggest that 
we can evaluate 
the $\beta$ function and the anomalous dimension at 
$\sigma^g(s,u_0')=\Sigma^g(s,u_0',0)$ 
with $\Sigma^{g,\phi}_k(u_0')$ determined by the fit. 
We evaluate the expectation values by 
\begin{eqnarray}
  -s\,\frac{\partial\,    \Sigma^g   (s,u_0',0)}{\partial s}
  &=&
  -s\,
  \sum_k\, 
  \frac{\partial\,W_k(s,0)                 }{\partial s}\,
  \Sigma^g_k(u_0')\ ,
\\
  -s\,\frac{\partial\,\ln \Sigma^\phi(s,u_0',0)}{\partial s}
  &=&
  -\frac{s}{\Sigma^\phi(s,u_0',0)}
  \sum_k\,
  \frac{\partial\,W_k(s,0)                 }{\partial s}\,
  \Sigma^\phi_\phi(u_0')\ .
\end  {eqnarray}
The statistical error of $-s\,\partial\,\Sigma^g(s,u_0',0)/\partial s$ 
is evaluated by 
\begin{equation}
  \Delta
  \left(-s\,\frac{\partial\,\Sigma^g(s,u_0',0)}{\partial s}\right)^2
  =
  s^2\,
  \sum_{k,\,l}\,
  \left(A^g(u_0')\right)^{-1}_{kl}\,
  \frac{\partial\,W_k(s,0)}{\partial s}\,
  \frac{\partial\,W_l(s,0)}{\partial s}\ .
\label{eqn:stat_err2}
\end  {equation}
On the other hands, 
to evaluate 
the statistical error of $-s\,\partial\,\ln\Sigma^\phi(s,u_0',0)/\partial s$, 
we use an approximation $(\ln\Sigma^\phi)\simeq \Sigma^\phi-1$\,, 
and calculate 
\begin{eqnarray}
  \Delta
  \left(-s\,\frac{\partial\,\ln\Sigma^\phi(s,u_0',0)}{\partial s}\right)^2
  &\simeq &
  \left(-s\,\frac{\partial\left(\,\Sigma^\phi(s,u_0',0)-1\,\right)}
                 {\partial s}\right)^2
\nonumber  \\
  &=&
  s^2\,
  \sum_{k,\,l}\,
  \left(A^\phi(u_0')\right)^{-1}_{kl}\,
  \frac{\partial\,W_k(s,0)}{\partial s}\,
  \frac{\partial\,W_l(s,0)}{\partial s}\ .
\label{eqn:stat_err3}
\end  {eqnarray}
From Tables \ref{tbl:raw_dat2_A}, 
            \ref{tbl:raw_dat2_B}, 
            \ref{tbl:raw_dat2_C}, 
            \ref{tbl:raw_dat2_D} and 
            \ref{tbl:raw_dat2_E}, 
we find $|\Sigma^\phi-1|<0.23$ in $u_0'=0.6755-1.0595$\,. 
To compensate the underestimation due to the approximation, 
we multiply the factor $1+0.23^2=1.0529$ in the evaluation of 
$\Delta\left(-s\,\partial\,\ln\Sigma^\phi(s,u_0',0)/\partial s\right)$\,.

In Table \ref{tbl:fit_result}, 
we give the fitting results to Eqs. 
(\ref{eqn:fit_form_Sigma_g2}) and (\ref{eqn:fit_form_Sigma_ph}) 
with $(i_{\rm max},j_{\rm max})=(2,2)$\,. 
We show 
$\sigma^g   (2,u_0')=\Sigma^g   (2,u_0',0)$ and 
$\sigma^\phi(2,u_0')=\Sigma^\phi(2,u_0',0)$ for each $u_0'$\,. 
We also list the $\beta$ function and the anormalous dimension 
at $\sigma^g(2,u_0')=\Sigma^g(2,u_0',0)$\,. 
%
%---------------------------------
%
\begin{table}[p]
\begin{center}
\caption{The fitting results to Eqs. 
(\ref{eqn:fit_form_Sigma_g2}) and (\ref{eqn:fit_form_Sigma_ph}) 
with $(i_{\rm max},j_{\rm max})=(2,2)$\,. 
We show 
$\sigma^g   (2,u_0')=\Sigma^g   (2,u_0',0)$ and 
$\sigma^\phi(2,u_0')=\Sigma^\phi(2,u_0',0)$ for each $u_0'$\,. 
We also list the $\beta$ function and the anormalous dimension 
at $\sigma^g(2,u_0')=\Sigma^g(2,u_0',0)$\,. }
\medskip
\begin{tabular}{ccccccc}
\hline
\hline
$u_0'$                       & $\qquad$ & $0.6755$        & $0.7383$        & $0.8166$        & $0.9176$        & $1.0595$         \\
\hline
$\chi^2/N_{\rm df}$ for $\Sigma^g$    & & $ 1.57$         & $ 1.20$         & $ 1.15$         & $ 1.33$         & $ 1.60$          \\
$\chi^2/N_{\rm df}$ for $\Sigma^\phi$ & & $ 1.33$         & $ 1.69$         & $ 2.07$         & $ 1.31$         & $ 2.18$          \\
$\sigma^g   (2,u_0')$                 & & $ 0.73932 (59)$ & $ 0.81688 (68)$ & $ 0.91793 (83)$ & $ 1.05414 (95)$ & $ 1.2716  (11)$  \\
$\sigma^\phi(2,u_0')$                 & & $ 0.85580 (14)$ & $ 0.84272 (16)$ & $ 0.82635 (21)$ & $ 0.80520 (23)$ & $ 0.77470 (24)$  \\
$ \beta_{\rm FV}(\sigma^g(2,u_0'))$   & & $-0.0939  (37)$ & $-0.1251  (43)$ & $-0.1595  (53)$ & $-0.2310  (61)$ & $-0.3741  (69)$  \\
$\gamma_{\rm FV}(\sigma^g(2,u_0'))$   & & $ 0.23698 (93)$ & $ 0.2593  (11)$ & $ 0.2943  (14)$ & $ 0.3342  (16)$ & $ 0.4031  (16)$  \\
\hline
\hline
\end  {tabular}
\label{tbl:fit_result}
\end  {center}
\end  {table}
%
% @@@ =========================================================================
%
\subsection{Scale dependence}
\label{ssec:Scale_dependence}
\begin{figure}[p]
\begin{center}
\includegraphics[width=115mm]{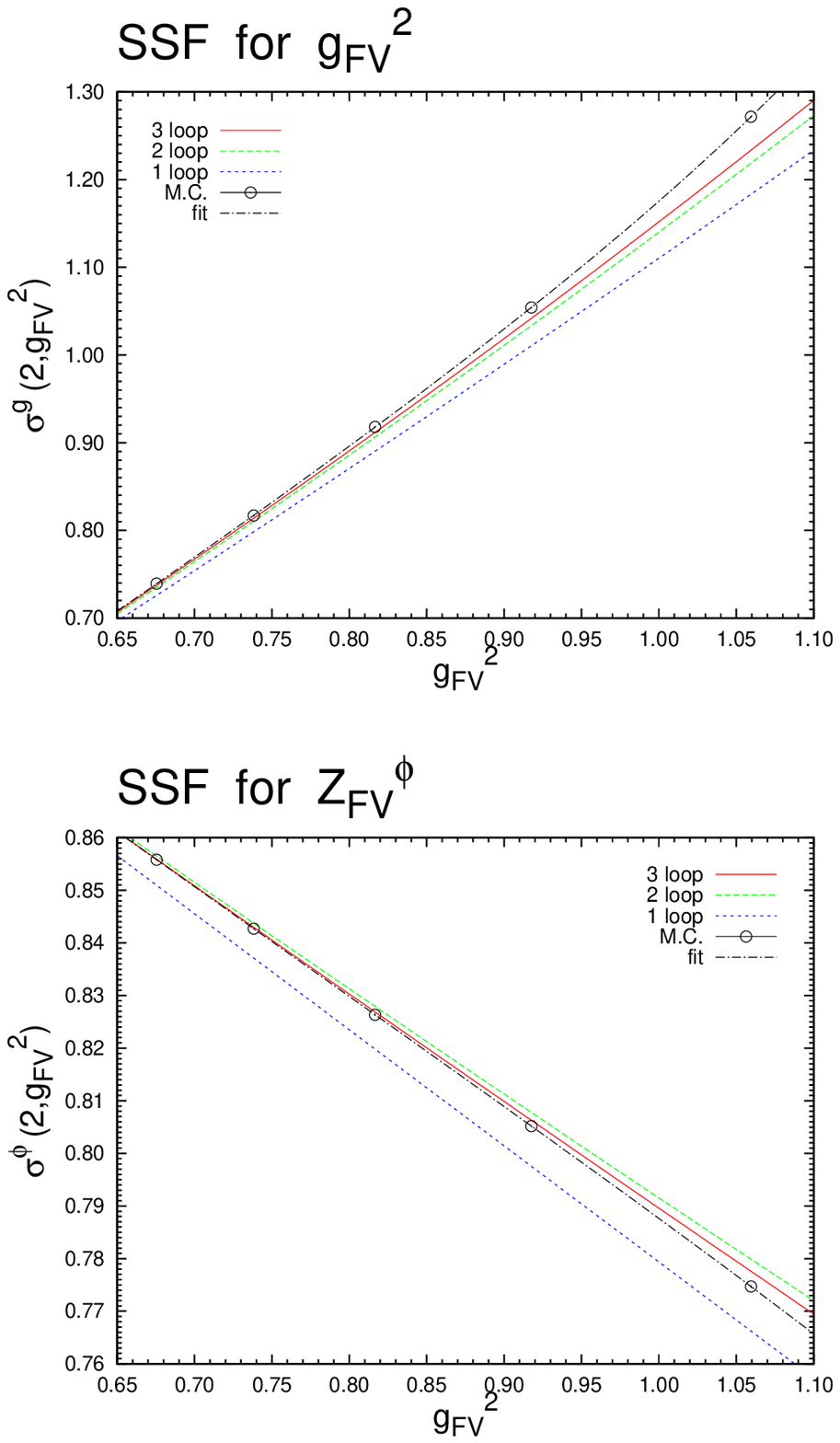}
\caption{
SSFs obtained from the Monte Carlo simulation and perturbative evaluation. 
$\sigma^g   (2,g_{\rm FV}^2)$ is shown in the top    panel, and 
$\sigma^\phi(2,g_{\rm FV}^2)$          in the bottom panel. 
The $1$-loop evaluation is described by the blue  dotted curve, 
the $2$-loop one                     by the green dashed curve, and 
the $3$-loop one                     by the red   solid  curve. 
The result fitted to Eqs. (\ref{eqn:ssf_fit1}) or (\ref{eqn:ssf_fit2}) 
is given by the black dashed-dotted curve. 
}
\label{fig:ssf}
\end  {center}
\end  {figure}

We discuss the continuum SSFs with $s=2$, 
which are obtained in the analysis of Sec. \ref{ssec:Continuum_limit}. 
In Fig. \ref{fig:ssf}, 
we give a comparison 
between the Monte Carlo simulation and the perturbative evaluation. 
$\sigma^g   (2,g_{\rm FV}^2)$ is shown in the top    panel, and 
$\sigma^\phi(2,g_{\rm FV}^2)$          in the bottom panel. 
It can be observed that 
the perturbative evaluation approaches the result by Monte Carlo simulation 
with an increase in the order. 
Moreover, we fit 
\begin{eqnarray}
  & &
  \sigma_{\rm F}^g   (2,u)
  \equiv
    u
  + u^2\left(\frac{ n-2   }{2\pi  }(\ln 2)\right)
  + u^3\left(\frac{ n-2   }{4\pi^2}(\ln 2)
            +\frac{(n-2)^2}{4\pi^2}(\ln 2)^2\right)
  +    \sum_{i=2}^5\,u^{i+2}\,\sigma_i^g\ ,\qquad
\label{eqn:ssf_fit1}
\\
  & &
  \sigma_{\rm F}^\phi(2,u)
  \equiv
    1
  + u \left(-\frac{n-1}{2\pi}(\ln 2)\right)
  +   \sum_{i=1}^4\,u^{i+1}\,\sigma_i^\phi\ ,
\label{eqn:ssf_fit2}
\end  {eqnarray}
to the Monte Carlo data, 
where $\sigma_i^g   $ ($2\le i\le5$)  
and   $\sigma_i^\phi$ ($1\le i\le4$)  
are free parameters in the fit. 
We use the universal forms independent of the renormalization shceme 
for the first three terms of $\sigma_{\rm F}^g   (2,g_{\rm FV}^2)$\,, 
and the first two   terms of $\sigma_{\rm F}^\phi(2,g_{\rm FV}^2)$\,. 
$\chi^2/N_{\rm df}$ in the fit is 
$0.40$  for $\sigma^g   (2,g_{\rm FV}^2)$\,, and 
$0.024$ for $\sigma^\phi(2,g_{\rm FV}^2)$\,.
The fitting result is also shown in Fig. \ref{fig:ssf}. 
%
%---------------------------------
%
%
%---------------------------------
%

\begin{figure}[p]
\begin{center}
\includegraphics[width=115mm]{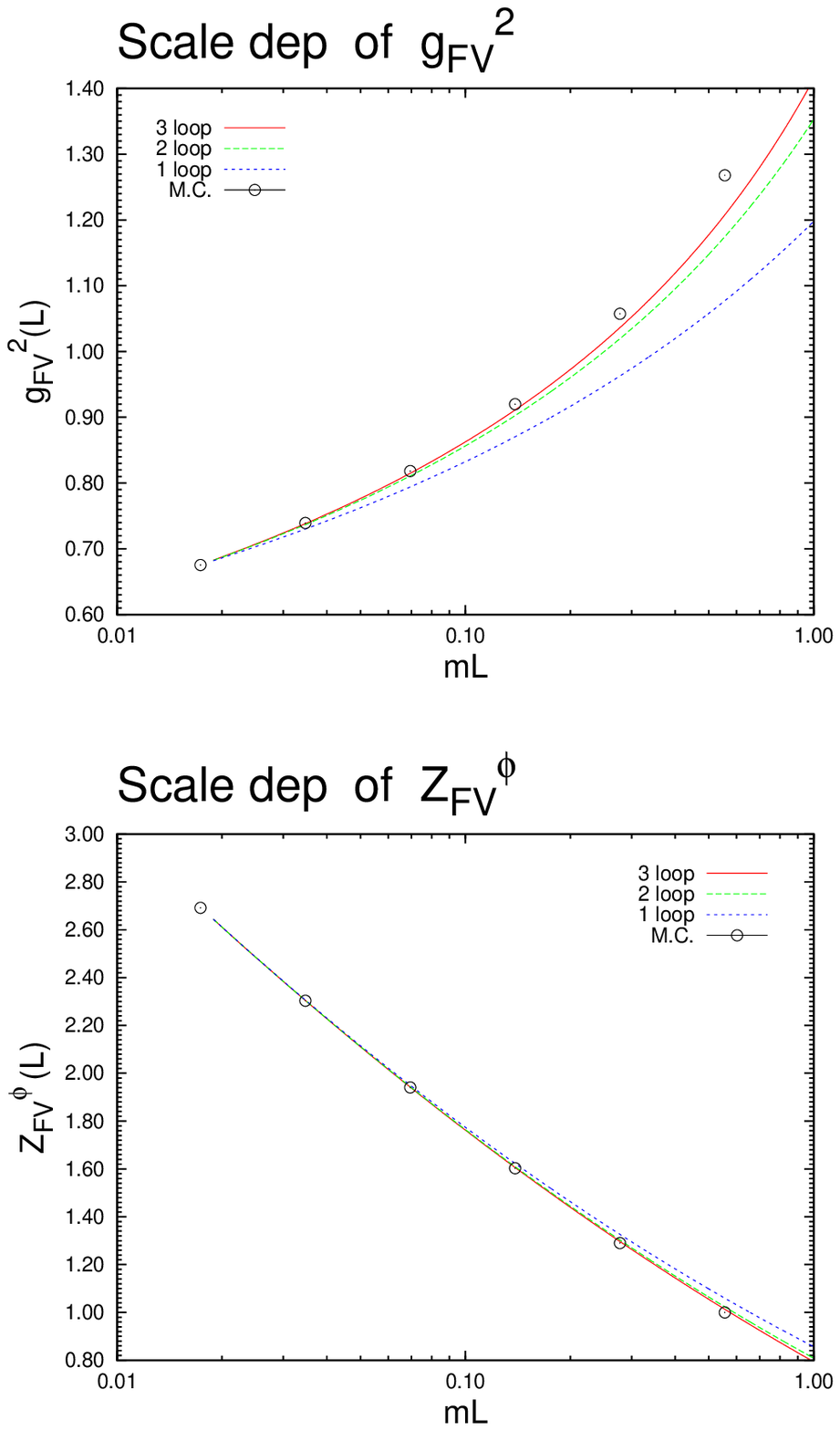}
\caption{
Scale dependence of $g_{\rm FV}^2$ and $Z_{\rm FV}^\phi$\,. 
In addition to the results by the Monte Carlo simulation, 
we also plot the results obtained by numerically integrating 
Eqs. (\ref{eqn:RG2_bt}) and (\ref{eqn:RG2_gm}) from $mL=0.017366$ 
with the perturbative 
$\beta_{\rm FV}(g_{\rm FV}^2)$ and $\gamma_{\rm FV}(g_{\rm FV}^2)$\,. 
The $1$-loop evaluation is described by the blue  dotted curve, 
the $2$-loop one                     by the green dashed curve, and 
the $3$-loop one                     by the red   solid  curve. 
}
\label{fig:running}
\end  {center}
\end  {figure}

We consider the scale dependence of $g_{\rm FV}^2$ and $Z_{\rm FV}^\phi$\,. 
The SSFs are determined by the fit to 
Eqs. (\ref{eqn:ssf_fit1}) and (\ref{eqn:ssf_fit2}), and 
expected to be sufficiently precise 
in the range of $g_{\rm FV}^2=0.67-1.27$\,. 
With the SSFs, 
we determine 
$g_{\rm FV}^2   (2^kL_{\rm min})$ and 
$Z_{\rm FV}^\phi(2^kL_{\rm min})$ $(k=0,\cdots,5)$ by 
Eqs. (\ref{eqn:def_SSF_g2}) and 
     (\ref{eqn:def_SSF_ph}). 
Here 
$L_{\rm min}$ is defined by 
$L_{\rm min}=2^{-5}L_{\rm max}$ with $mL_{\rm max}=0.5557(13)$\,. 
$m$ is the mass gap in the infinite volume. 
The value of $mL_{\rm max}$ 
corresponds to $g_{\rm FV}^2(L_{\rm max})=1.2680$, 
and is referred from Ref. \cite{Luscher.rc}. 
In addition, we set $Z_{\rm FV}^\phi(L_{\rm max})=1.0$\,.  
Note that the values of 
$g_{\rm FV}^2   (L_{\rm max})$ and 
$Z_{\rm FV}^\phi(L_{\rm max})$ are set without statistical errorrs. 
Ths statistical errors of of 
$g_{\rm FV}^2(L)$ and $Z_{\rm FV}^\phi(L)$ are recursively estimated by 
\begin{eqnarray}
  \Delta(g_{\rm FV}^2   (L))
  &=&
  \frac{1}{[\,\partial \sigma_{\rm F}^g(2,u)/\partial u\,]_{u=g_{\rm FV}^2(L)}}
  \sqrt{\Big[\, \Delta(g_{\rm FV}^2(2L))                    \,\Big]^2
       +\Big[\, \Delta(\sigma_{\rm F}^g(2,g_{\rm FV}^2(L))) \,\Big]^2
       }\ ,
\\
  \Delta(Z_{\rm FV}^\phi(L))
  &=&
  \frac{1}{\sigma_{\rm F}^\phi(2,g_{\rm FV}^2(L))}
  \sqrt{ \Big[\, \Delta(Z_{\rm FV}^\phi(2L))\,\Big]^2
       +\left[\, \frac{\Delta(\sigma_{\rm F}^\phi(2,g_{\rm FV}^2(L)))}
                      {       \sigma_{\rm F}^\phi(2,g_{\rm FV}^2(L))}\,
                 Z_{\rm FV}^\phi(2L)\,\right]^2
       }\ .\qquad
\end  {eqnarray}
$\Delta(\sigma_{\rm F}^{g,\phi}(2,g_{\rm FV}^2(L)))$ denotes 
the statistical error due to ones of the fitting parameters. 
For the estimation, 
we include a contribution from the correlation between the parameters 
as we have done in Sec. \ref{ssec:Continuum_limit}. 
In Table \ref{tbl:running}, the values of 
$m\times         2^kL_{\rm min} $, 
$g_{\rm FV}^2   (2^kL_{\rm min})$ and 
$Z_{\rm FV}^\phi(2^kL_{\rm min})$ $(k=0,\cdots,5)$ are listed. 
In Fig. \ref{fig:running}, these data are plotted. 
We also show the results obtained by numerically integrating 
Eqs. (\ref{eqn:RG2_bt}) and (\ref{eqn:RG2_gm}) 
from $mL_{\rm min}=0.017366$ 
with the perturbative 
$\beta_{\rm FV}(g_{\rm FV}^2)$ and $\gamma_{\rm FV}(g_{\rm FV}^2)$\,. 
We can confirm a reasonable behavior of the Monte Carlo data 
in comparison with the perturbative evaluation. 
%
%---------------------------------
%
\begin{table}
\begin{center}
\caption{
$g_{\rm FV}^2   (2^kL_{\rm min})$ and 
$Z_{\rm FV}^\phi(2^kL_{\rm min})$ $(k=0,\cdots,5)$ 
determined with the SSFs by using 
Eqs. (\ref{eqn:def_SSF_g2}) and 
     (\ref{eqn:def_SSF_ph}). 
We set 
$g_{\rm FV}^2   (L_{\rm max})=1.2680$ and 
$Z_{\rm FV}^\phi(L_{\rm max})=1.0   $ without statistical errors 
at $L_{\rm max}=2^5L_{\rm min}$\,. 
$g_{\rm FV}^2   (L_{\rm max})=1.2680$ 
corresponds to $mL_{\rm max}=0.5557(13)$\,. 
}
\medskip
\begin{tabular}{ccccccc}
\hline
\hline
$k$ & $\qquad$ & $m\times 2^k L_{\rm min}$ & $\qquad$ & $g_{\rm FV}^2(2^k L_{\rm min})$ & $\qquad$ & $Z_{\rm FV}^\phi(2^k L_{\rm min})$ \\
\hline
$0$ & & $0.017366(41)$ & & $0.67548 (76)$ & & $2.6916  (13)$  \\
$1$ & & $0.034731(81)$ & & $0.73917 (74)$ & & $2.3035  (11)$  \\
$2$ & & $0.06946 (16)$ & & $0.81829 (80)$ & & $1.94076 (87)$  \\
$3$ & & $0.13893 (33)$ & & $0.91982 (81)$ & & $1.60311 (66)$  \\
$4$ & & $0.27785 (65)$ & & $1.05738 (63)$ & & $1.29005 (39)$  \\
$5$ & & $0.5557  (13)$ & & $1.2680$       & & $1.0$           \\
\hline
\hline
\end  {tabular}
\label{tbl:running}
\end  {center}
\end  {table}
%
%---------------------------------

%---------------------------------
%
%
%---------------------------------
%

\begin{figure}[p]
\begin{center}
\includegraphics[width=115mm]{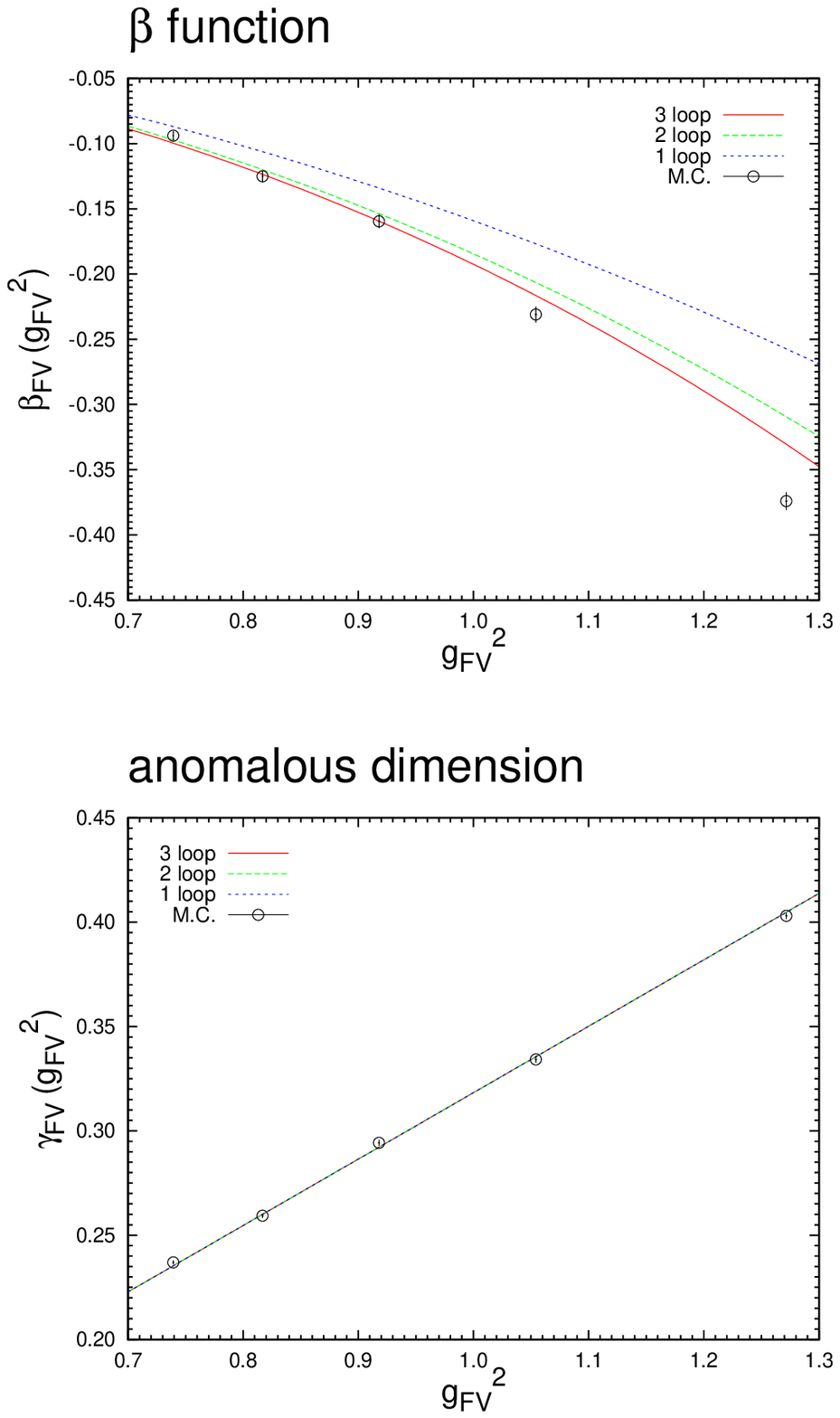}
\caption{
$ \beta_{\rm FV}(\sigma^g(2,u_0'))$ and 
$\gamma_{\rm FV}(\sigma^g(2,u_0'))$ determined in 
Sec. \ref{ssec:Continuum_limit}. 
The former is shown in the top    panel, and 
the latter          in the bottom panel. 
The perturbative evaluation is also plotted for a comparison. 
The $1$-loop evaluation is described by the blue  dotted curve, 
the $2$-loop one                     by the green dashed curve, and 
the $3$-loop one                     by the red   solid  curve. 
}
\label{fig:bet_gam}
\end  {center}
\end  {figure}

Finally, we discuss the $\beta$ function and the anomalous dimension. 
In principle, we can obtain them 
by differentiating $g_{\rm FV}^2(L)$ and $Z_{\rm FV}^\phi(L)$ 
with respect to $L$\,. 
However, 
the function forms of $g_{\rm FV}^2(L)$ and $Z_{\rm FV}^\phi(L)$\ 
are complicated, 
so that doing the differentiation numerically seems to be difficult. 
We alternatiely use the data of 
$ \beta_{\rm FV}(\sigma^g(2,u_0'))$ and 
$\gamma_{\rm FV}(\sigma^g(2,u_0'))$ determined in 
Sec. \ref{ssec:Continuum_limit}. 
In Fig. \ref{fig:bet_gam}, 
the results by the Monte Carlo simulation are shown. 
The perturbative evaluation is also described for a comparison. 
We can observe the reasonable behaviors again 
although the statistical errors are relatively large 
compared to ones of the SSFs or the renormalized parameters. 
It is possible to 
fit $\beta_{\rm FV}(g_{\rm FV}^2)$ and $\gamma_{\rm FV}(g_{\rm FV}^2)$, and 
determine the coefficients. 
However, we have only five data points in the present study. 
It is difficult to determine them with a sufficient statistical precision, 
so that we do not perform the fit. 
We leave the precise determination of the coefficients as a future task. 
%
%---------------------------------
%
%
%---------------------------------
%
% @@ ==================================================================
%
\section  {Conclusion}
\label{sec:Conclusion}
We have studied the finite box-size effects on 
the wave-function renormalization of the $2$-dimensional $O(3)$ sigma model. 
We have analyzed the step-scaling functions (SSF), 
which was proposed in a successful analysis 
of the renormalized coupling of the same model 
\cite{Luscher.rc}.
The SSF of the wave-function renormalization factor, $Z_{\rm FV}^\phi$, 
is determined with a sufficient precision and its scale dependence is studied.
We have compared the results with the perturbative evaluation 
and found a good agreement between them.
The $\beta$ function and the anomalous dimension are determined 
and it is found that
their behaviors are also consistent with the perturbative evaluation.

Our analysis supports that 
the RG description of $Z_{\rm FV}^\phi$ works well,
and thus it gives another evidence for the existence of RG equation, 
which describes the box-size dependence of the $N$-point Green function. 
The concrete construction of the RG equation will be addressed 
in the forthcoming report 
\cite{forthcoming}.

At the end, we give prospective views. 
The RG equation in few-body quantum systems will be a useful tool 
for the analysis of spatially extended objects, 
such as (loosely) bound two or a few-body systems 
and of the Efimov-like critical behaviors of the system 
as we have mentioned in Sec. 
\ref{sec:Introduction}. 
The analyses of the finite box-size effects 
should also have realistic physical meaning,
not being an artifact or an systematic error. 
A well-known example is the finite temperature system,
where the finiteness in the imaginary time direction plays a key role. 
There the RG equation for the box-size parameter 
must be a new powerful tool for studying the temperature dependences.
%
% @@@ =================================================================
%
\section*{Acknowledgments}
A part of the numerical calculations 
was carried out on the super parallel computers, 
CRAY XC40 at YITP in Kyoto University. 
%
% @@ ==================================================================
%
\appendix
%
% @@ ==================================================================
%
\section  {Perturbative evaluation}
\label{app:Perturbative_evaluation}
%
% @@@ =========================================================================
%
\subsection{Preparation}
\label{ssec:Preparation}
To clarify the notation, 
we give a brief description 
of the action, Feynman rules, Green functions and boundary condition.

We use the dimensional regularization, 
so that Eq. (\ref{eqn:2d_action}) must be extended to $d$-dimensional action, 
\begin{equation}
  S[{\bm \phi}]
  =
  \frac{1}{2g^2}\,\int_\Lambda d^d z\ 
  \partial_\mu {\bm \phi}(z)\cdot\partial_\mu {\bm \phi}(z)\quad
  (\mu=0,\cdots,d-1)\ .
\label{eqn:dd_action}
\end  {equation}
Here the system is put on 
\begin{equation}
  \Lambda
  =
  \Big\{\,
    z\,
    \Big|\,
    z_0\in [-T/2,T/2],\ 
    z_i\in [0,L]\ \ \mbox{for}\ \ i=1,\cdots,d-1\,
  \Big\}\ .
\end  {equation}

As we have said in Sec. \ref{sec:Introduction}, 
a paticular attention must be paid for the treatment of zero mode. 
The zero mode occurs from the degree of freedom 
where the ${\bm \phi}$ field is rotated by a same matrix 
over all points of the time and space. 
To separate it, 
we use an $O(n)$ rotation matrix $\Omega$ 
which is independent of the space-time points, 
and parametrize the ${\bm \phi}$ field as 
\begin{equation}
  {\bm \phi}(z)
  =
  \Omega
  \left(\,\sqrt{1-g^2{\bm \pi}^2(z)},\,g{\bm \pi}(z)\,\right)^T\ .
\label{eqn:parametrization}
\end  {equation}
${\bm \pi}(z)=(\pi_i(z),i=1,\cdots,n-1)$ is the $(n-1)$-component field. 
In the following, we use the bullet point symbol 
also for the scalar product of $(n-1)$-component vectors. 
As long as there is no confusion, 
we use the abbreviation such as ${\bm \pi}^2={\bm \pi}\cdot{\bm \pi}$\,. 
According to Ref. \cite{Hasenfratz:1984jk,Hasenfratz:1989pk}, 
we consider the identity, 
\begin{equation}
  1
  =
  \int d^n{\bm m}\ \delta^n
  \left(
    {\bm m} - \frac{1}{TL^{d-1}} \int d^d z\ {\bm \phi}(z)
  \right)\ .
\label{eqn:identity1}
\end  {equation}
Substituting Eq. (\ref{eqn:parametrization}) to Eq. (\ref{eqn:identity1}), 
we obtain 
\begin{equation}
  1
  =
  S_{n-1}
  \left[\,
    \prod_{i=1}^{n-1}
    \delta\left(-\frac{g}{TL^{d-1}}\int d^d z\ \pi_i\right)\,
  \right]
  \exp
  \left[\,
    -\frac{(n-1)g^2}{2TL^{d-1}}\int d^dz\ {\bm \pi}^2\,
  \right]\ ,
\label{eqn:identity2}
\end  {equation}
where $S_{n-1}$ is the surface area of $(n-1)$-dimensional unit sphere. 
Note that the identity (\ref{eqn:identity2}) is satisfied 
even in the interior of path integration 
with respect to the ${\bm \pi}$ field. 
By applying Eq. (\ref{eqn:identity2}) to the partition function, 
we have 
\begin{eqnarray}
  {\cal Z}
  &=&
  \int\ \Big[\,\delta({\bm \phi}^2(z)-1)\,d{\bm \phi}(z)\,\Big]\ 
  {\rm e}^{-S[{\bm \phi}]}
\nonumber  \\
  &=&
  \int\ \left[\,\frac{g\,d{\bm \pi}(z)}{\sqrt{1-g^2{\bm \pi}^2(z)}}\,\right]\ 
  \exp \left[\,-\int d^d z\,
       \left\{\,
           \frac{1}{2}\partial_\mu{\bm \pi}\cdot
                      \partial_\mu{\bm \pi}
         + \frac{g^2}{8}
           \frac{(\partial_\mu{\bm \pi}^2)^2}{1-g^2{\bm \pi}^2}\,
       \right\}\,
       \right]
\nonumber  \\
  &=&
  \int\ \left[\,g\,d{\bm \pi}(z)\,\right]\ 
  \exp \left[\,-\int d^d z\,
       \left\{\,
           \frac{1}{2}\partial_\mu{\bm \pi}\cdot
                      \partial_\mu{\bm \pi}
         + \frac{g^2}{8}
           \frac{(\partial_\mu{\bm \pi}^2)^2}{1-g^2{\bm \pi}^2}
         + \frac{1}{2}\delta^d(0)\,\ln(1-g^2{\bm \pi}^2)
       \right\}\,
       \right]
\nonumber  \\
  &=&
  S_{n-1}
  \int\ \left[\,g\,d{\bm \pi}(z)\,\right]\ 
  \left[\,
    \prod_{i=1}^{n-1}
    \delta\left(-\frac{g}{TL^{d-1}}\int d^d z\ \pi_i\right)\,
  \right]
\nonumber  \\
  & &
  \times
  \exp \left[\,-\int d^d z\,
       \left\{\,
           \frac{1}{2}\partial_\mu{\bm \pi}\cdot
                      \partial_\mu{\bm \pi}
         + \frac{g^2}{8}
           \frac{(\partial_\mu{\bm \pi}^2)^2}{1-g^2{\bm \pi}^2}
         + \frac{1}{2}\delta^d(0)\,\ln(1-g^2{\bm \pi}^2)
         + \frac{(n-1)g^2}{2TL^{d-1}}{\bm \pi}^2\,
       \right\}\,
       \right] .
\nonumber  \\
  & &
\label{eqn:part_func}
\end  {eqnarray}
The factor of delta functions 
excludes the zero mode of ${\bm \pi}$ field. 
As the compensation, 
we must include an extra interaction term, 
$\displaystyle \frac{(n-1)g^2}{2TL^{d-1}}{\bm \pi}^2$.

Due to Eq. (\ref{eqn:parametrization}), 
${\bm \pi}$ must satisfy the condition, $|{\bm \pi}|\le 1/g$. 
However, 
starting from the final expression of Eq. (\ref{eqn:part_func}), 
we are no longer constrained by the condition. 
It can be also understood 
from the free-field part of Eq. (\ref{eqn:part_func}) that 
$|{\bm \pi}|\le  1  $ gives a main contribution to the path integration, 
and 
$|{\bm \pi}|\sim 1/g$ does an exponentially small contribution. 
Thus, 
the range of integration can be safely extended to $|{\bm \pi}|<\infty$. 
See Ref. \cite{ZJ:Book} for more details.

\begin{figure}[htbp]
\begin{center}
\includegraphics[width=160mm]{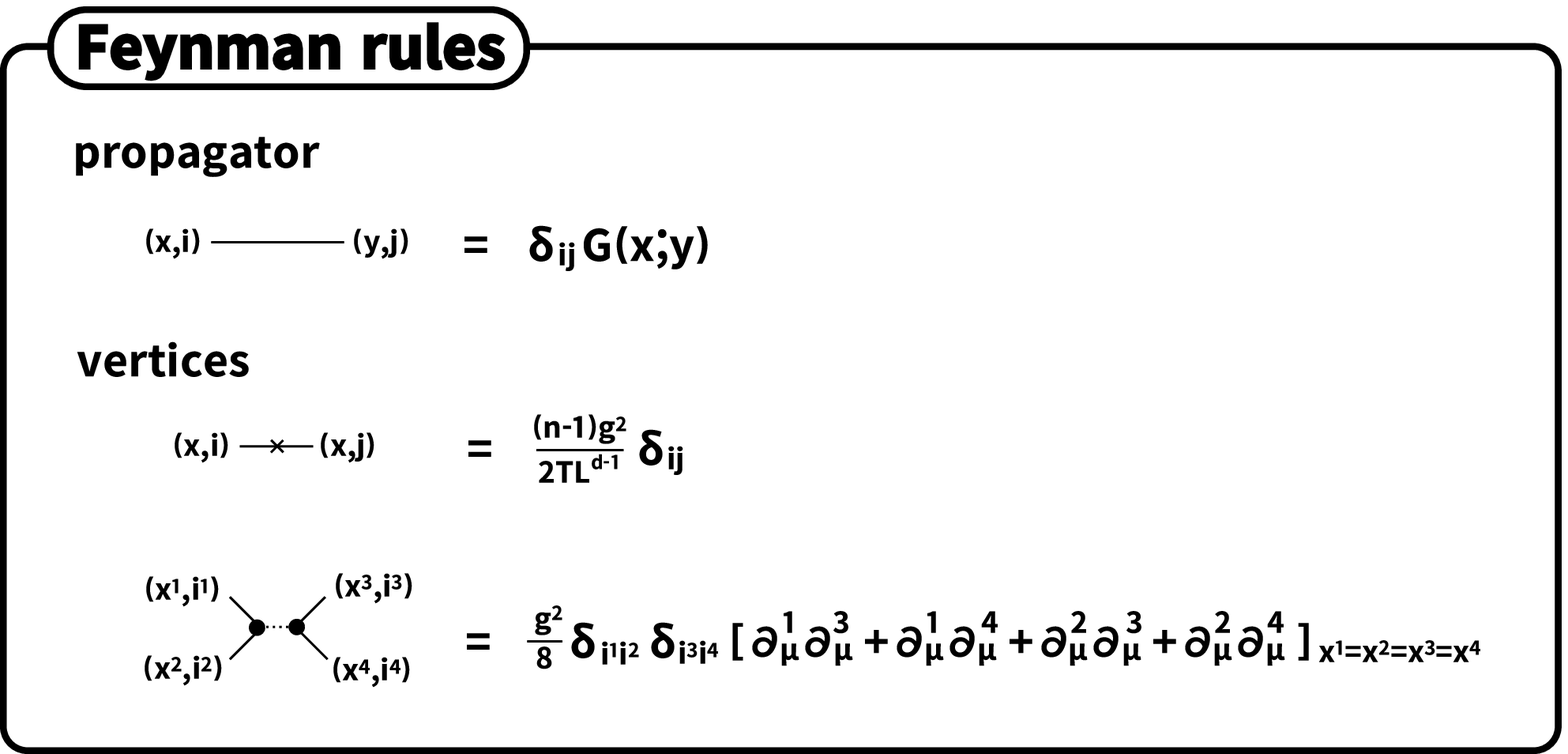}
\caption{Feynman rules read off 
  from the action (\ref{eqn:action_pi}) up to ${\cal O}(g^2)$. 
  The $x$ and $y$ represent the coordinate. 
  The $i$ and $j$ do the number of component for the ${\bm \pi}$ field. 
  The cross symbol denotes a vertex due to the zero mode. 
  The filled circle does a vertex due to the term 
  $\displaystyle \frac{g^2}{8}
   \frac{(\partial_\mu{\bm \pi}^2)^2}{1-g^2{\bm \pi}^2}$. 
  The dotted lines means that 
  vertices linked by them are located at the same coordinate. 
}
\label{fig:Feyn_rule}
\end  {center}
\end  {figure}

We have $\delta^d(0)=0$ with the dimensional regularization\footnote{
We can write as 
$\displaystyle 
 \delta^d(0)
=\int_{-\infty}^\infty \frac{d^dk}{(2\pi)^d}\,1
=\int_{-\infty}^\infty \frac{d^dk}{(2\pi)^d}\,\frac{k^2+\alpha^2}{k^2+\alpha^2}
=\int_{-\infty}^\infty \frac{d^dk}{(2\pi)^d}\ (k^2+\alpha^2)
 \int_0^\infty dt\ {\rm e}^{-(k^2+\alpha^2)\,t}
$ 
with an arbitrary non-zero parameter $\alpha$\,. 
After changing the order of integration 
and doing the $k$-integration, we have 
$\displaystyle 
 \delta^d(0)
=\frac{1}{2^d\pi^{d/2}}
 \left[\,
   \frac{d}{2}\int_0^\infty dt\ {\rm e}^{-\alpha^2 t}t^{ -d/2   -1}
 + \alpha^2   \int_0^\infty dt\ {\rm e}^{-\alpha^2 t}t^{(-d/2+1)-1}\,
 \right].
$
When ${\rm Re}(d)<0$, these integrations converge, 
and we obtain $\delta^d(0)=0$\,. 
Considering $\delta^d(0)$ as an analytical function of $d$, 
we can continuate it to ${\rm Re}(d)\ge 0$\,, analytically. 
Then, we have $\delta^d(0)=0$ for any complex dimension $d$\,. 
}. 
Thus, we need not consider the interaction due to the integral measure, 
$\displaystyle \frac{1}{2}\delta^d(0)\,\ln(1-g^2{\bm \pi}^2)$, 
in the following discussion. 
Note that if we use the lattice regularization, 
the IR cutoff mass $g/a$ is introduced due to $\delta^2(0)=1/a^2$ 
and enables to isolate the momentum zero mode in a finite volume. 
The action for the ${\bm \pi}$ field can be written as 
\begin{eqnarray}
  S[{\bm \pi}]
  &=&
  \int d^d z\ 
  \left[\,
      \frac{1}{2}\partial_\mu{\bm \pi}\cdot\partial^\mu{\bm \pi}
    + \frac{(n-1)g^2}{2TL^{d-1}}{\bm \pi}^2
    + \frac{g^2}{8}
      \frac{(\partial_\mu{\bm \pi}^2)^2}{1-g^2{\bm \pi}^2}
  \right]
\nonumber  \\
  &=&
  \int d^d z\ 
  \left[\,
      \frac{1}{2}\partial_\mu{\bm \pi}\cdot\partial^\mu{\bm \pi}
    + \frac{(n-1)g^2}{2TL^{d-1}}{\bm \pi}^2
    + \frac{g^2}{8}(\partial_\mu{\bm \pi}^2)^2
    + {\cal O}(g^4)\,
  \right] .
\label{eqn:action_pi}
\end  {eqnarray}
In Fig. \ref{fig:Feyn_rule}, we summalize the Feynman rules 
which we can interpret from this action. 
The details of the ${\bm \pi}$ propagator, $G(x,y)$, 
are discussed a bit later. 
%
%---------------------------------
%
%
%---------------------------------

We define the expectation value of an operator $O$ as 
\begin{eqnarray}
  \langle\,O\,\rangle
  &=&
  \frac{S_{n-1}}{\cal Z}
  \int\ \left[\,g\,d{\bm \pi}(z)\,\right]\ 
  \left[\,
    \prod_{i=1}^{n-1}
    \delta\left(-\frac{g}{TL^{d-1}}\int d^d z\ \pi_i\right)\,
  \right]
  {\rm e}^{-S_{\bm \pi}}\,
  \times\,O
\nonumber  \\
  &=&
  \frac{S_{n-1}}{\cal Z}
  \int\ \left[\,g\,d{\bm \pi}(z)\,\right]\ 
  \left[\,
    \prod_{i=1}^{n-1}
    \delta\left(-\frac{g}{TL^{d-1}}\int d^d z\ \pi_i\right)\,
  \right]
  {\rm e}^{-\int d^d z\,
           \frac{1}{2}\partial_\mu{\bm \pi}
                 \cdot\partial_\mu{\bm \pi}}
\nonumber  \\
  & &
  \times\ O\,
  \left\{\,
      1
    - g^2 \int d^d z\,
      \left[\, \frac{(n-1)}{2TL^{d-1}}{\bm \pi}^2
             + \frac{1}{8}(\partial_\mu{\bm \pi}^2)^2\,
      \right]
    + {\cal O}(g^4)\,
  \right\} .
\label{eqn:exp_val}
\end  {eqnarray}
In the following, 
the coefficient of $(g^2)^i$ is referred by $\langle\,O\,\rangle_i$ 
with a non-negative integer $i$.

From Eq. (\ref{eqn:exp_val}), 
the free ${\bm \pi}$ propagator, 
\begin{equation}
  \langle\,\pi_i(x)\,\pi_j(y)\,\rangle_0 = \delta_{ij}\,G(x; y)\ ,
\end  {equation}
satisfies\footnote{
The constant term, 
$\displaystyle \frac{1}{TL^{d-1}}$, 
appears due to the exclusion of zero mode. 
Consider a formal solution 
$\displaystyle 
\bar{G}(z)
= \frac{1}{TL^{d-1}} \sum_{p\ne 0}\,\frac{{\rm e}^{ip\cdot z}}{p^2}
$\ .
One can confirm the appearance of the constant term from the calculation of 
$
\displaystyle 
\Box_z\,\bar{G}(z)
=   \frac{1}{TL^{d-1}} \sum_{p\ne 0}\,\frac{\Box_z\,{\rm e}^{ip\cdot z}}{p^2}
= - \frac{1}{TL^{d-1}} \sum_{p     }\,              {\rm e}^{ip\cdot z}
  + \frac{1}{TL^{d-1}} \sum_{p=   0}\,              {\rm e}^{ip\cdot z}
= - \delta^d(z) + \frac{1}{TL^{d-1}}
$\ .
}
\begin{equation}
  \Box_x\,G(x;y) = - \delta^d(x-y) + \frac{1}{TL^{d-1}}\ .
\end  {equation}
To determine $G(x;y)$, the boundary condition must be set. 
We adopt 
the NBC for the temporal direction, and 
the PBC for the spatial  direction, 
\begin{eqnarray}
  & &
  \frac{\partial}{\partial x_0}\,G(x_0,{\bm x};y_0,{\bm y}) = 0\quad
  (\,x_0=\pm T/2\,)\ ,\quad
  \frac{\partial}{\partial y_0}\,G(x_0,{\bm x};y_0,{\bm y}) = 0\quad
  (\,y_0=\pm T/2\,)\ ,
\nonumber  \\
  & &
  G(x_0,{\bm x}+L{\bm n}_x;y_0,{\bm y}+L{\bm n}_y)
  =
  G(x_0,{\bm x};y_0,{\bm y})\quad
  (\,{\bm n}_x,{\bm n}_y\in\mathbb{Z}^{d-1}\,)\ .
\label{eqn:bc2}
\end  {eqnarray}
Then, $G(x;y)$ can be written as 
\begin{equation}
  G(x;y)
  =
  \sum_{(m,{\bm n})\ne (0,{\bm 0})}
  \frac{1}{\lambda_{m{\bm n}}^2}\,
  \phi_{m{\bm n}}^*(y)\,\phi_{m{\bm n}}(x)\ ,
\end  {equation}
with 
\begin{equation}
  \lambda_{m{\bm n}}^2 = p_m^2 + {\bm q}_{\bm n}^2\quad
  \left(\,
    p_m             = \frac{ \pi m}{T}\,,\ 
    {\bm q}_{\bm n} = \frac{2\pi{\bm n}}{L}\ \Big|\ 
    m\in\mathbb{Z}\,,\ 
    {\bm n}\in\mathbb{Z}^{d-1}\,
  \right)\ 
\end  {equation}
and 
\begin{equation}
  \phi_{m{\bm n}}(x)
  = 
  \left\{
  \begin{array}{lll}
    \sqrt{1/(TL^{d-1})}\,
    {\rm e}^{i{\bm q}_{\bm n}\cdot{\bm x}}                &
    $\quad$ & (\,m=0\,) 
\\
    \sqrt{2/(TL^{d-1})}\,
    {\rm e}^{i{\bm q}_{\bm n}\cdot{\bm x}}\,\cos(p_m x_0) &
            & (\,m:\,{\rm even\ but\ not\ zero}\,) 
\\
    \sqrt{2/(TL^{d-1})}\,
    {\rm e}^{i{\bm q}_{\bm n}\cdot{\bm x}}\,\sin(p_m x_0) &
            & (\,m:\,{\rm odd}\,) 
\\
  \end  {array}
  \right.\ .
\end  {equation}
$\lambda_{m{\bm n}}^2$ and $\phi_{m{\bm n}}(x)$ are
are the eigenvalue and eigenfunction of 
\begin{equation}
  \Box_x    \,\phi(x) = - \lambda^2\,\phi(x)\ 
\label{eqn:wave_eq}
\end  {equation}
with the boundary condition 
\begin{equation}
  \frac{\partial}{\partial x_0}\,\phi(x_0,{\bm x}) = 0\quad
  (\,x_0=\pm T/2\,)\ ,\quad
  \phi(x_0,{\bm x}+L{\bm n}_x)
  =
  \phi(x_0,{\bm x})\quad
  (\,{\bm n}_x\in\mathbb{Z}^{d-1}\,)\ .
\label{eqn:bc3}
\end  {equation}
The normalization of $\varphi_{m{\bm n}}(x)$ is taken such that 
the orthnormal condition 
\begin{equation}
  \int_\Lambda d^d x\ 
  \phi_{m{\bm n}}^*(x)\,\phi_{m'{\bm n}'}(x)
  =
  \delta_{mm'}\delta_{{\bm n}{\bm n}'}
\end  {equation}
is satisfied.

It is useful to separate $G(x;y)$ as 
\begin{equation}
  G(x;y) = G_{\rm Z}(x_0;y_0) + G_{\rm N}(x;y)\ ,
\end  {equation}
where $G_{\rm Z}(x_0;y_0)$ is a contribution 
from the momentum zero mode (\,$m\ne0$, ${\bm n}={\bm 0}$\,) 
\begin{eqnarray}
  G_{\rm Z}(x_0;y_0)
  &\equiv &
  \sum_{m\ne 0}
  \frac{1}{\lambda_{m{\bm 0}}^2}\,
  \phi_{m{\bm 0}}^*(y)\,\phi_{m{\bm 0}}(x)\ ,
\label{eqn:zr_mode}
\end  {eqnarray}
and $G_{\rm N}(x;y)$ is one 
from the momentum non-zero mode (\,${\bm n}\ne {\bm 0}$\,) 
\begin{eqnarray}
  G_{\rm N}(x;y)
  &\equiv &
  \sum_{m}  \sum_{{\bm n}\ne {\bm 0}}
  \frac{1}{\lambda_{m{\bm n}}^2}\,
  \phi_{m{\bm n}}^*(y)\,\phi_{m{\bm n}}(x)\ .
\label{eqn:nz_mode}
\end  {eqnarray}
After some calculations, we obtain 
\begin{eqnarray}
  G_{\rm Z}(x_0;y_0)
  &=&
  \frac{1}{L^{d-1}}
  \left(\,
    - \frac{|x_0-y_0|}{2}
    + \frac{x_0^2+y_0^2}{2T}
    + \frac{T}{12}\,
  \right)\ ,
\label{eqn:GZ}
\\
  G_{\rm N}(x;y)
  &=&
  \sum_{m=-\infty}^\infty
  \{\,
    R(x_0-y_0+ 2m   T,{\bm x}-{\bm y})
    +
    R(x_0+y_0+(2m+1)T,{\bm x}-{\bm y})\,
  \}\ ,
\qquad
\label{eqn:GN}
\end  {eqnarray}
with the non-zero mode propagator in the infinite temporal extent, 
\begin{equation}
  R(z)
  \equiv
  \frac{1}{2L^{d-1}}
  \sum_{{\bm q}_{\bm n}\ne {\bm 0}}
  \frac{1}{|{\bm q}_{\bm n}|}\, 
  {\rm e}^{-|{\bm q}_{\bm n}||z_0|+i{\bm q}_{\bm n}\cdot{\bm z}}\ .
\label{eqn:def_Rz}
\end  {equation}
Eq. (\ref{eqn:GN}) can be interpreted that 
$R(z)$ is padded in the temporal extent $T$ 
in such a way as to satisfy the NBC. 
In addition, 
if we use the zero mode propagator in the infinite temporal extent, 
\begin{equation}
  r(z_0)
  =
  \frac{1}{2L^{d-1}}
  \lim_{\omega\to 0^+}\frac{1}{\omega}\,{\rm e}^{-\omega |z_0|}\ ,
\end  {equation}
instead of $R(z)$ in Eq. (\ref{eqn:GN}), 
we can reproduce Eq. (\ref{eqn:GZ}).

The $O(n)$-invariant 2-point Green function 
defined by Eq. (\ref{eqn:def_inv_Green}) can be written as 
\begin{eqnarray}
  G_{\rm inv}(x;y)
  &=&
  1
  + g^2
  \left(
                 \langle\,{\bm \pi}  (x)\cdot{\bm \pi}  (y)\,\rangle
    - \frac{1}{2}\langle\,{\bm \pi}^2(x)                   \,\rangle
    - \frac{1}{2}\langle\,                   {\bm \pi}^2(y)\,\rangle
  \right)
\nonumber  \\
  & &
  + g^4
  \left(
      \frac{1}{4}\langle\,{\bm \pi}^2(x)     {\bm \pi}^2(y)\,\rangle
    - \frac{1}{8}\langle\,{\bm \pi}^4(x)                   \,\rangle
    - \frac{1}{8}\langle\,                   {\bm \pi}^4(y)\,\rangle
  \right)
  + {\cal O}(g^6)
\nonumber  \\
  &=&
  1
  + g^2
  \left(
                 \langle\,{\bm \pi}  (x)\cdot{\bm \pi}  (y)\,\rangle_0
    - \frac{1}{2}\langle\,{\bm \pi}^2(x)                   \,\rangle_0
    - \frac{1}{2}\langle\,                   {\bm \pi}^2(y)\,\rangle_0
  \right)
\nonumber  \\
  & &
  + g^4
  \left(
                 \langle\,{\bm \pi}  (x)\cdot{\bm \pi}  (y)\,\rangle_1
    - \frac{1}{2}\langle\,{\bm \pi}^2(x)                   \,\rangle_1
    - \frac{1}{2}\langle\,                   {\bm \pi}^2(y)\,\rangle_1
  \right.
\nonumber  \\
  & &
  \qquad
  \left.
    + \frac{1}{4}\langle\,{\bm \pi}^2(x)     {\bm \pi}^2(y)\,\rangle_0
    - \frac{1}{8}\langle\,{\bm \pi}^4(x)                   \,\rangle_0
    - \frac{1}{8}\langle\,                   {\bm \pi}^4(y)\,\rangle_0
  \right)
  + {\cal O}(g^6)\ .
\qquad
\label{eqn:inv_Green1}
\end  {eqnarray}
The final expression of Eq. (\ref{eqn:inv_Green1}) is evaluated 
in Sec. \ref{ssec:Evaluation_at_O_g2} and \ref{ssec:Evaluation_at_O_g4}, 
perturbatively. 
%
% @@@ =========================================================================
%
\subsection{Evaluation at ${\cal O}(g^2)$}
\label{ssec:Evaluation_at_O_g2}
\begin{figure}[htbp]
\begin{center}
\includegraphics[width=140mm]{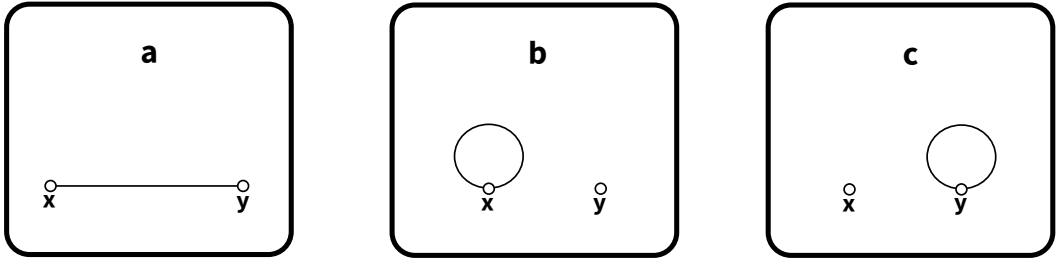}
\caption{The diagramatic description of 
the $O(n)$-invariant $2$-point Green function at ${\cal O}(g^2)$. 
The open circle means the external space-time point. }
\label{fig:g2_diag}
\end  {center}
\end  {figure}

It is instructive to follow 
the derivation of the ${\cal O}(g^2)$ contribution 
to $O(n)$-invariant $2$-point Green function. 
The contribution is given 
by the second term in Eq. (\ref{eqn:inv_Green1}). 
Corresponding to 
$\langle\,{\bm \pi}  (x)\cdot{\bm \pi}  (y)\,\rangle_0$\,, 
$\langle\,{\bm \pi}^2(x)                   \,\rangle_0$\, and 
$\langle\,                   {\bm \pi}^2(y)\,\rangle_0$\,, 
we refer these contributions by ``a'', ``b'' and ``c'', respectively. 
The diagramatic description is given in Fig. \ref{fig:g2_diag}. 
We project the total momentum in the external line to zero. 
Each contribution can be calculated from 
\begin{eqnarray}
  P_{\rm a}(\tau)
  &=&
  \frac{1}{L^{2(d-1)}}\int d^{d-1}{\bm x}\int d^{d-1}{\bm y}\ 
  G(x;y)\,|_{x_0=-\tau,\,y_0=+\tau}\ ,
\nonumber  \\
  P_{\rm b}(\tau)
  &=&
  \frac{1}{L^{2(d-1)}}\int d^{d-1}{\bm x}\int d^{d-1}{\bm y}\ 
  G(x;x)\,|_{x_0=-\tau}\ ,
\nonumber  \\
  P_{\rm c}(\tau)
  &=&
  \frac{1}{L^{2(d-1)}}\int d^{d-1}{\bm x}\int d^{d-1}{\bm y}\ 
  G(y;y)\,|_{y_0=+\tau}\ ,
\label{eqn:g2_cont}
\end  {eqnarray}
except for 
the coefficient in              Eq. (\ref{eqn:inv_Green1}), 
the coefficient in the brace of Eq. (\ref{eqn:exp_val}), 
the multiplicity of the spin component and 
the statistical factor. 
Here we adopt $x_0=-\tau$ and $y_0=+\tau$\,. 
%
%---------------------------------
%
%
%---------------------------------

We need not consider to the momentum non-zero mode for $P_{\rm a}(\tau)$, 
\begin{equation} 
  P_{\rm a}(\tau)
  =
  G_{\rm Z}(x_0;y_0)\,|_{x_0=-\tau,\,y_0=+\tau}
  =
  \frac{1}{L^{d-1}}
  \left(\,
    - |\tau|
    + \frac{\tau^2}{T}
    + \frac{T}{12}\,
  \right)\ .
\end  {equation}
On the other hands, 
the momentum non-zero mode is needed 
for $P_{\rm b}(\tau)$ and $P_{\rm c}(\tau)$. 
Dropping the terms which are exponentially small in $T\to\infty$, 
we have 
\begin{eqnarray} 
  & &
  P_{\rm b}(\tau)
  =
  G_{\rm Z}(x_0;x_0)\,|_{x_0=-\tau}
  +
  G_{\rm N}(x  ;x  )\,|_{x_0=-\tau}
  =
  \frac{1}{ L^{d-1}}
  \left(\,
      \frac{\tau^2}{T}
    + \frac{T}{12}\,
  \right)
  +
  R(0)\ ,
\label{eqn:Pb}
\\
  & &
  P_{\rm c}(\tau)
  =
  G_{\rm Z}(y_0;y_0)\,|_{y_0=+\tau}
  +
  G_{\rm N}(y  ;y  )\,|_{y_0=+\tau}
  =
  \frac{1}{ L^{d-1}}
  \left(\,
      \frac{\tau^2}{T}
    + \frac{T}{12}\,
  \right)
  +
  R(0)\ .
\label{eqn:Pc}
\end  {eqnarray}
In the derivation of Eqs. (\ref{eqn:Pb}) and (\ref{eqn:Pc}), 
the condition $-T/2<\tau<T/2$ is used.

The factors multiplying to each $P(\tau)$ are given by 
\begin{eqnarray}
  F_{\rm a}
  &=&
  g^2     \times 1\times (n-1)\times 1\ ,
\nonumber  \\
  F_{\rm b}
  &=&
  (-g^2/2)\times 1\times (n-1)\times 1\ ,
\nonumber  \\
  F_{\rm c}
  &=&
  (-g^2/2)\times 1\times (n-1)\times 1\ .
\end  {eqnarray}
Thus, 
the total contribution at ${\cal O}(g^2)$ can be written as 
\begin{eqnarray}
  & &
  g^2\,
  \frac{1}{L^{2(d-1)}}\int d^{d-1}{\bm x}\int d^{d-1}{\bm y}\ 
  \langle\,{\bm \phi}(x)\cdot{\bm \phi}(y)\,\rangle_1
\nonumber  \\
  &=&
    F_{\rm a}\,P_{\rm a}(\tau)
  + F_{\rm b}\,P_{\rm b}(\tau)
  + F_{\rm c}\,P_{\rm c}(\tau)
\nonumber  \\
  &=&
  g^2\,
  \left[\,
    - (n-1)\,R(0)
    - (n-1)\left(\,\frac{|\tau|}{L^{d-1}}\,\right)\,
  \right] .\qquad
\label{eqn:inv_Green1_g2}
\end  {eqnarray}
Note that the terms proportional to $T$ cancel. 
This fact means that 
the divergence at $T\to\infty$ (IR divergence) vanishes 
by treating the $O(n)$-invariant Green function. 
It should be also noted that 
the terms proportional to $\tau^2$ cancel. 
The appearance of the $\tau^2$ terms 
in $P_{\rm b}(\tau)$ and $P_{\rm c}(\tau)$ is due to the use of NBC. 
Such a cancellation does not occur 
if we use the periodic boundary condition for the temporal direction. 
%
% @@@ =========================================================================
%
\subsection{Evaluation at ${\cal O}(g^4)$}
\label{ssec:Evaluation_at_O_g4}
\begin{figure}[p]
\begin{center}
\includegraphics[width=160mm]{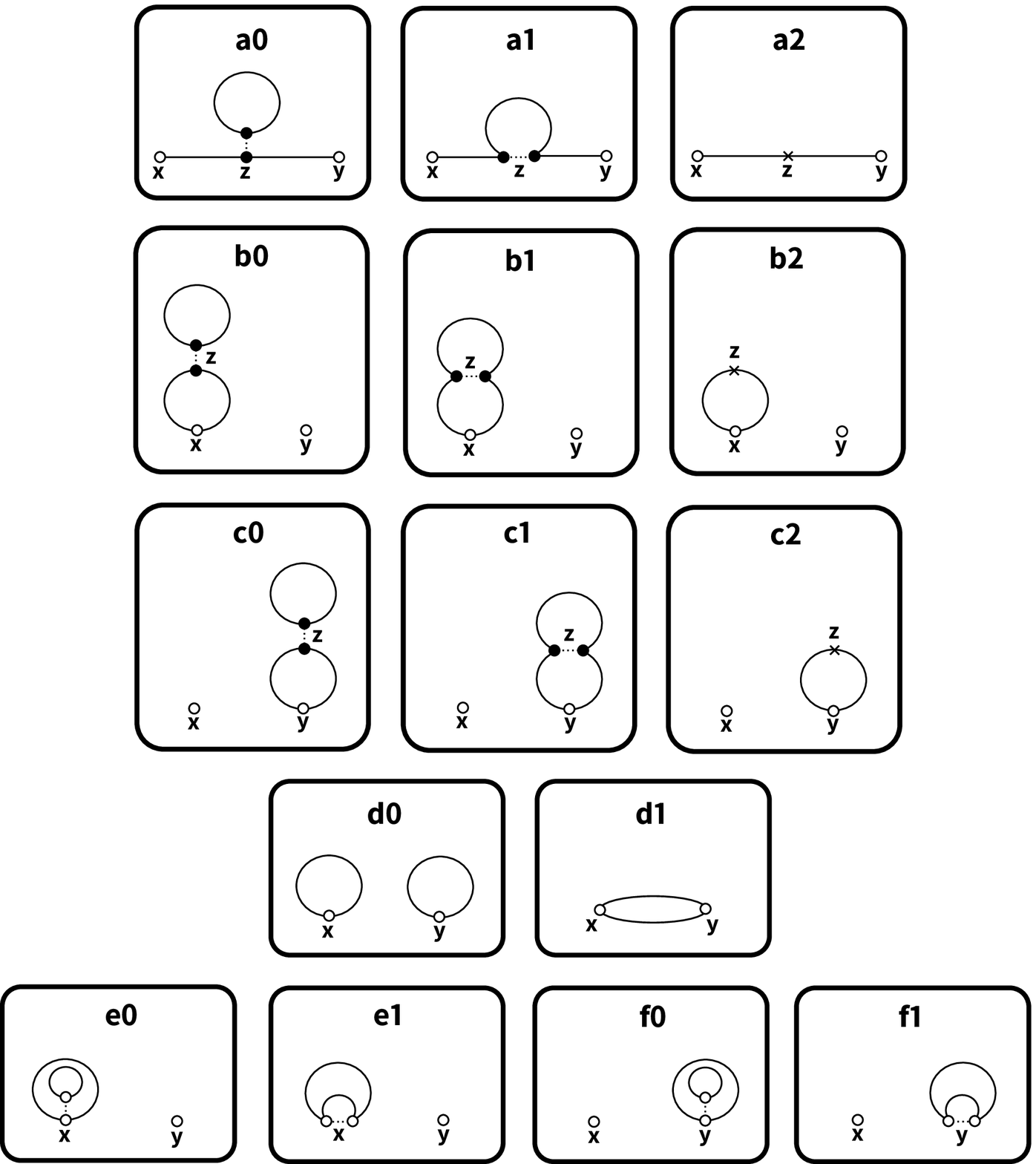}
\caption{The diagramatic description of 
the $O(n)$-invariant $2$-point Green function at ${\cal O}(g^4)$.} 
\label{fig:g4_diag}
\end  {center}
\end  {figure}

The ${\cal O}(g^4)$ contribution to $O(n)$-invariant $2$-point Green function 
is given by the third term in Eq. (\ref{eqn:inv_Green1}). 
Corresponding to 
$\langle\,{\bm \pi}  (x)\cdot{\bm \pi}  (y)\,\rangle_1$\,,
$\langle\,{\bm \pi}^2(x)                   \,\rangle_1$\,,
$\langle\,                   {\bm \pi}^2(y)\,\rangle_1$\,,
$\langle\,{\bm \pi}^2(x)     {\bm \pi}^2(y)\,\rangle_0$\,,
$\langle\,{\bm \pi}^4(x)                   \,\rangle_0$ and 
$\langle\,                   {\bm \pi}^4(y)\,\rangle_0$\,,
we refer these contributions by 
``a'', ``b'', ``c'', ``d'', ``e'' and ``f'', respectively. 
Moreover, we subdivide each contribution into some groups 
based on the type of interaction, or the pattern for contraction. 
The diagramatic description is given in Fig. \ref{fig:g4_diag}. 
We project the total momentum in the external line to zero, again, 
and introduce the abridged notation, 
\begin{eqnarray}
  {\rm Int}_2[\,f(x,y)\,]
  & \equiv &
  \frac{1}{L^{2(d-1)}}\int d^{d-1}{\bm x}\int d^{d-1}{\bm y}\ 
  [\,f(x,y)\,]_{x_0=-\tau,\,y_0=+\tau}\ ,
\nonumber  \\
  {\rm Int}_3[\,f(x,y,z)\,]
  & \equiv &
  \frac{1}{L^{2(d-1)}}\int d^{d-1}{\bm x}\int d^{d-1}{\bm y}\int d^d z\ 
  [\,f(x,y,z)\,]_{x_0=-\tau,\,y_0=+\tau}\ .
\end  {eqnarray}
Then, each contribution can be calculated from 
\begin{eqnarray}
  P_{\rm a0}(\tau)
  &=&
  {\rm Int}_3 
  [\,  \partial_\mu^2               G(x;z)\,
                                    G(z;y)\,
       \partial_\mu^1               G(z;z)\,
    +\,\partial_\mu^2               G(x;z)\,
                                    G(z;y)\,
       \partial_\mu^2               G(z;z)
\nonumber  \\
  & &
    \quad\,
    +\,                             G(x;z)\,
       \partial_\mu^1               G(z;y)\,
       \partial_\mu^1               G(z;z)\,
    +\,                             G(x;z)\,
       \partial_\mu^1               G(z;y)\,
       \partial_\mu^2               G(z;z)\,
    ]
\nonumber  \\
  P_{\rm a1}(\tau)
  &=&
  {\rm Int}_3
  [\,  \partial_\mu^2               G(x;z)\,
                                    G(z;y)\,
       \partial_\mu^2               G(z;z)\,
    +\,\partial_\mu^2               G(x;z)\,
       \partial_\mu^1               G(z;y)\,
                                    G(z;z)
\nonumber  \\
  & &
    \quad\,
    +\,                             G(x;z)\,
                                    G(z;y)\,
       \partial_\mu^1\partial_\mu^2 G(z;z)\,
    +\,                             G(x;z)\,
       \partial_\mu^1               G(z;y)\,
       \partial_\mu^1               G(z;z)\,
  ]
\nonumber  \\
  P_{\rm a2}(\tau)
  &=&
  {\rm Int}_3
  [\,                               G(x;z)\,
                                    G(z;y)\,
  ]
\nonumber  \\
  P_{\rm b0}(\tau)
  &=&
  {\rm Int}_3
  [\,  \partial_\mu^2               G(x;z)\,
                                    G(z;x)\,
       \partial_\mu^1               G(z;z)\,
    +\,\partial_\mu^2               G(x;z)\,
                                    G(z;x)\,
       \partial_\mu^2               G(z;z)
\nonumber  \\
  & &
    \quad\,
    +\,                             G(x;z)\,
       \partial_\mu^1               G(z;x)\,
       \partial_\mu^1               G(z;z)\,
    +\,                             G(x;z)\,
       \partial_\mu^1               G(z;x)\,
       \partial_\mu^2               G(z;z)\,
    ]
\nonumber  \\
  P_{\rm b1}(\tau)
  &=&
  {\rm Int}_3
  [\,  \partial_\mu^2               G(x;z)\,
                                    G(z;x)\,
       \partial_\mu^2               G(z;z)\,
    +\,\partial_\mu^2               G(x;z)\,
       \partial_\mu^1               G(z;x)\,
                                    G(z;z)
\nonumber  \\
  & &
    \quad\,
    +\,                             G(x;z)\,
                                    G(z;x)\,
       \partial_\mu^1\partial_\mu^2 G(z;z)\,
    +\,                             G(x;z)\,
       \partial_\mu^1               G(z;x)\,
       \partial_\mu^1               G(z;z)\,
  ]
\nonumber  \\
  P_{\rm b2}(\tau)
  &=&
  {\rm Int}_3
  [\,                               G(x;z)\,
                                    G(z;x)\,
  ]
\nonumber  \\
  P_{\rm c0}(\tau)
  &=&
  {\rm Int}_3
  [\,  \partial_\mu^2               G(y;z)\,
                                    G(z;y)\,
       \partial_\mu^1               G(z;z)\,
    +\,\partial_\mu^2               G(y;z)\,
                                    G(z;y)\,
       \partial_\mu^2               G(z;z)
\nonumber  \\
  & &
    \quad\,
    +\,                             G(y;z)\,
       \partial_\mu^1               G(z;y)\,
       \partial_\mu^1               G(z;z)\,
    +\,                             G(y;z)\,
       \partial_\mu^1               G(z;y)\,
       \partial_\mu^2               G(z;z)\,
  ]
\nonumber  \\
  P_{\rm c1}(\tau)
  &=&
  {\rm Int}_3
  [\,  \partial_\mu^2               G(y;z)\,
                                    G(z;y)\,
       \partial_\mu^2               G(z;z)\,
    +\,\partial_\mu^2               G(y;z)\,
       \partial_\mu^1               G(z;y)\,
                                    G(z;z)
\nonumber  \\
  & &
    \quad\,
    +\,                             G(y;z)\,
                                    G(z;y)\,
       \partial_\mu^1\partial_\mu^2 G(z;z)\,
    +\,                             G(y;z)\,
       \partial_\mu^1               G(z;y)\,
       \partial_\mu^1               G(z;z)\,
  ]
\nonumber  \\
  P_{\rm c2}(\tau)
  &=&
  {\rm Int}_3
  [\,                               G(y;z)\,
                                    G(z;y)\,
  ]
\nonumber  \\
  P_{\rm d0}(\tau)
  &=&
  {\rm Int}_2
  [\,                               G(x;x)\,
                                    G(y;y)\,
  ]
\nonumber  \\
  P_{\rm d1}(\tau)
  &=&
  {\rm Int}_2
  [\,                               G(x;y)\,
                                    G(x;y)\,
  ]
\nonumber  \\
  P_{\rm e0}(\tau)
  &=&
  {\rm Int}_2
  [\,                               G(x;x)\,
                                    G(x;x)\,
  ]
\nonumber  \\
  P_{\rm e1}(\tau)
  &=&
  {\rm Int}_2
  [\,                               G(x;x)\,
                                    G(x;x)\,
  ]
\nonumber  \\
  P_{\rm f0}(\tau)
  &=&
  {\rm Int}_2
  [\,                               G(y;y)\,
                                    G(y;y)\,
  ]
\nonumber  \\
  P_{\rm f1}(\tau)
  &=&
  {\rm Int}_2
  [\,                               G(y;y)\,
                                    G(y;y)\,
  ]
\label{eqn:g4_contribution}
\end  {eqnarray}
except for 
the coefficient in              Eq. (\ref{eqn:inv_Green1}), 
the coefficient in the brace of Eq. (\ref{eqn:exp_val}), 
the multiplicity of the spin component and 
the statistical factor. 
Here the superscript in partial differential symbol means 
\begin{equation}
  \partial_\mu^1 G(u,v) 
  \equiv 
  \frac{\partial\,G(u,v)}{\partial\,u^\mu}\ ,
\quad
  \partial_\mu^2 G(u,v) 
  \equiv 
  \frac{\partial\,G(u,v)}{\partial\,v^\mu}\ .
\end  {equation}
%
%---------------------------------
%
%
%---------------------------------
%

The explicit forms of 
each $P(\tau)$ in Eq. (\ref{eqn:g4_contribution}) 
and the factor to be multiplied $F$ 
are summarized in Table \ref{tbl:NLO_calc}. 
Note that 
$R(x_0+y_0\pm T,{\bm x}-{\bm y})$ in Eq. (\ref{eqn:nz_mode})
cannot be ignored for the derivation of some of $P(\tau)$ at this order. 
In addition, 
it should be also noted that 
$\displaystyle 
 \partial_0^1\partial_0^2G_Z(x_0;y_0)\,|_{x_0=y_0}
 =\frac{1}{L^{d-1}}\delta(0)$ 
vanishes due to $\delta(0)=0$ with the dimensional regularization. 
%
%---------------------------------
%
\begin{table}[p]
\begin{center}
\caption{
The explicit forms of 
each $P(\tau)$ in Eq. (\ref{eqn:g4_contribution}) 
and the factor to be multiplied $F$. 
The prime in $\sum_{\bm p}'$ means that 
the ${\bm p}={\bm 0}$ contribution is excluded from the summation. 
Note that 
$\displaystyle \frac{1}{2L^{d-1}}
 \Big({\textstyle \sum_{\bm p}'}\frac{1}{{\bm p}^2}\Big)$ 
is nothing less than $R(0)$\,.
}
\medskip
\begin{tabular}{cclcc}
\hline
\hline
diagram & $\qquad$ & 
\multicolumn{1}{c}{$P(\tau)$} 
& $\qquad$ & $F$ \\
\hline
a0 & & 
$\displaystyle 
\frac{1}{L^{2(d-1)}}
\left[\ 
  \left(
      \frac{7\tau^4}{ 6T^2}
    - \frac{4\tau^3}{ 3T  }
    + \frac{ \tau^2}{12   }
    + \frac{    T^2}{1440 }
  \right)
\right.
$
& & \\
    & & 
$\displaystyle 
\qquad\qquad
\left.
+ \left(
    {\textstyle \sum_{\bm p}'}\frac{1}{{\bm p}^2}
  \right)
  \left(
    - \frac{ \tau^2}{ 2T^2} 
    + \frac{1      }{24   }
  \right)
+ \left(
    {\textstyle \sum_{\bm p}'}\frac{1}{{\bm p}^4}
  \right)
  \left(
    - \frac{3      }{ 4T^2}
  \right)\ 
\right]
$
& & $\displaystyle -\frac{(n-1)^2g^4}{2}$ 
\\
a1 & & 
$\displaystyle 
\frac{1}{L^{2(d-1)}}
\left[\ 
  \left(
      \frac{5\tau^4}{ 6T^2}
    - \frac{ \tau^3}{  T  }
    + \frac{5\tau^2}{12   }
    - \frac{T\tau  }{12   }
    + \frac{11  T^2}{1440 }
  \right)
\right.
$
& & \\
    & & 
$\displaystyle 
\qquad\qquad
\left.
+ \left(
    {\textstyle \sum_{\bm p}'}\frac{1}{|{\bm p}|}
  \right)
  \left(
      \frac{ \tau^2}{ 2T  }
    - \frac{ \tau  }{ 2   }
    + \frac{T      }{24   }
  \right)
+ \left(
    {\textstyle \sum_{\bm p}'}\frac{1}{{\bm p}^4}
  \right)
  \left(
    \frac{1      }{ 4T^2}
  \right)\ 
\right]
$
& & $\displaystyle -(n-1)g^4$ 
\\
a2 & & 
$\displaystyle 
\frac{1}{L^{d-1}}
\left(
  - \frac{ \tau^4}{ 3T}
  + \frac{2\tau^3}{ 3 }
  - \frac{T\tau^2}{ 3 }
  + \frac{T^3    }{720}
\right)
$
& & $\displaystyle -\frac{(n-1)^2g^4}{TL^{d-1}}$ 
\\
b0 & & 
$\displaystyle 
\frac{1}{L^{2(d-1)}}
\left[\ 
  \left(
      \frac{7\tau^4}{ 6T^2}
    + \frac{ \tau^2}{12   }
    + \frac{    T^2}{1440 }
  \right)
+ \left(
    {\textstyle \sum_{\bm p}'}\frac{1}{ {\bm p}^2 }
  \right)
  \left(
    - \frac{ \tau^2}{ 2T^2}
    + \frac{1      }{24   }
  \right)
\right.
$
& & \\
    & & 
$\displaystyle 
\qquad\qquad
\left.
+ \left(
    {\textstyle \sum_{\bm p}'}\frac{1}{|{\bm p}|^3}
  \right)
  \left(
    - \frac{1      }{ 2T  }
  \right)
+ \left(
    {\textstyle \sum_{\bm p}'}\frac{1}{ {\bm p}^4 }
  \right)
  \left(
    - \frac{3      }{ 4T^2}
  \right)\ 
\right]
$
& & $\displaystyle +\frac{(n-1)^2g^4}{4}$ 
\\
b1 & & 
$\displaystyle 
\frac{1}{L^{2(d-1)}}
\left[\ 
  \left(
      \frac{5\tau^4}{ 6T^2}
    - \frac{ \tau^2}{12   }
    + \frac{11  T^2}{1440 }
  \right)
+ \left(
    {\textstyle \sum_{\bm p}'}\frac{1}{|{\bm p}|}
  \right)
  \left(
      \frac{ \tau^2}{  T  }
    + \frac{T      }{12   }
  \right)
\right.
$
& & \\
    & & 
$\displaystyle 
\qquad\qquad
\left.
+ \left(
    {\textstyle \sum_{\bm p}'}\frac{1}{ {\bm p}^4}
  \right)
  \left(
    \frac{1      }{ 4T^2}
  \right)
+ \left(
    {\textstyle \sum_{\bm p}'}\frac{1}{|{\bm p}|}
  \right)^2
  \left(
    \frac{1      }{ 4   }
  \right)\ 
\right]
$
& & $\displaystyle +\frac{(n-1)g^4}{2}$
\\
b2 & & 
$\displaystyle 
\frac{1}{L^{d-1}}
\left[\ 
  \left(
    - \frac{ \tau^4}{ 3T}
    + \frac{T\tau^2}{ 6 }
    + \frac{T^3    }{720}
  \right)
+ \left(
    {\textstyle \sum_{\bm p}'}\frac{1}{|{\bm p}|^3}
  \right)
  \left(
    \frac{1      }{ 4 }
  \right)\ 
\right]
$
& & $\displaystyle +\frac{(n-1)^2g^4}{2TL^{d-1}}$ 
\\
c0 & & 
$\displaystyle 
\frac{1}{L^{2(d-1)}}
\left[\ 
  \left(
      \frac{7\tau^4}{ 6T^2}
    + \frac{ \tau^2}{12   }
    + \frac{    T^2}{1440 }
  \right)
+ \left(
    {\textstyle \sum_{\bm p}'}\frac{1}{ {\bm p}^2 }
  \right)
  \left(
    - \frac{ \tau^2}{ 2T^2}
    + \frac{1      }{24   }
  \right)
\right.
$
& & \\
    & & 
$\displaystyle 
\qquad\qquad
\left.
+ \left(
    {\textstyle \sum_{\bm p}'}\frac{1}{|{\bm p}|^3}
  \right)
  \left(
    - \frac{1      }{ 2T  }
  \right)
+ \left(
    {\textstyle \sum_{\bm p}'}\frac{1}{ {\bm p}^4 }
  \right)
  \left(
    - \frac{3      }{ 4T^2}
  \right)\ 
\right]
$
& & $\displaystyle +\frac{(n-1)^2g^4}{4}$ 
\\
c1 & & 
$\displaystyle 
\frac{1}{L^{2(d-1)}}
\left[\ 
  \left(
      \frac{5\tau^4}{ 6T^2}
    - \frac{ \tau^2}{12   }
    + \frac{11  T^2}{1440 }
  \right)
+ \left(
    {\textstyle \sum_{\bm p}'}\frac{1}{|{\bm p}|}
  \right)
  \left(
      \frac{ \tau^2}{  T  }
    + \frac{T      }{12   }
  \right)
\right.
$
& & \\
    & & 
$\displaystyle 
\qquad\qquad
\left.
+ \left(
    {\textstyle \sum_{\bm p}'}\frac{1}{ {\bm p}^4}
  \right)
  \left(
    \frac{1      }{ 4T^2}
  \right)
+ \left(
    {\textstyle \sum_{\bm p}'}\frac{1}{|{\bm p}|}
  \right)^2
  \left(
    \frac{1      }{ 4   }
  \right)\ 
\right]
$
& & $\displaystyle +\frac{(n-1)g^4}{2}$
\\
c2 & & 
$\displaystyle 
\frac{1}{L^{d-1}}
\left[\ 
  \left(
    - \frac{ \tau^4}{ 3T}
    + \frac{T\tau^2}{ 6 }
    + \frac{T^3    }{720}
  \right)
+ \left(
    {\textstyle \sum_{\bm p}'}\frac{1}{|{\bm p}|^3}
  \right)
  \left(
    \frac{1      }{ 4 }
  \right)\ 
\right]
$
& & $\displaystyle +\frac{(n-1)^2g^4}{2TL^{d-1}}$
\\
d0 & & 
$\displaystyle 
\frac{1}{L^{2(d-1)}}
\left[\ 
  \left(
      \frac{ \tau^2}{  T  }
    + \frac{    T  }{12   }
  \right)
+ \left(
    {\textstyle \sum_{\bm p}'}\frac{1}{|{\bm p}|}
  \right)
  \left(
      \frac{1      }{ 2   }
  \right)\ 
\right]^2
$
& & $\displaystyle +\frac{(n-1)^2g^4}{4}$
\\
d1 & & 
$\displaystyle 
\frac{1}{L^{2(d-1)}}
\left(
      \frac{ \tau^2}{  T  }
    - \tau
    + \frac{    T  }{12   }
\right)^2
$
& & $\displaystyle +\frac{(n-1)g^4}{2}$
\\
e0 & & 
$\displaystyle 
\frac{1}{L^{2(d-1)}}
\left[\ 
  \left(
      \frac{ \tau^2}{  T  }
    + \frac{    T  }{12   }
  \right)
+ \left(
    {\textstyle \sum_{\bm p}'}\frac{1}{|{\bm p}|}
  \right)
  \left(
      \frac{1      }{ 2   }
  \right)\ 
\right]^2
$
& & $\displaystyle -\frac{(n-1)^2g^4}{8}$ 
\\
e1 & & 
$\displaystyle 
\frac{1}{L^{2(d-1)}}
\left[\ 
  \left(
      \frac{ \tau^2}{  T  }
    + \frac{    T  }{12   }
  \right)
+ \left(
    {\textstyle \sum_{\bm p}'}\frac{1}{|{\bm p}|}
  \right)
  \left(
      \frac{1      }{ 2   }
  \right)\ 
\right]^2
$
& & $\displaystyle -\frac{(n-1)g^4}{4}$
\\
f0 & & 
$\displaystyle 
\frac{1}{L^{2(d-1)}}
\left[\ 
  \left(
      \frac{ \tau^2}{  T  }
    + \frac{    T  }{12   }
  \right)
+ \left(
    {\textstyle \sum_{\bm p}'}\frac{1}{|{\bm p}|}
  \right)
  \left(
      \frac{1      }{ 2   }
  \right)\ 
\right]^2
$
& & $\displaystyle -\frac{(n-1)^2g^4}{8}$
\\
f1 & & 
$\displaystyle 
\frac{1}{L^{2(d-1)}}
\left[\ 
  \left(
      \frac{ \tau^2}{  T  }
    + \frac{    T  }{12   }
  \right)
+ \left(
    {\textstyle \sum_{\bm p}'}\frac{1}{|{\bm p}|}
  \right)
  \left(
      \frac{1      }{ 2   }
  \right)\ 
\right]^2
$
& & $\displaystyle -\frac{(n-1)g^4}{4}$
\\
\hline
\hline
\label{tbl:NLO_calc}
\end  {tabular}
\end  {center}
\end  {table}
%
%---------------------------------

The total contribution at ${\cal O}(g^4)$ can be written as 
\begin{eqnarray}
  & &
  g^4\,
  \frac{1}{L^{2(d-1)}}\int d^{d-1}{\bm x}\int d^{d-1}{\bm y}\ 
  \langle\,{\bm \phi}(x)\cdot{\bm \phi}(y)\,\rangle_1
\nonumber  \\
  &=&
  g^4\,
  \left[\,
      \frac{(n-1)}{2}\,R(0)^2
    +       (n-1)    \,R(0)
                       \left(\,\frac{|\tau|}{L^{d-1}}\,\right)
    + \frac{(n-1)^2}{2}\left(\,\frac{|\tau|}{L^{d-1}}\,\right)^2\,
  \right]\ .
\label{eqn:inv_Green1_g4}
\end  {eqnarray}
%
% @@@ =========================================================================
%
\subsection{Evaluation of $R(0)$}
\label{ssec:Evaluation_of_R0}
We evaluate $R(0)$ which appears in 
Eqs. (\ref{eqn:inv_Green1_g2}) and (\ref{eqn:inv_Green1_g4}). 
Discussion in this subsection is based on Ref. \cite{Niedermayer:2016ilf}. 
From Eq. (\ref{eqn:def_Rz}), we have 
\begin{eqnarray}
  R(z)
  &=&
  \frac{1}{L^{d-1}}\sum_{{\bm q}_{\bm n}\ne{\bm 0}}
  \int_{-\infty}^{\infty} \frac{dq_0}{2\pi}\,
  \frac{1}{q_0^2+{\bm q}_{\bm n}^2}\,
  {\rm e}^{iq_0z_0+i{\bm q}_{\bm n}\cdot{\bm z}}
\nonumber  \\
  &=&
  \int_0^\infty d\lambda\ 
  \left[\,
    \int_{-\infty}^{\infty} \frac{dq_0}{2\pi}\,
    {\rm e}^{-\lambda q_0^2+iq_0z_0}\,
  \right]
  \times
  \left[\,
    \frac{1}{L^{d-1}}\sum_{{\bm q}_{\bm n}\ne{\bm 0}}
    {\rm e}^{-\lambda{\bm q}_{\bm n}^2+i{\bm q}_{\bm n}\cdot{\bm z}}\,
  \right]
\nonumber  \\
  &=&
  \int_0^\infty d\lambda\ 
  \left[\,
    (4\pi\lambda)^{-\frac{1}{2}}\,
    {\rm e}^{-\frac{z_0^2}{4\lambda}}\,
  \right]
  \times
  \left[\,
    (4\pi\lambda)^{-\frac{d-1}{2}}\,
    \sum_{{\bm w}\in\mathbb{Z}^{d-1}}
    {\rm e}^{-\frac{({\bm z}+L{\bm w})^2}{4\lambda}}
    -
    \frac{1}{L^{d-1}}\,
  \right]\ .
\end  {eqnarray}
We change the variable from $\lambda$ to $u\equiv 4\pi\lambda/L^2$. 
Then, we have 
\begin{equation}
  R(z)\,\mu^{-(d-2)}
  =
  \frac{1}{4\pi (\mu L)^{d-2}}
  \int_0^\infty du\ 
  u^{-\frac{1}{2}}\,{\rm e}^{-\frac{\pi z_0^2}{L^2 u}}\,
  \left[\,
    u^{-\frac{d-1}{2}}
    \prod_{\mu=1}^{d-1} 
    \Big\{\,
      \sum_{w_\mu=-\infty}^\infty
      {\rm e}^{-\pi \frac{(z_\mu/L+w_\mu)^2}{u}}\,
    \Big\}
    - 1\,
  \right]\ ,
\end  {equation}
where arbitrary scale $\mu$ with a mass dimension is introduced 
to make $R(z)$ dimensionless. 
$\mu$ is called renormalization scale.

We consider the case of $z=0$\,. 
We define the function 
\begin{equation}
  S(u) \equiv \sum_{n=-\infty}^\infty {\rm e}^{-\pi u n^2}\ . 
\end  {equation}
Using the relation $S(u) = u^{-1/2}\,S(u^{-1})$,  
we obtain 
\begin{equation}
  R(0)\,\mu^{-(d-2)}
  =
  \frac{1}{4\pi (\mu L)^{d-2}}
  \int_0^\infty du\ u^{-1/2}\,\Big[\,S(u)^{d-1}-1\,\Big]\ .
\label{eqn:R0_1}
\end  {equation}
Moreover, we introduce the notation 
\begin{equation}
  [\,f(u)\,]_{\rm sub}
  =
  \left\{
  \begin{array}{lll}
    f(u) - [\,f(u)\,]_0      &\quad & (0<u<1)  \\
    f(u) - [\,f(u)\,]_\infty &      & (1<u  )  \\
  \end  {array}
  \right.\ ,
\end  {equation}
where 
$[\,f(u)\,]_0$ and $[\,f(u)\,]_\infty$ 
denote the leading asymptotic parts of $f(u)$ 
at $u\to 0$ and $u\to\infty$, respectively.
Then, Eq. (\ref{eqn:R0_1}) can be rewritten as 
\begin{equation}
  R(0)\,\mu^{-(d-2)}
  =
  \frac{1}{2\pi (\mu L)^{d-2}}
  \left[\,
    - \frac{1}{d-2}
    - 1
    + \frac{1}{2}
      \int_0^\infty du\ u^{-1/2}\,\Big[\,S(u)^{d-1}\,\Big]_{\rm sub}\,
  \right]\ .
\label{eqn:R0_2}
\end  {equation}
We expand $S(u)^{d-1}$ with respect to $(d-2)$, and obtain 
\begin{equation}
  R(0)\,\mu^{-(d-2)}
  =
  \frac{1}{2\pi (\mu L)^{d-2}}
  \left\{\,
    - \frac{1}{d-2}
    - 1
    + \sum_{j=0}^\infty\,(d-2)^j\,\frac{X_j}{j\,!}\,
  \right\}\ ,
\end  {equation}
with 
\begin{equation}
  X_j
  \equiv
  \frac{1}{2}\,
  \int_0^\infty du\ u^{-1/2}\,
  \Big[\,
    S(u)\,(\ln S(u))^j\,
  \Big]_{\rm sub}\ .
\end  {equation}
As we will see later, 
$X_0$ and $X_1$ do not appear 
in the final expression of the $\beta$ function and anomalous dimension 
up to the ${\cal O}(g^6)$ order of the renormalization factor. 
Thus, we do not give the numerical values of $X_0$ and $X_1$\,. 
We add that $X_0$ can be written in an analytical form 
\begin{equation}
  X_0 = 1 - \frac{1}{2}\, \Big(\,\ln 4\pi+\Gamma'(1)\,\Big)\ .
\end  {equation}
Finally, expanding $(\mu L)^{-(d-2)}$ with respect to $(d-2)$, 
we obtain 
\begin{equation}
  R(0)\,\mu^{-(d-2)}
  =
  \frac{1}{2\pi}
  \left\{\,
    - \frac{1}{d-2}
    +        Y_0(\mu L)
    + (d-2)\,Y_1(\mu L)
    + {\cal O}((d-2)^2)
  \right\}\ ,
\end  {equation}
where we use the following functions, 
\begin{eqnarray}
  Y_0(\mu L) &\equiv &       (X_0-1) + (\ln\mu L)\ ,
\\
  Y_1(\mu L) &\equiv & X_1 - (X_0-1)   (\ln\mu L) - \frac{1}{2}(\ln\mu L)^2\ .
\end  {eqnarray}
%
% @@@ =========================================================================
%
\subsection{Renormalization in MS scheme}
\label{ssec:Renormalization_in_MS_scheme}
In Eq.  (\ref{eqn:inv_Green1}) with 
   Eqs. (\ref{eqn:inv_Green1_g2}) and 
        (\ref{eqn:inv_Green1_g4}), 
the $O(n)$-invariant $2$-point Green function was given 
as a series of the bare coupling $g^2$. 
It might be preferable to rewrite in terms of the renormalized coupling. 
As the UV property is not affected by properties of the box 
such as the size or the boundary condition, 
it is possible to use the MS scheme for the renormalization. 
The ${\cal O}(g^6)$ renormalization factor on the $O(n)$ sigma model 
has been already given in Ref. \cite{Hikami:1977vr}. 
The renormalization is done by the replacement 
\begin{eqnarray}
  g^2\mu^{d-2}  &=&  Z_{\rm MS}^g\,          g_{\rm MS}^2          \ ,
\label{eqn:g2_renom_MS}
\\
  {\bm \phi}(z) &=& (Z_{\rm MS}^\phi)^{1/2}\,{\bm \phi}_{\rm MS}(z)\ ,
\label{eqn:ph_renom_MS}
\end  {eqnarray}
with 
\begin{eqnarray}
  Z_{\rm MS}^g
  &=&
  1
  +
  \frac{n-2}{2\pi(d-2)}\,
  g_{\rm MS}^2
  +
  \left[\,
      \frac{ n-2   }{8\pi^2(d-2)  }
    + \frac{(n-2)^2}{4\pi^2(d-2)^2}\,
  \right]
  g_{\rm MS}^4
\nonumber  \\
  & &
  +
  \left[\,
      \frac{ (n-2)(n+2)}{96\pi^3(d-2)  }
    + \frac{7(n-2)^2   }{48\pi^3(d-2)^2}\,
    + \frac{ (n-2)^3   }{ 8\pi^3(d-2)^3}\,
  \right]
  g_{\rm MS}^6
  +
  {\cal O}(g_{\rm MS}^8)\ ,\qquad
\label{eqn:g2_Z_MS}
\\
  Z_{\rm MS}^\phi
  &=&
  1
  +
  \frac{n-1}{2\pi(d-2)}\,
  g_{\rm MS}^2
  +
  \frac{(n-1)(n-\frac{3}{2})}{4\pi^2(d-2)^2}\,
  g_{\rm MS}^4
\nonumber  \\
  & &
  +
  \left[\,
      \frac{(n-1)(n-2)}{32\pi^3(d-2)  }
    + \frac{(n-1)(n-2)}{24\pi^3(d-2)^2}\,
    + \frac{(n-1)(n^2-\frac{19}{6}n+\frac{5}{2})}{ 8\pi^3(d-2)^3}\,
  \right]
  g_{\rm MS}^6
  +
  {\cal O}(g_{\rm MS}^8)\ .
\nonumber  \\
  & &
\label{eqn:ph_Z_MS}
\end  {eqnarray}
$\mu$, which was introduced in Sec. \ref{ssec:Evaluation_of_R0}, 
is also used to make $g_{\rm MS}^2$ dimensionless. 
Note that the MS scheme focuses on only the elimination of UV divergence 
and thus the coefficients 
in Eqs. (\ref{eqn:g2_Z_MS}) and (\ref{eqn:ph_Z_MS}) 
contain only the pole terms.

Using Eqs. (\ref{eqn:g2_renom_MS}) and (\ref{eqn:g2_Z_MS}), 
the zero-momentum projected $O(n)$-invariant Green function 
$G_{\rm inv}(x_0;y_0)\,|_{x_0=-\tau,\,y_0=+\tau}$ can be rewritten as 
\begin{eqnarray}
  & &
  1 
  + 
  g_{\rm MS}^2
  \left[\,
    \left(\,
        \frac{      n-1 }{2\pi(d-2)}
      - \frac{      n-1 }{2\pi     }\,Y_0(\mu L)
      - \frac{(d-2)(n-1)}{2\pi     }\,Y_1(\mu L)\,
    \right)
  \right.
\nonumber  \\
  & &
  \qquad
  \left.
    +\,
    \Big(\,
      - (n-1)
      + (d-2)(n-1)(\ln\mu L)\,
    \Big)
    \left(
      \frac{|\tau|}{L}
    \right)\,
  \right]
\nonumber  \\
  & &
  +
  g_{\rm MS}^4
  \left[\,
    \left(\,
        \frac{(n-1)(n-\frac{3}{2})}{4\pi^2(d-2)^2}
      - \frac{(n-1)^2             }{4\pi^2(d-2)  }\,Y_0(\mu L)
      + \frac{ n-1                }{8\pi^2       }\,Y_0(\mu L)^2
      - \frac{(n-1)^2             }{4\pi^2       }\,Y_1(\mu L)\,
    \right)
  \right.
\nonumber  \\
  & &
  \qquad
  \left.
    +
    \left(\,
      - \frac{(n-1)^2}{2\pi(d-2)}
      + \frac{ n-1   }{2\pi}\,Y_0(\mu L)
      + \frac{(n-1)^2}{2\pi}\,(\ln\mu L)\,
    \right)
    \left(
      \frac{|\tau|}{L}
    \right)\,
    +
    \frac{(n-1)^2}{2}
    \left(
      \frac{|\tau|}{L}
    \right)^2\,
  \right]
\nonumber  \\
  & &
  + {\cal O}(g_{\rm MS}^6)\ .
\label{eqn:inv_Green2}
\end  {eqnarray}
In Eq. (\ref{eqn:inv_Green2}), 
we abbreviate 
${\cal O}((d-2)^2)$ terms in the coefficient of $g_{\rm MS}^2$\,, and 
${\cal O} (d-2)   $ terms in ones            of $g_{\rm MS}^4$\,.

For the later discussion, 
we summalize the $\beta$ function and the anomalous dimension. 
In general, the perturbative expression in R scheme can be written as 
\begin{eqnarray}
   \beta_{\rm R}(g_{\rm R}^2)
  &=&
  (d-2)\,g_{\rm R}^2
  - 
  g_{\rm R}^4\,
  \sum_{i=0}^\infty\,g_{\rm R}^{2i}\, \beta_{{\rm R},i}\ ,
\label{eqn:bt_exp_dim}
\\
  \gamma_{\rm R}(g_{\rm R}^2)
  &=&
  - 
  g_{\rm R}^2\,
  \sum_{i=0}^\infty\,g_{\rm R}^{2i}\,\gamma_{{\rm R},i}\ ,
\label{eqn:gm_exp_dim}
\end  {eqnarray}
with the dimensional regularization. 
For the MS scheme, each coefficient are given by 
\begin{eqnarray}
  & &
   \beta_{{\rm MS},0} =   \frac{  n-2      }{ 2\pi  }\ ,\quad
   \beta_{{\rm MS},1} =   \frac{  n-2      }{ 4\pi^2}\ ,\quad
   \beta_{{\rm MS},2} =   \frac{ (n-2)(n+2)}{32\pi^3}\ ,
\label{eqn:bt_coeff_MS}
\\
  & &
  \gamma_{{\rm MS},0} = - \frac{n-1}{2\pi}\ ,\quad
  \gamma_{{\rm MS},1} = 0\ ,\quad
  \gamma_{{\rm MS},2} = - \frac{3(n-1)(n-2)}{32\pi^3}\ ,
\label{eqn:gm_coeff_MS}
\end  {eqnarray}
up to the ${\cal O}(g_{\rm MS}^6)$ order of the renormalization factor. 
$ \beta_{{\rm MS},2}$ and 
$\gamma_{{\rm MS},2}$ have been first derived in 
Ref. \cite{Hikami:1977vr}. 
The derivation is straightforward from 
Eqs. (\ref{eqn:g2_renom_MS}), 
     (\ref{eqn:g2_Z_MS}) and 
     (\ref{eqn:ph_Z_MS}). 
%
% @@@ =========================================================================
%
\subsection{Renormalization in FV scheme}
\label{ssec:Renormalization_in_FV_scheme}
The zero-momentum projected $O(n)$-invariant Green function 
can be re-expressed with the mass gap $M$ and the amplitude $A$ as 
\begin{equation}
  G_{\rm inv}(x_0;y_0)\,|_{x_0=-\tau,\,y_0=+\tau}
  =
  A\,{\rm e}^{-2|\tau|M}\ . 
\label{eqn:inv_Green3}
\end  {equation}
Using the expansion 
\begin{equation}
  M
  =
  \sum_{i=1}^\infty\,g_{\rm MS}^{2i}\,M_i\ ,\quad 
  A
  =
  1 + 
  \sum_{i=1}^\infty\,g_{\rm MS}^{2i}\,A_i\ ,
\label{eqn:expns_MZ}
\end  {equation}
the right hand side of Eq. (\ref{eqn:inv_Green3}) can be expand as 
\begin{eqnarray}
  A\,{\rm e}^{-2|\tau|M}
  &=&
    1
  + g_{\rm MS}^2
    \left[\,
      A_1 - 2LM_1 \left(\frac{|\tau|}{L}\right)\,
    \right]
\nonumber  \\
  & &
  \quad
  + g_{\rm MS}^4
    \left[\,
      A_2 - 2L  (A_1M_1+M_2) \left(\frac{|\tau|}{L}\right)
          + 2L^2 M_1^2       \left(\frac{|\tau|}{L}\right)^2\,
    \right]
  + {\cal O}(g_{\rm MS}^6)\ .\qquad
\label{eqn:inv_Green4}
\end  {eqnarray}
Comparing Eqs. 
(\ref{eqn:inv_Green2}) and 
(\ref{eqn:inv_Green4}), 
we obtain 
\begin{eqnarray}
  & &
  M_1 = \frac{      n-1 }{2L}
      - \frac{(d-2)(n-1)}{2L}\,(\ln\mu L)\ ,\quad
  M_2 = \frac{(n-1)(n-2)}{2L}\times\frac{1}{2\pi}\,Y_0(\mu L)\ ,
\label{eqn:coeff_M}
\\
  & &
  A_1 = \frac{      n-1           }{2\pi(d-2)    }
      - \frac{      n-1           }{2\pi         }\,Y_0(\mu L)
      - \frac{(d-2)(n-1)          }{2\pi         }\,Y_1(\mu L)\ ,\quad
\nonumber  \\
  & &
  A_2 = \frac{(n-1)(n-\frac{3}{2})}{4\pi^2(d-2)^2}
      - \frac{(n-1)^2             }{4\pi^2(d-2)  }\,Y_0(\mu L)
      + \frac{ n-1                }{8\pi^2       }\,Y_0(\mu L)^2
      - \frac{(n-1)^2             }{4\pi^2       }\,Y_1(\mu L)\ .\qquad
\label{eqn:coeff_A}
\end  {eqnarray}
We abbreviate 
${\cal O}((d-2)^2)$ terms in $M_1$ and $A_1$\,, and 
${\cal O} (d-2)   $ terms in $M_2$ and $A_2$\,.

Until now, 
the perturbative evaluation of the mass gap in a finite box 
has been given at 
the ${\cal O}(g_{\rm MS}^4)$ order 
\cite{Luscher:1982uv,Brezin:1985xx}, 
the ${\cal O}(g_{\rm MS}^6)$ order 
\cite{Floratos:1984bz,Luscher.rc,Niedermayer:2016yll}, and 
the ${\cal O}(g_{\rm MS}^8)$ order 
\cite{Shin:1996gi}. 
The ${\cal O}(g_{\rm MS}^6)$ expression is 
\begin{equation}
  M
  =
  \frac{n-1}{2L}
  \left[\,
      g_{\rm MS}^2
    + g_{\rm MS}^4\,C_1
    + g_{\rm MS}^6\,C_2
    + {\cal O}(g_{\rm MS}^8)\,
  \right]\ ,
\end  {equation}
with
\begin{equation}
  C_1 = \frac{n-2}{2\pi}\,Y_0(\mu L)\,,\quad 
  C_2 = C_1^2 + \frac{C_1}{2\pi} + \frac{3(n-2)}{16\pi^2}\ .
\label{eqn:coeff_M_previous}
\end  {equation}
Our evaluation in Eq. (\ref{eqn:coeff_M}) is consistent with 
$C_1$ in Eq. (\ref{eqn:coeff_M_previous}) at $d=2$.

The mass gap $M$ does not depend on arbitrarily introduced $\mu$\,. 
This fact means that 
the $\mu$-dependence in $C_i$ cancels with one in $g_{\rm MS}^2$\,, and 
$M$ leaves only the $L$-dependence. 
The situation is same for the amplitude $A$.

We introduce the FV scheme. 
The $\mu$- and $L$-dependences in each coefficient $C_i$
appear only through the form of $\mu L$ even in a higher order. 
Thus, 
if we set the renormalization scale by $\mu=1/L$\,, 
the coefficient $c_i\equiv C_i\,|_{\mu=1/L}$\, 
is a constant independent of $\mu$ and $L$\,. 
Then, we can consider the new renormalized coupling 
\begin{equation}
  g_{\rm FV}^2
  =
  g_{\rm MS}^2
  \left(\,
  1
  +
  \sum_{i=1}^\infty 
  g_{\rm MS}^{2i}\,c_i\,
  \right)\ .
\label{eqn:def_FV_g2}
\end  {equation}
Moreover, 
with the another constant coefficient $a_i\equiv A_i\,|_{\mu=1/L}$\,, 
we can introduce the new wave-function renormalization 
\begin{equation}
  Z_{\rm FV}^\phi
  =
  1
  +
  \sum_{i=1}^\infty 
  g_{\rm MS}^{2i}\,
  a_i\ .
\label{eqn:def_FV_ph}
\end  {equation}
These are nothing less than the renormalization in the FV scheme, 
which has been discussed in Sec. \ref{ssec:Renormalization_scheme}. 
%
% @@@ =========================================================================
%
\subsection{$\beta$ function and anomalous dimension in FV scheme}
\label{ssec:beta_function_and_anomalous_dimension_in_FV_scheme}
We discuss the $\beta$ function and the anomalous dimension in the FV scheme. 
The $\beta$ function  in the FV scheme 
is conberted from one in the MS scheme by 
\begin{eqnarray}
  \beta_{\rm FV}(g_{\rm FV}^2)
  &=&
  \beta_{\rm MS}(g_{\rm MS}^2)\,
  \frac{d g_{\rm FV}^2}{d g_{\rm MS}^2}\ .
\label{eqn:bt_conbert}
\end  {eqnarray}
Substituting 
Eqs. (\ref{eqn:bt_exp_dim}) and 
     (\ref{eqn:def_FV_g2}) to 
Eq.  (\ref{eqn:bt_conbert}), 
at $d=2$, we obtain 
\begin{equation}
  \beta_{{\rm FV},0} = \beta_{{\rm MS},0}\ ,\quad 
  \beta_{{\rm FV},1} = \beta_{{\rm MS},1}\ ,\quad 
  \beta_{{\rm FV},2} = \beta_{{\rm MS},2}
                     - \beta_{{\rm MS},1}\, c_1
                     - \beta_{{\rm MS},0}\,(c_2-c_1^2)\ ,
\end  {equation}
up to the ${\cal O}(g_{\rm MS}^6)$ of the renormalization factor. 
Then, we have 
\begin{equation}
  \beta_{{\rm FV},0} = \frac{      n-2 }{2\pi  }\ ,\quad 
  \beta_{{\rm FV},1} = \frac{      n-2 }{4\pi^2}\ ,\quad 
  \beta_{{\rm FV},2} = \frac{(n-1)(n-2)}{8\pi^3}\ .
\label{eqn:bt_coeff_FV}
\end  {equation}
with Eqs. (\ref{eqn:bt_coeff_MS}) and (\ref{eqn:coeff_M_previous}). 
We add Eq. (\ref{eqn:bt_coeff_FV}) has been first given in 
Ref. \cite{Luscher.rc}.

The anomalous dimension in the FV scheme 
is converted from one   in the MS scheme by 
\begin{eqnarray}
  \gamma_{\rm FV}(g_{\rm FV}^2)
  &=&
  \gamma_{\rm MS}(g_{\rm MS}^2)\,
  +
   \beta_{\rm MS}(g_{\rm MS}^2)\,
  \frac{d}{d g_{\rm MS}^2}
  \ln\eta(g_{\rm MS}^2)
\label{eqn:gm_conbert}
\end  {eqnarray}
with 
$\eta\equiv Z_{\rm FV}^\phi/Z_{\rm MS}^\phi$\,. 
The difference of the renormalization scheme affects only the finite part. 
Thus, in the expansion of 
\begin{equation}
  \eta
  = 
  1
  + 
  \sum_{i=1}^\infty g_{\rm MS}^{2i}\,\eta_i\ ,
\label{eqn:exp_eta}
\end  {equation}
each coefficient $\eta_i$ has no pole terms at $d=2$. 
In fact, the coefficients are 
\begin{equation}
  \eta_1 = - \frac{n-1}{2\pi  }(X_0-1)\ ,\quad
  \eta_2 =   \frac{n-1}{8\pi^2}(X_0-1)^2\ ,
\label{eqn:exp_eta2}
\end  {equation}
up to ${\cal O}(g_{\rm MS}^4)$
from Eqs. (\ref{eqn:ph_Z_MS}) and (\ref{eqn:def_FV_ph}). 
Substituting 
Eqs. (\ref{eqn:bt_exp_dim}), 
     (\ref{eqn:gm_exp_dim}), 
     (\ref{eqn:def_FV_g2}) and
     (\ref{eqn:exp_eta}) to 
Eq.  (\ref{eqn:gm_conbert}), 
we obtain 
\begin{eqnarray}
  & &
  \gamma_{{\rm FV},0} =  \gamma_{{\rm MS},0}\ ,\quad 
  \gamma_{{\rm FV},1} =  \gamma_{{\rm MS},1}
                      -  \gamma_{{\rm MS},0}\,c_1
                      +   \beta_{{\rm MS},0}\,\eta_1\ ,
\nonumber  \\
  & &
  \gamma_{{\rm FV},2} =  \gamma_{{\rm MS},2}
                      - 2\gamma_{{\rm MS},1}\,c_1
                      -  \gamma_{{\rm MS},0}\,(c_2-2c_1^2)
                      +   \beta_{{\rm MS},1}\,\eta_1
                      +   \beta_{{\rm MS},0}\,
                         (2\eta_2-\eta_1^2-2\eta_1 c_1)\ ,\qquad
\end  {eqnarray}
up to the ${\cal O}(g_{\rm MS}^6)$ of the renormalization factor. 
Then, we have 
\begin{equation}
  \gamma_{{\rm FV},0} = -\frac{n-1}{2\pi}\ ,\quad 
  \gamma_{{\rm FV},1} = 0\ ,\quad 
  \gamma_{{\rm FV},2} = 0\ .
\label{eqn:gm_coeff_FV}
\end  {equation}
with Eqs. 
(\ref{eqn:bt_coeff_MS}), 
(\ref{eqn:gm_coeff_MS}), 
(\ref{eqn:coeff_M_previous}) and 
(\ref{eqn:exp_eta2}).

\clearpage
%
% @@ ==================================================================
%

%

\begin{thebibliography}{99}
%
%---------------------------------------------------------
\bibitem{Luscher.ps}
%--------------------
M.~L\"uscher,
%``Volume Dependence of the Energy Spectrum 
%  in Massive Quantum Field Theories. 
%  2. Scattering States,''
Commun.\ Math.\ Phys.\  {\bf 105} (1986) 153;\ 
%
%--------------------
%``Two Particle States On A Torus
%  And Their Relation To The Scattering Matrix,''
Nucl.\ Phys.\ B {\bf 354} (1991) 531.
%
%---------------------------------------------------------
\bibitem{Luscher.rc}
%--------------------
M.~L\"uscher, P.~Weisz and U.~Wolff,
%``A Numerical method to compute the running coupling 
%  in asymptotically free theories,''
Nucl.\ Phys.\ B {\bf 359} (1991) 221.
%
%---------------------------------------------------------
\bibitem{Efimov:1970zz}
%--------------------
V.~Efimov,
%``Energy levels arising form the resonant two-body forces 
%  in a three-body system,''
Phys.\ Lett.\  {\bf 33B} (1970) 563.
%
%---------------------------------------------------------
\bibitem{GellMann:1960np}
%--------------------
M.~Gell-Mann and M.~Levy,
%``The axial vector current in beta decay,''
Nuovo Cim.\  {\bf 16} (1960) 705.
%
%---------------------------------------------------------
\bibitem{Polyakov:1975rr}
%--------------------
A.~M.~Polyakov,
%``Interaction of Goldstone Particles in Two-Dimensions. 
%  Applications to Ferromagnets and Massive Yang-Mills Fields,''
Phys.\ Lett.\  {\bf 59B} (1975) 79.
%
%---------------------------------------------------------
\bibitem{Migdal:1975zf}
%--------------------
A.~A.~Migdal,
%``Gauge Transitions in Gauge and Spin Lattice Systems,''
Sov.\ Phys.\ JETP {\bf 42} (1975) 743. 
%
%---------------------------------------------------------
\bibitem{Brezin_Zinn-Justin}
%--------------------
E.~Br\'ezin and J.~Zinn-Justin,
%``Renormalization of the nonlinear sigma model 
%  in 2 + epsilon dimensions. Application to the Heisenberg ferromagnets,''
Phys.\ Rev.\ Lett.\  {\bf 36} (1976) 691;\ 
%
%--------------------
%``Spontaneous Breakdown of Continuous Symmetries Near Two-Dimensions,''
Phys.\ Rev.\ B {\bf 14} (1976) 3110.
%
%---------------------------------------------------------
\bibitem{Brezin:1976ap}
%--------------------
E.~Br\'ezin, J.~Zinn-Justin and J.~C.~Le Guillou,
%``Renormalization of the Nonlinear Sigma Model 
%  in (Two + Epsilon) Dimension,''
Phys.\ Rev.\ D {\bf 14} (1976) 2615.
%
%---------------------------------------------------------
\bibitem{tHooft:1973mfk}
%--------------------
G.~'t Hooft,
%``Dimensional regularization and the renormalization group,''
Nucl.\ Phys.\ B {\bf 61} (1973) 455.
%
%---------------------------------------------------------
\bibitem{Mermin:1966fe}
%--------------------
N.~D.~Mermin and H.~Wagner,
%``Absence of ferromagnetism or antiferromagnetism 
%  in one-dimensional or two-dimensional isotropic Heisenberg models,''
Phys.\ Rev.\ Lett.\  {\bf 17} (1966) 1133.
%
%---------------------------------------------------------
\bibitem{Elitzur:1978ww}
%--------------------
S.~Elitzur,
%``The Applicability of Perturbation Expansion 
%  to Two-dimensional Goldstone Systems,''
Nucl.\ Phys.\ B {\bf 212} (1983) 501.
%
%---------------------------------------------------------
\bibitem{David:1980rr}
%--------------------
F.~David,
%``Cancellations of Infrared Divergences 
%  in the Two-dimensional Nonlinear Sigma Models,''
Commun.\ Math.\ Phys.\  {\bf 81} (1981) 149.
%
%---------------------------------------------------------
\bibitem{Luscher:1982uv}
%--------------------
M.~L\"uscher,
%``A New Method to Compute the Spectrum of Low Lying States 
%  in Massless Asymptotically Free Field Theories,''
Phys.\ Lett.\  {\bf 118B} (1982) 391.
%
%---------------------------------------------------------
\bibitem{Hasenfratz:1984jk}
%--------------------
P.~Hasenfratz,
%``Perturbation Theory and Zero Modes in O($N$) Lattice $\sigma$ Models,''
Phys.\ Lett.\  {\bf 141B} (1984) 385.
%
%---------------------------------------------------------
\bibitem{Floratos:1984bz}
%--------------------
E.~G.~Floratos and D.~Petcher,
%``A Two Loop Calculation of the Mass Gap 
%  for the 0($N$) Model in Finite Volume,''
Nucl.\ Phys.\ B {\bf 252} (1985) 689.
%
%---------------------------------------------------------
\bibitem{Brezin:1985xx}
%--------------------
E.~Br\'ezin and J.~Zinn-Justin,
%``Finite Size Effects in Phase Transitions,''
Nucl.\ Phys.\ B {\bf 257} (1985) 867.
%
%---------------------------------------------------------
\bibitem{Shin:1996gi}
%--------------------
D.~S.~Shin,
%``A Determination of the mass gap in the O(n) sigma model,''
Nucl.\ Phys.\ B {\bf 496} (1997) 408.
%
%---------------------------------------------------------
\bibitem{Niedermayer:2010mx}
%--------------------
F.~Niedermayer and C.~Weiermann,
%``The rotator spectrum in the delta-regime 
%  of the O(n) effective field theory in 3 and 4 dimensions,''
Nucl.\ Phys.\ B {\bf 842} (2011) 248. 
%
%---------------------------------------------------------
\bibitem{Niedermayer:2016yll}
%--------------------
F.~Niedermayer and P.~Weisz,
%``Matching effective chiral Lagrangians with dimensional 
%  and lattice regularizations,''
JHEP {\bf 1604} (2016) 110. 
%
%---------------------------------------------------------
\bibitem{Niedermayer:2016ilf}
%--------------------
F.~Niedermayer and P.~Weisz,
%``Massless sunset diagrams in finite asymmetric volumes,''
JHEP {\bf 1606} (2016) 102. 
%
%---------------------------------------------------------
\bibitem{Capitani:1998mq}
%--------------------
S.~Capitani, M.~L\"uscher, R.~Sommer and H.~Wittig,
%``Non-perturbative quark mass renormalization in quenched lattice QCD,''
Nucl.\ Phys.\ B {\bf 544} (1999) 669, 
Erratum: [Nucl.\ Phys.\ B {\bf 582} (2000) 762]
%
%---------------------------------------------------------
\bibitem{forthcoming}
%--------------------
S.~Calle~Jimenez, M.~Oka and K.~Sasaki, 
in progress. 
%
%---------------------------------------------------------
\bibitem{Hasenfratz:1989pk}
%--------------------
P.~Hasenfratz and H.~Leutwyler,
%``Goldstone Boson Related Finite Size Effects 
%  in Field Theory and Critical Phenomena With O($N$) Symmetry,''
Nucl.\ Phys.\ B {\bf 343} (1990) 241.
%
%---------------------------------------------------------
\bibitem{ZJ:Book}
%--------------------
J.~Zinn-Justin,
%
Quantum Field Theory and Critical Phenomena, Fourth Edition 
(Clarendon Press, Oxford, 2002). 
%
%---------------------------------------------------------
\bibitem{Hikami:1977vr}
%--------------------
S.~Hikami and E.~Br\'ezin,
%``Three Loop Calculations in the Two-Dimensional Nonlinear Sigma Model,''
J.\ Phys.\ A {\bf 11} (1978) 1141.
%
%---------------------------------------------------------
%
\end{thebibliography}
\end  {document}